\documentclass[onecolumn]{IEEEtran} % letterpaper is the default
\IEEEoverridecommandlockouts
\usepackage{geometry}
\usepackage[utf8]{inputenc}   % omit with LuaLaTeX/XeLaTeX
\usepackage[T1]{fontenc}      % better hyphenation/fonts
\usepackage[english]{babel}

\usepackage{amsmath,amssymb,amsthm,bm}
\usepackage{mathtools}
\usepackage{booktabs}
\usepackage{array,tabularx}
\usepackage{rotating}         % for sidewaystable
\usepackage{enumitem}
\usepackage[acronym]{glossaries}
\usepackage{algorithm}
\usepackage[noend]{algpseudocode}
\usepackage{dsfont}
\usepackage{makecell}
\usepackage[dvipsnames]{xcolor}
\usepackage{mathrsfs} % enables \mathscr
\usepackage{cite}

\usepackage{nomencl}
\renewcommand{\baselinestretch}{1.5}

\usepackage{hyperref}
\hypersetup{colorlinks=true,linkcolor=blue,citecolor=blue,urlcolor=blue}

\usepackage{tocloft}
\AtBeginDocument{%
  \algrenewcommand\algorithmicrequire{\textbf{Input:}}%
  \algrenewcommand\algorithmicensure{\textbf{Solve:}}%
  \let\Input\Require
  \let\Solve\Ensure
}

\makeatletter
\providecommand{\subparagraph}{\@startsection{subparagraph}{6}{\parindent}%
  {3.25ex \@plus1ex \@minus.2ex}%
  {-1em}%
  {\normalfont\normalsize\bfseries}}
\makeatother

\newacronym{5g}{5G}{Fifth-Generation Wireless Systems}

\newacronym{ai}{AI}{Artificial Intelligence}
\newacronym{ann}{ANN}{Artificial Neural Network}
\newacronym{ask}{ASK}{amplitude-shift keying}
\newacronym[plural={AWGN}, \glsshortpluralkey={AWGN}]{awgn}{AWGN}{additive white Gaussian noise}

\newacronym{bbcs}{BBCs}{bad bit-channels}
\newacronym{bch}{BCH}{Bose-Chaudhuri-Hocquenghem}
\newacronym{bec}{BEC}{binary erasure channel}
\newacronym{ber}{BER}{bit error rate}
\newacronym{bimc}{BIMSC}{binary input memoryless symmetric channel}
\newacronym{bler}{BLER}{block error rate}
\newacronym{bpsk}{BPSK}{binary phase-shift keying}
\newacronym{bp}{BP}{belief propagation}
\newacronym{bsc}{BSC}{binary symmetric channel}
\newacronym{cdf}{CDF}{cumulative distribution function}
\newacronym{club}{CLUB}{Contrastive Log-ratio Upper Bound}
\newacronym{clt}{CLT}{central limit theorem}
\newacronym{cnbmm}{CNBMM}{Conditional Neural Bernoulli-Mixture Model}

\newacronym{cnn}{CNN}{convolutional neural network}
\newacronym{csi}{CSI}{channel state information}

\newacronym{dmc}{DMC}{discrete memoryless channel}
\newacronym{dp}{DP}{dynamic programming}

\newacronym{ecc}{ECC}{error-correcting code}
\newacronym{ecdf}{ECDF}{empirical distribution function}
\newacronym{ee}{EE}{energy efficiency}
\newacronym{epi}{EPI}{entropy power inequality}
\newacronym{evd}{EVD}{eigenvalue decomposition}

\newacronym{fbl}{FBL}{finite blocklength}
\newacronym{ff}{FF}{feed-forward}
\newacronym{ffnn}{FF-NN}{feed-forward neural network}
\newacronym{fsk}{FSK}{frequency-shift keying}

\newacronym{gbcs}{GBCs}{good bit-channels}
\newacronym{gec}{GEC}{generalized erasure channel }

\newacronym{gm}{GM}{Gaussian mixture}
\newacronym{gwtc}{GWTC}{Gaussian wiretap channel}

\newacronym{iid}{i.i.d.}{independent and identically distributed}
\newacronym{infonce}{ InfoNCE}{Information Noise-Contrastive Estimation}
\newacronym{ir}{IR}{information reconciliation}
\newacronym{il}{IL}{information leakage}

\newacronym{iot}{IoT}{internet of things}

\newacronym{kld}{KL divergence}{Kullback-Leibler divergence}

\newacronym{lb}{LB}{lower bound}
\newacronym{lcm}{lcm}{least common multiple}

\newacronym{ldpc}{LDPC}{low density parity check}
\newacronym{lhl}{LHL}{leftover hash lemma}
\newacronym{lhs}{LHS}{left hand side}
\newacronym{llr}{LLR}{log likelihood ratio}
\newacronym{los}{LoS}{line-of-sight}
\newacronym{lvs}{LVS}{latent variable secrecy}

\newacronym{mac}{MAC}{multiple access channel}
\newacronym{map}{MAP}{maximum a posteriori}
\newacronym{mwis}{MWIS}{maximum-weight independent set}

\newacronym{mc}{MC}{Monte Carlo}
\newacronym{mdl}{MDL}{mode-dependent loss}
\newacronym{mi}{M.I.}{mutual information}
\newacronym{milp}{MILP}{mixed-integer linear programming}
\newacronym{mimo}{MIMO}{multiple-input multiple-output}
\newacronym{mimome}{MIMOME}{multiple-input multiple-output and multiple antennas at eavesdropper}
\newacronym{mine}{MINE}{mutual information neural estimator}
\newacronym{miso}{MISO}{multiple-input single-output}
\newacronym{ml}{ML}{machine learning}
\newacronym{mlp}{MLP}{multi-layer perceptron}
\newacronym{mmf}{MMF}{multi-mode fiber}
\newacronym{mmwave}{mmWave}{millimeter wave}
\newacronym{mop}{MOP}{multi-objective programming problem}
\newacronym{mcm}{MCM}{Monte-Carlo method}
\newacronym{mrc}{MRC}{maximum ratio combining}
\newacronym{msc}{MSC}{modulation and coding scheme}
\newacronym{mse}{MSE}{mean squared error}
\newacronym{mvb}{MVB}{multivariate Bernoulli }

\newacronym{nlos}{NLoS}{non-line-of-sight}
\newacronym{nn}{NN}{neural network}
\newacronym{nve}{NVE}{normalized validation error}
\newacronym{nwj}{NWJ}{Nguyen-Wainwright-Jordan bound}

\newacronym{ot}{OT}{Oblivious transfer}

\newacronym{pdf}{PDF}{probability density function}
\newacronym{pls}{PLS}{Physical layer security}
\newacronym{pmf}{PMF}{probability mass function}
\newacronym{poset}{poset}{partial order set}
\newacronym{pwc}{PWC}{polar wiretap code}

\newacronym{qam}{QAM}{quadrature amplitude modulation}

\newacronym{ra}{RA}{rearrangement algorithm}
\newacronym{rbf}{RBF}{radial basis function}
\newacronym{relu}{ReLU}{rectified linear unit}
\newacronym{rhs}{RHS}{right hand side}
\newacronym{ris}{RIS}{reconfigurable intelligent surface}
\newacronym{rnn}{RNN}{recurrent neural network}
\newacronym{scd}{SCD}{successive cancellation deocder}
\newacronym{sdm}{SDM}{space-division multiplexing}
\newacronym{sfa}{SfA}{Secrecy for Alice}
\newacronym{sfb}{SfB}{Secrecy for Bob}
\newacronym{sgd}{SGD}{stochastic gradient descent}
\newacronym{sgp}{SGP}{secure goodput}
\newacronym{sic}{SIC}{successive interference cancellation}
\newacronym{simo}{SIMO}{single-input multiple-output}
\newacronym{sinr}{SINR}{signal-to-interference-plus-noise ratio}
\newacronym{siso}{SISO}{single-input single-output}
\newacronym{skg}{SKG}{secret key generation}
\newacronym{snr}{SNR}{signal-to-noise ratio}
\newacronym{sop}{SOP}{secrecy outage probability}
\newacronym{svm}{SVM}{support vector machine}
\newacronym{swc}{SWC}{Slepian-Wolf coding}

\newacronym{tvd}{TVD}{total variation distance}
\newacronym{tuba}{TUBA}{total Correlation Upper Bound Approximation}

\newacronym{uav}{UAV}{unmanned aerial vehicle}
\newacronym{ub}{UB}{upper bound}
\newacronym{uhf}{UHF}{universal hash families}
\newacronym{upo}{UPO}{universal partial order}
\newacronym{urllc}{URLLC}{ultra-reliable low latency communication}

\newacronym{vclub}{vCLUB}{variational CLUB}
\newacronym{vd}{VD}{variational distance}

\newacronym{wtc}{WTC}{wiretap channel}
\newacronym{wrt}{w.r.t.}{with respect to}
\newacronym{zoc}{ZOC}{zero-outage capacity}
\newacronym{zosc}{ZOSC}{zero-outage secrecy capacity}
\newacronym{zosr}{ZOSR}{zero-outage secrecy rate}
\usepackage{amsmath,amssymb,amsfonts,amsthm,dsfont}
\usepackage{babel}
\usepackage{graphicx}
\usepackage{textcomp}
\usepackage{xcolor}
\def\BibTeX{{\rm B\kern-.05em{\sc i\kern-.025em b}\kern-.08em
    T\kern-.1667em\lower.7ex\hbox{E}\kern-.125emX}}

\usepackage{tikz}
\usetikzlibrary{arrows.meta, positioning, shapes.geometric, calc}
\usepackage{tabularx}\label{Table_comparison_use_of_polar_codes}
\usepackage{pgfplots}
\usepackage{pgfplotstable} % For loading CSV files
\pgfplotsset{compat=1.18}

\newcommand{\Input}{\Require}
\newcommand{\Solve}{\Ensure}

\definecolor{Red}{RGB}{190,30,60}
\definecolor{Black}{RGB}{0,0,0}
\definecolor{Gray}{RGB}{128,128,128}
\definecolor{LightGray}{RGB}{160,160,160}
\definecolor{White}{RGB}{255,255,255}

\definecolor{LightYellowOrange}{RGB}{255,200,42}
\definecolor{OrangeBright}{RGB}{225,109,0}
\definecolor{DarkBordeaux}{RGB}{113,28,47}

\definecolor{SkyBlue}{RGB}{102,180,211}
\definecolor{BlueLight}{RGB}{0,112,155}
\definecolor{BlueDark}{RGB}{0,63,87}

\definecolor{GreenLight}{RGB}{172,193,58}
\definecolor{GreenMid}{RGB}{109,131,0}
\definecolor{GreenDark}{RGB}{0,83,74}

\definecolor{VioletLight}{RGB}{138,48,127}
\definecolor{VioletMid}{RGB}{81,18,70}
\definecolor{VioletDark}{RGB}{76,24,48}

\usepackage{tcolorbox}

\newcounter{boxcounterR} % Define a new counter
\newcounter{boxcounterG} % Define a new counter
\newcounter{boxcounterB} % Define a new counter

\newcommand{\Var}{\operatorname{Var}}

\newcommand{\bA}{\mathbf A}
\newcommand{\bP}{\mathbf P}

\newcommand{\bF}{\mathbf F}
\newcommand{\bI}{\mathbf I}
\newcommand{\bQ}{\mathbf Q}

\newcommand{\bS}{\mathbf S}
\newcommand{\bU}{\mathbf U}

\newcommand{\prefixnumber}[1]{\renewcommand{\theALG@line}{#1.\arabic{ALG@line}}}
\newcommand{\resetnumbering}{\renewcommand{\theALG@line}{\arabic{ALG@line}}}
\makeatother
\DeclareMathOperator{\Aut}{Aut}

\theoremstyle{boldremark}
\newtheorem{remark}{Remark}
\newtheorem{corollary}{Corollary}

\DeclareMathOperator{\supp}{supp}

\newtheorem{thm}{Theorem}
\newtheorem{lemma}{Lemma}
\newtheorem{definition}{Definition}
\newtheorem{proposition}{Proposition}
\newtheorem{example}{Example}
\newcommand{\bT}{\mathbf{T}}

\DeclareMathOperator{\Unif}{Unif}
\newcommand{\g}[1]{\textcolor{ForestGreen}{\textbf{#1}}}
\newcommand{\y}[1]{\textcolor{orange}{\textbf{#1}}}

\newcommand{\be}{\mathbf e}
\newcommand{\pperp}{\mathrel{\perp\!\!\!\perp}}

\begin{document}

\title{Implementation of Oblivious Transfer over Binary-Input AWGN Channels by Polar Codes \\
% \thanks{The work of Eduard Axel Jorswieck was supported in part by the Federal Ministry of Education and
% Research (BMBF), Germany, through the Program of Souverän, Digital, and Vernetzt Joint Project 6G Research and Innovation Cluster (6G-RIC) under Grant 16KISK031 and the work of Pin-Hsun Lin was supported in part by BMBF-Projektes QuaPhySI under Grant 16KIS2234 and COST ACTION CA22168 6G-PHYSEC.}
%   \thanks{
%  \textcolor{blue}{Vague text, focusing on ideas/brainstorming. Missing citations.}
%  \textcolor{red}{Refined text, might be missing citations. Ready for partial internal review.}
%  \textcolor{darkgreen}{Further refined text with citations. Ready for final internal review.}
%  Final version. Will not be changed.
% }
}

\author{
\IEEEauthorblockN{
Pin-Hsun Lin\IEEEauthorrefmark{1},
Hadi Aghaee\IEEEauthorrefmark{1},
Christian Deppe\IEEEauthorrefmark{1},
Eduard A. Jorswieck\IEEEauthorrefmark{1},
Marcel Mross\IEEEauthorrefmark{1},
Holger Boche\IEEEauthorrefmark{2}
}\\
\IEEEauthorblockA{\IEEEauthorrefmark{1}Institute for Communications Technology, Technische Universit\"at Braunschweig, Braunschweig, Germany}
\IEEEauthorblockA{\IEEEauthorrefmark{2}Institute of Theoretical Information Technology, TUM School of Computation, Information and Technology, Technical University of Munich, Germany}
% optional (if you want emails):
% \IEEEauthorblockA{Email: \{p.lin, hadi.aghaee, christian.deppe, e.jorswieck\}@tu-braunschweig.de, holger.boche@tum.de}
}

\maketitle

\tableofcontents

\begin{abstract}
We develop a $\binom{2}{1}$--oblivious transfer (OT) protocol over the binary-input additive white Gaussian noise (BI–AWGN) channel using polar codes. The scheme uses two decoder views linked by automorphisms of the polar transform and publicly draws the encoder at random from the corresponding automorphism group. This yields perfect receiver privacy at any finite blocklength, since the public encoder distribution is independent of the receiver’s choice bit. Sender privacy is obtained asymptotically via channel polarization combined with privacy amplification. Because the construction deliberately injects randomness on selected bad bit-channels, we derive a relaxed reliability criterion and evaluate the finite-blocklength performance. Finally, we characterize the polar-transform automorphisms as bit-level permutations of bit-channel indices, and exploit this structure to derive and optimize an achievable finite-blocklength OT rate.
\end{abstract}

\section{Introduction}
Oblivious transfer (OT) is a fundamental building block for two-party secure computation: once $\binom{2}{1}$--OT is available, any polynomial-time functionality can be securely computed by standard composition, so OT sits neatly between low-level communication and high-level privacy services \cite{Kilian88,CrepeauKilian88}. In practice, OT powers private set intersection, private information retrieval, commit-and-reveal / contract-signing, and private learning and inference. It can be realized in two qualitatively different ways: (i) information-theoretically, from noisy communication resources without computational assumptions; and (ii) computationally, from a small public-key “seed’’ plus fast symmetric-key extensions. This dual nature makes OT central to both cryptography and communications \cite{Beaver95,NascimentoWinter2008}.

At the same time, OT cannot be constructed from scratch using arbitrary weak randomness, reflecting a central impossibility phenomenon for interactive cryptographic protocols based on indistinguishability---including bit commitment, zero-knowledge, secret sharing, and secure two-party computation---even against efficient adversaries. Dodis et al.\ show that OT cannot be realized using imperfect randomness from any weak entropy source, including slightly imperfect Santha--Vazirani (SV) sources \cite{DodisOngPrabhakaranSahai04,SanthaVazirani86}. Roughly, any two functions that produce computationally indistinguishable outputs from such sources must agree on almost all inputs, contradicting the inherent unpredictability required for OT. Thus, one must either assume stronger (e.g., cryptographic) primitives or exploit richer physical randomness such as channel noise.

Classical work defined and related several OT variants---Rabin’s OT \cite{Rabin81}, Even--Goldreich--Lempel’s $1$-out-of-$2$ OT, and Cr{\'e}peau’s equivalence between forms of OT \cite{EGL85,Crepeau87}---and Cr{\'e}peau--Kilian showed how to achieve OT from weakened assumptions \cite{CrepeauKilian88}. Beaver’s “random OT’’ cleanly separated the generation of an offline OT correlation from its online consumption \cite{Beaver95}. Moving into \emph{information-theoretic} OT, noisy-channel realizations appeared first: Cr{\'e}peau gave protocols from noisy channels \cite{Crepeau1997Noisy}, and efficient unconditional OT from (almost) any noisy channel was later obtained by Cr{\'e}peau, Morozov, and Wolf \cite{CrepeauMorozovWolf2005}. For specific channel models, Stebila--Wolf proved feasibility for nontrivial binary symmetric channels (BSCs) \cite{StebilaWolf2002ISIT}, and Imai, Morozov, and Nascimento studied the oblivious transfer capacity of erasure channels \cite{ImaiMorozovNascimento2006}. The landscape was further clarified by symmetry-based characterizations \cite{WolfWullschleger2006Symmetric} and models with weak/noisy assumptions \cite{Wullschleger2009TCC}.

Noisy channels provide a much richer entropy source than weak randomness: although their overall behavior can be characterized statistically, the exact noise pattern varies unpredictably across transmissions, making repeated outputs almost never identical. This insight motivated a capacity-oriented viewpoint. Ahlswede and Csisz{\'a}r initiated the study of OT capacity from noisy resources \cite{AhlswedeCsiszar2009}, while Nascimento and Winter gave a general formulation of OT capacity for noisy correlations, with both achievability and converse bounds \cite{NascimentoWinter2006,NascimentoWinter2008}. Their results were refined for generalized-erasure-type models and related settings \cite{AhlswedeCsiszar2013OTCap,ImaiMorozovNascimento2006}. Beyond point-to-point links, shared noisy channels---where multiple users interact with a common medium---remain comparatively less explored, but they offer correlated noise that can enhance privacy, reduce communication cost, and support joint encoding/decoding, making them promising for extending OT to broadcast and multiple-access scenarios. In the context of symmetric private information retrieval (SPIR), this line of research has been studied under the OT framework, where a noisy channel between the parties is leveraged to achieve information-theoretic security; see, for example, \cite{amir1, amir2, amir3}. These works also develop achievability schemes that exploit physical randomness to induce information asymmetry between the parties.

More recently, these information-theoretic limits have been revisited with practical, coding-based protocols. Oggier and Z{\'e}mor derive an explicit coding-theoretic OT protocol over binary symmetric channels that is secure against malicious behavior \cite{OggierZemor22}. Their constructions achieve a concrete positive rate using linear codes whose Schur squares are asymptotically good, relying on standard building blocks such as polar codes for reliability. A recent \gls{gec}-based OT protocol construction \cite{SudaWatanabe25} starts from discrete channels with finite alphabets and emulates generalized
erasures via alphabet extension and subspace/complement labeling induced by linear constraints from the polar
transform. In contrast, we consider BI--\gls{awgn} and use genuine polarization into \gls{gbcs} $\mathcal{G}$ and \gls{bbcs} $\mathcal{B}$. This yields an erasure-like abstraction and an automorphism-driven two-view design with a computable finite-$n$ OT-rate criterion.
Moreover, the bit-channel mutual information translates the secrecy constraints \gls{sfb} and \gls{sfa} into explicit index-set constraints and yields a structured finite-blocklength optimization for OT-rate maximization. 

Our main contributions are summarized as follows.
\begin{itemize}
  \item[(1)] We introduce a polar-code-based \gls{ot} framework that views the
  underlying binary-input \gls{awgn} channel through a \emph{virtual} \gls{bec} way:
  indices in the good set $\mathcal{G}$ play the role of reliable and
  non-erased positions for Bob's chosen message; indices in the
  bad set $\mathcal{B}$ are used as positions that should convey essentially
  no information about the unchosen message.
  This \gls{bec} emulation is defined in an information-theoretic sense and is
  tailored specifically to the analysis of \gls{ot}. Note that our emulation of \gls{bec} is essentially different from the alphabet extension/GEC (labeling induced by linear constraints from the polar
  transform) as \cite{SudaWatanabe25}.

  \item[(2)] To mitigate leakage from the publicly shared information, we introduce a special set of permutations---\textit{automorphisms} $\Aut(\bT)$ of the
  polar transform $\bT$---to generate different ``views'' of $\mathcal{G}$ and
  $\mathcal{B}$ at Alice and Bob. 
  We further trade a controlled amount of reliability for a more symmetric
  virtual \gls{bec} by letting a carefully selected small subset of \gls{bbcs} carry independent random bits unknown to Bob. From Bob's viewpoint, these bits behave as virtual erasures for the undesired message.
  Our construction achieves the desired \gls{bec}-like structure using only
  polarization and permutations from $\Aut(\bT)$.

  \item[(3)] We provide an information-theoretic security analysis of the
  resulting protocol, proving both \gls{sfa} and \gls{sfb}. We also introduce a relaxed reliability constraint that reflects the nonstandard two-view use of polar codes.
  To the best of our knowledge, this is the first polar-code-based \gls{ot} protocol for a binary-input \gls{awgn} channel that leverages automorphisms.

  \item[(4)] To make the permutation step in our OT construction explicit, efficiently
  implementable, and analytically tractable, we provide a complete
  characterization of the automorphism group of the polar transform
  $\bT=\bT_0^{\otimes m}$, where $\bT_0$ is Arikan's 2-by-2 fundamental polarization matrix. Concretely, we show that every $\mathbf P\in\Aut(\bT)$ is
  induced by a unique permutation of the $m$ bit positions and $|\Aut(\bT)|=m!$.
  This characterization yields two practical benefits: (i) it provides a
  complete search space of permissible permutations for hiding the
  \gls{gbcs} and \gls{bbcs} structure without breaking the polar transform and, (ii) it enables
  uniform sampling and enumeration of automorphisms rather than relying on ad-hoc
  permutations.

 \item[(5)] We develop an explicit finite-blocklength OT-rate optimization
  framework that jointly selects (i) the permutation from $\Aut(\bT)$
  and (ii) paired index sets $(\mathcal G,\mathcal B)$, to maximize the OT payload under
  finite-$n$, leakage, and reliability constraints.
  Using Gaussian-approximation (GA) recursion to compute bit-channel mutual
  informations at the operating SNR, we formulate a discrete optimization over
  $\sigma$ and $\mathcal G$, and show that for each fixed $\sigma$ the inner
  problem admits a closed-form max-$k$ selection rule,
  yielding a low complexity procedure for OT-rate evaluation and design.
\end{itemize}

Compared with the constructions in \cite{OggierZemor22} and \cite{SudaWatanabe25}, our scheme uses polarization as the core mechanism. In \cite{OggierZemor22}, polar codes are used only as a good BSC code to reach reliability. In contrast, the \gls{sfa} and \gls{sfb} are provided by the Schur-square construction and privacy amplification, not by polarization. Besides, \cite{SudaWatanabe25} uses the parity check space of the polarization matrix to categorize the labels of alphabet extension. The Kronecker-product structure of the polarization gives a systematic, recursive parity space, which makes the GEC construction efficient. However, both of them do not use the intrinsic properties of \gls{gbcs} and \gls{bbcs} in polar codes. In contrast, we explicitly exploit the $\mathcal{G}/\mathcal{B}$ split and $\Aut(\mathbf{T})$ to emulate erasures, symmetrize Bob’s choice, and drive both reliability and secrecy on the same polarized index set. Furthermore, our construction targets general BIMCs (in particular BI--AWGN), avoids alphabet extension.

% In preamble

% \usepackage{dblfloatfix} % (optional) lets * floats appear at bottom too
% \usepackage{placeins}    % (optional) \FloatBarrier to force float placement

% Define these near your content (outside the caption)
\newcommand{\oz}{Oggier--Z{\'e}mor~\cite{OggierZemor22}}
\newcommand{\sw}{Suda--Watanabe~\cite{SudaWatanabe25}}
\newcommand{\ours}{This work}

\section{Preliminaries and system model}
\subsection{Notation}
Let $[n]:=\{1,\dots,n\}$. Define $x^n:=(x_1,\dots,x_n)$.
We use capital letters with normal font for random variables and lower-case
letters for their realizations. We use capital letters in sans-serif font as deterministic variables to be distinguished from random variables. We write $A\overset{d}{=}B$ to denote that $A$ and $B$ have the same distribution. Statistical independence between random variables $A$ and $B$ is denoted by $A\pperp B$. Sets are denoted by calligraphic letters. Let $\be_j\in\{0,1\}^n$ denote the $j$-th standard basis vector.
For any bijection $\sigma:[n]\to[n]$, the permutation matrix
$\bP_{\sigma}\in\{0,1\}^{n\times n}$ is defined by
$\bP_{\sigma}\be_j=\be_{\sigma(j)},\, j\in[n],$ equivalently, $(\bP_{\sigma})_{i,j}=1$ if and only if $i=\sigma(j)$. Define lcm(.) as the least common multiple operator. We use $u^n|_{\mathcal I}$ and $u^n_{\mathcal I}$ interchangeably to denote the subvector of $u^n$ obtained by selecting the components with indices in $\mathcal I\subseteq[n]$. We denote a truncation by $(\cdot)_{\downarrow \ell}$, which keeps $\ell$ entries.

A binary-input memoryless channel (BIMC) is denoted by $W:\{0,1\}\to\mathcal{Y}$,
and its $n$-fold extension is $W^n(y^n\mid x^n)=\prod_{i=1}^n W(y_i\mid x_i),\; x_i\in\{0,1\},\ y_i\in\mathcal{Y}.$ We define the mutual information of $W$ under uniform input as $I(W) := I(X;Y)$ for $X\sim \mathrm{Bern}\!\left(\tfrac12\right)$, $Y\sim W(\cdot\mid X)$. When the channel output $Y$ has conditional densities $f_{Y|X}(\cdot\mid x)$, the Bhattacharyya parameter is defined as $Z(W):=\int_{\mathcal{Y}}\sqrt{f_{Y|X}(y\mid 0)\,f_{Y|X}(y\mid 1)}\,dy$. 
For two distributions $P$ and $Q$ on the same alphabet, we write
\begin{align}\label{EQ_TVD}
    d_{\mathrm{var}}(P,Q)
    := \sup_{\mathcal A} |P(\mathcal A)-Q(\mathcal A)|
    =
    \begin{cases}
    \tfrac12\displaystyle\sum_{a\in\mathcal{A}} \bigl|P(a)-Q(a)\bigr|,
    & \text{discrete alphabet $\mathcal{A}$},\\[1.2em]
    \tfrac12\displaystyle\int_{\mathbb{R}^d} \bigl|p(y)-q(y)\bigr|\,dy,
    & \text{continuous alphabet $\mathcal{A}=\mathbb{R}^d$, $p\mbox{ and }q$ are the densities}.
    \end{cases}    
\end{align}

We split the indices into a \emph{good} set (a set of \gls{gbcs}) and a \emph{bad} set (a set of \gls{bbcs}) according to the intrinsic qualities of the polar bit-channels induced by the underlying binary-input memoryless channel $W$ and the fixed polar transform $\mathbf T$.
For this purpose, introduce an \emph{auxiliary test vector} $\bar U^n\sim\mathrm{Bern}(\tfrac12)^n$ and let $\bar X^n:=\bar U^n\mathbf T$, $\bar Y^n$ be the corresponding channel output over $W$.
For each $i\in[n]$, define the $i$-th bit-channel mutual information (i.e., the symmetric capacity of $W_n^{(i)}$) as
\[
I_i \;:=\; I\!\bigl(\bar U_i;\,\bar Y^n \,\big|\, \bar U^{i-1}\bigr),\qquad i\in[n].
\]
Given a threshold $\gamma_n\in[0,1]$ that vanishes with $n$, define
\begin{align}\label{EQ_GBC_BBC_def}
    \mathcal{I}_{\mathcal G}(\gamma_n)
    := \bigl\{\,i\in[n]:\ I_i \ge 1-\gamma_n\,\bigr\},
    \qquad
    \mathcal{I}_{\mathcal B}(\gamma_n)
    := \bigl\{\,i\in[n]:\ I_i \le \gamma_n\,\bigr\}.    
\end{align}

Note that the actual encoder input $U^n$ is not i.i.d.\ over $[n]$, but selects an information set $\mathcal A\subseteq \mathcal{I}_{\mathcal G}(\gamma_n)$ and freezes $\mathcal A^c$ (typically to zeros).

In the following, we introduce the \gls{upo} \cite{UPO_Schurch_ISIT16,UPO_He_GC17}, to better understand how much information Alice can know about the \gls{gbcs} and \gls{bbcs}, when Bob shares only a permuted polarization matrix to her, but not together with the channel distribution, which affects \gls{sfa} and \gls{sfb} by the proposed scheme.

% === Universal Partial Order (UPO) for polar codes ===
\begin{definition}
Let $n=2^m$. Index the polarized bit–channels by $i\in\{0,\dots,n-1\}$ and denote the $i$-th polarized bit–channel by $W_{n}^{(i)}$.
Let the binary label as $\widehat{i}=(i_0,i_1,\dots,i_{m-1})$ with the \emph{least significant bit on the left}. The operator $\preceq$ on indices is defined as follows:
\begin{align}\label{EQ_UPO_marjorization}
i \preceq j
\,\Longleftrightarrow\,
\sum_{t=0}^{r} i_t \le \sum_{t=0}^{r} j_t \;\; \text{for all } r=0,\dots,m-1.    
\end{align}
\end{definition}

% Note that \eqref{EQ_UPO_marjorization} is different from majorization, i.e., the condition $\sum_{t=0}^{m-1} i_t = \sum_{t=0}^{m-1}$ is not considered in \gls{upo}, which is stated as follows.

\begin{thm}[ \cite{UPO_Schurch_ISIT16}]
If $i \preceq j$, then for every \gls{bimc} channel $W$ and $n=2^m$, $I\!\big(W_{n}^{(i)}\big) \le I\!\big(W_{n}^{(j)}\big)
\text{ and }
Z\!\big(W_{n}^{(i)}\big) \ge Z\!\big(W_{n}^{(j)}\big).$
\end{thm}

Note that in the proposed protocol, we assume that Bob does not share the channel distribution but shares a permuted polarization matrix and a permuted $\mathcal{I}_{\mathcal G}$ and $\mathcal{I}_{\mathcal B}$. Even without knowing the channel law, \gls{upo} allows Alice to determine the reliability ordering of the synthesized bit-channels. Therefore, to achieve \gls{sfa} and \gls{sfb}, we further randomize the construction by applying additional permutations from automorphisms to the polarization matrix and to the index sets $\mathcal{I}_{\mathcal G}$ and $\mathcal{I}_{\mathcal B}$.

\subsection{Permutation and automorphism}\label{Sec_Permute_Aut}
Our proposed scheme relies on a subset of permutation matrices, namely, automorphism of the polarization matrix, defined as follows.

\begin{definition}\label{def:symmetric-group}
Let \(n\in\mathbb N\). The \emph{symmetric group} \(\mathcal{S}_n\) is a set of all bijections
\(\sigma:[n]\to[n]\).
For \(\sigma,\tau\in\mathcal{S}_n\), their composition is the bijection
\(\sigma\tau\in\mathcal{S}_n\) defined by $(\sigma\tau)(i):=\sigma(\tau(i)),\, i\in[n].$
\end{definition}

\begin{definition}[Automorphism]\label{Def_Aut}
For a polarization matrix $\mathbf{T}\in\mathds{F}_2^{n\times n}$, define its automorphism group as
\[
\Aut(\mathbf{T})
:=\Bigl\{\mathbf{P}\in\mathds{F}_2^{n\times n}:\ \mathbf{P}\ \text{is a permutation matrix and}\ \mathbf{P}^{\!\top}\mathbf{T}\mathbf{P}=\mathbf{T}\Bigr\}.
\]
\end{definition}
Equivalently, $\mathbf{P}\in\Aut(\mathbf{T})$ if and only if $\mathbf{P}=\mathbf{P}_\sigma$ for some $\sigma\in\mathcal S_n$ satisfying $\mathbf{P}_{\sigma}^{\!\top}\mathbf{T}\mathbf{P}_{\sigma}=\mathbf{T}$.
When convenient, we abuse notation and write $\sigma\in\Aut(\mathbf T)$ to mean that
$\mathbf P_\sigma\in\Aut(\mathbf T)$, and we use $\sigma(i)$ to denote the induced action on indices.

\begin{definition}[Cross-cut]\label{def:cross-cut}
Fix $\textsf{SNR}$ and let
$\mathcal{G}_{\rm sel}(\bT;\textsf{SNR})\subseteq[n]$ and
$\mathcal{B}_{\rm sel}(\bT;\textsf{SNR})\subseteq[n]$
denote the selected \gls{gbcs} and \gls{bbcs} under the polarization matrix $\bT$. For $\sigma\in\Aut(\bT)$, we say that \emph{$i$ is cross-cut paired under $\sigma$}
if
\[
i\in\mathcal{G}_{\rm sel}(\bT;\textsf{SNR})
\quad\text{and}\quad
\sigma(i)\in\mathcal{B}_{\rm sel}(\bT;\textsf{SNR}).
\]
A selected set $\mathcal{G}\subseteq[n]$ satisfies the \emph{cross-cut constraint} if
\[
\mathcal{G}\subseteq\mathcal{G}_{\rm sel}(\bT;\textsf{SNR})
\quad\text{and}\quad
\sigma(\mathcal{G})\subseteq\mathcal{B}_{\rm sel}(\bT;\textsf{SNR}).
\]
\end{definition}

Our later arguments rely on algebraic manipulation of permutation matrices, so we
recall a basic property.

\begin{lemma}\label{lem:aut-inverse}
Let $\bP\in\mathds{F}_2^{n\times n}$ be a permutation matrix. Then $\bP^\top=\bP^{-1}$.
Moreover, if $\bP\in\Aut(\bT)$, then $\bP^\top\in\Aut(\bT)$.
\end{lemma}

% \begin{proof}
% For $i,k\in[n]$,
% \[
%  (\bP^\top \bP)_{ik}=\sum_{j=1}^n \bP_{ji}\bP_{jk}.
% \]
% If $i\neq k$, each row $j$ has at most one entry equal to $1$, hence $\bP_{ji}\bP_{jk}=0$ for all $j$, so $(\bP^\top \bP)_{ik}=0$.
% If $i=k$, there is exactly one row $j^\ast$ with $\bP_{j^\ast i}=1$, so $(\bP^\top \bP)_{ii}=1$.
% Thus $\bP^\top \bP=\bI_n$, i.e., $\bP$ is orthogonal, and therefore $\bP^{-1}=\bP^\top$.
% \end{proof}

% \begin{proof}
% Since $\bP\in\Aut(\bT)$, $\bP^\top\bT\bP=\bT$, left–multiply by $\bP$ and right–multiply by $\bP^\top$ to get
% $\bT=\bP \bT\bP^\top$, which is equivalent to $\bT(\bP^\top)^\top \bT\bP^\top=\bT$, i.e., $\bP \bT \bP^\top=\bT$.
% \end{proof}

\begin{definition}[Partially ordered set]\label{Def_POSET}
Let $\mathcal{P}$ be a nonempty set. We say that $\le$ is a \emph{partial order} on $\mathcal{P}$
if, for all $x,y,z\in\mathcal{P}$, the following properties hold:
\begin{enumerate}
  \item {Reflexivity:} $x \le x$.
  \item {Antisymmetry:} if $x \le y$ and $y \le x$, then $x = y$.
  \item {Transitivity:} if $x \le y$ and $y \le z$, then $x \le z$.
\end{enumerate}
In this case, the pair $(\mathcal{P},\le)$ is called a partially ordered set
(or \emph{poset}).
\end{definition}
In this work we use the poset $(\mathcal{X},\le)$ with
$\mathcal{X} := \{0,1\}^m$, where $\le$ is the bit-wise order
$x\le y \mbox{ if and only if } x_i \le y_i$ for all $i\in[m]$.

% \begin{definition}[Cycles and cycle decomposition]
% Let $[n]\triangleq\{1,\dots,n\}$. 
% A \emph{$k$-cycle} on $[n]$ is a permutation $C:[n]\to[n]$ for which there exist
% pairwise distinct indices $a_1,\dots,a_k\in[n]$ such that
% \[
%   C(a_j)=a_{j+1}\quad\text{for } j=1,\dots,k-1,
%   \qquad
%   C(a_k)=a_1,
% \]
% and $ C(i)=i \quad \text{for all } i\in [n]\setminus\{a_1,\dots,a_k\},$ i.e., no permutation.
% We denote such a permutation in cycle notation by $C = (a_1\,a_2\,\dots\,a_k),$
% and call $k$ the \emph{length} of the cycle.
% Every permutation $\pi\in S_n$ can be written (uniquely up to the order of the
% factors) as a product of pairwise disjoint cycles $\pi = C_1 C_2 \cdots C_r,$
% where each $C_i$ is a cycle in the above sense.  If $C_i$ moves the indices
% $a_{i,1},\dots,a_{i,\ell_i}$, we write $  C_i = (a_{i,1}\,a_{i,2}\,\dots\,a_{i,\ell_i}),$
% and call $\ell_i$ the \emph{length} of $C_i$.  Since the cycles are disjoint,
% powers of $\pi$ act on each cycle separately:
% $ \pi^t = C_1^{\,t} C_2^{\,t} \cdots C_r^{\,t}  \;\text{for all } t\in\mathbb N.$
% For each $i$ we have $C_i^{\,t}=\mathrm{id}$ if and only if $\ell_i$ divides
% $t$, and therefore
% \[
%   \pi^t = \mathrm{id}
%   \mbox{ iff }
%   \ell_i \mid t \text{ for all } i=1,\dots,r.
% \]
% If $\mathbf A$ is a permutation matrix, we denote by $\pi_{\mathbf A}$ the
% corresponding permutation of $[n]$, defined via
% \[
%   \mathbf A e_j = e_{\pi_{\mathbf A}(j)}\qquad\text{for }j=1,\dots,n,
% \]
% where $e_j$ is the $j$-th standard basis vector.
% \end{definition}

\begin{definition}\label{def:cycle-decomp-type}
Let $\pi\in\mathcal{S}_n$ be a permutation and $\pi$ can be decomposed into \emph{disjoint cycles}:
there exist permutations $c_1,\dots,c_r\in\mathcal{S}_n$ such that
$\pi = c_1 c_2 \cdots c_r,$ where the composition follows Definition \ref{def:symmetric-group}, and $c_j$ are pairwise disjoint, which map disjoint subsets of $[n]$.
% This decomposition is unique in the following sense: the set of cycles is determined uniquely by $\pi$,
% except that (i) the disjoint cycles may be written in any order, and (ii) each individual cycle may be
% written starting from any element of that cycle (a cyclic shift).
A permutation $c\in\mathcal{S}_n$ is called a
\emph{$k$-cycle} if there exist pairwise distinct indices $a_1,\dots,a_k\in[n]$
such that
\[
  c(a_j)=a_{j+1},\ \ j=1,\dots,k-1,\qquad
  c(a_k)=a_1,\qquad
  c(i)=i,\ \ \forall\, i\notin\{a_1,\dots,a_k\}.
\]
We write such a cycle as $c=(a_1\,a_2\,\cdots\,a_k)$ and call $k$ its \emph{length}. The \emph{cycle type} of $\pi$ is the multiset of cycle lengths $\{\ell_1,\dots,\ell_r\}$, where $\ell_j$ is the length of $c_j$.
The \emph{order} of $\pi$ is $\mathrm{ord}(\pi):=\operatorname{lcm}(\ell_1,\dots,\ell_r)$, i.e., the smallest $t\ge 1$ such that $\pi^t=\mathrm{id}$.
If $\bA\in\{0,1\}^{n\times n}$ is a permutation matrix, we denote by $\pi_{\bA}$
the induced permutation of $[n]$ defined by
\[
  \bA \be_j \;=\; \be_{\pi_{\bA}(j)},\qquad j=1,\dots,n,
\]
where $\be_j$ is the $j$-th standard basis vector.
\end{definition}

We write permutations in disjoint cycle notation. For example,
$(a\ b)(c\ d)$ denotes the permutation that swaps $a$ with $b$ and swaps $c$ with $d$,
while leaving all other indices fixed. Since the cycles are disjoint, their composition is order-independent.

\begin{example}\label{EX_cycle}
    Let $n=6$ and $\pi_{\mathbf A}=(1\,4\,3)(2\,5)(6)$.
Then $\pi_{\mathbf A}(1)=4,\ \pi_{\mathbf A}(4)=3,\ \pi_{\mathbf A}(3)=1,
\pi_{\mathbf A}(2)=5,\ \pi_{\mathbf A}(5)=2,
\pi_{\mathbf A}(6)=6.$
The cycle lengths are $\ell_1=3$, $\ell_2=2$, $\ell_3=1$.
Thus $\pi_{\mathbf A}^6=\mathrm{id}$ and no smaller $1\le k<6$ gives identity.
For the associated permutation matrix $\mathbf A$, we have $\mathbf A^6=\mathbf I$.
\end{example} 

\begin{definition}[Group isomorphism \cite{DummitFoote2004}]\label{Def_group_isomorphism}
Let $(\mathcal{G},\circ)$ and $(\mathcal{H},\ast)$ be two groups.
A map $\varphi:\mathcal{G}\to\mathcal{H}$ is called a \emph{group isomorphism} if the following two conditions hold:
\begin{enumerate}
\item {Homomorphism property:} for all $g_1,g_2\in\mathcal{G}$, $\varphi(g_1\circ g_2) \;=\; \varphi(g_1)\ast\varphi(g_2).$
\item {Bijection:} $\varphi$ is a bijective map from $\mathcal{G}$ onto $\mathcal{H}$.
\end{enumerate}
If such a map $\varphi$ exists, we write $\mathcal{G} \cong \mathcal{H}.$
\end{definition}

\subsection{Universal Hash Family}
\begin{definition}[Universal hash family (UHF)]
Let $\mathcal{X}$ be a finite set and let $\ell\in\mathbb{N}$.
A family of functions $\mathcal{F}=\{f:\mathcal{X}\to\{0,1\}^{\ell}\}$ is called
\emph{universal} if for all distinct $x\neq x'\in\mathcal{X}$,
\[
\Pr\big(F(x)=F(x')\big)\le 2^{-\ell},
\]
where $F$ is drawn uniformly at random from $\mathcal{F}$.
\end{definition}

\begin{definition} (Smooth conditional min-entropy) 
For distributions \( P_{XZ} \) and \( Q_Z \), and smoothing parameter \( 0 \leq \epsilon < 1 \), the smooth conditional min-entropy of \( P_{XZ} \) given \( Q_Z \) is defined as
\begin{align}\label{Def_smooth_min_entropy1}
    H_{\min}^{\epsilon}(P_{XZ} \mid Q_Z) := \sup_{P_{\tilde{X}\tilde{Z}} \in B_{\epsilon}(P_{XZ})} H_{\min}(P_{\tilde{X}\tilde{Z}} \mid Q_Z),    
\end{align}

where \( B_{\epsilon}(P_{XZ}) \) is the set of subdistributions \( P_{\tilde{X}\tilde{Z}} \) that are within \( \epsilon \)-variational distance from \( P_{XZ} \). 

The smooth conditional min-entropy of \( P_{XZ} \) given \( Z \) is then defined as
\begin{align}\label{Def_smooth_min_entropy2}
H_{\min}^{\epsilon}(P_{XZ} \mid Z) := \sup_{Q_Z} H_{\min}^{\epsilon}(P_{XZ} \mid Q_Z).
\end{align}
\end{definition}

In addition to the adversary's original observation $Z$, the adversary may also observe
an additional side-information $V$ that is leaked by the legitimate parties
during the protocol. Then we can have a general \gls{lhl} \cite[Sec. 7.5]{Watanabe} as follows:

\begin{corollary}(General \gls{lhl}) \label{Corollary_LHL}
For a given distribution \( P_{XVZ} \) on $\mathcal{X}\times\mathcal{V}\times\mathcal{Z}$, and for a mapping \( F \sim\mbox{Unif}(\mathcal F)\), define \( K := F(X) \). Then for any $0\le \epsilon<1$,
\begin{align}\label{EQ_LHL}
    &d_{\text{var}}(P_{KVZF}, \Unif(\{0,1\}^\ell) \times P_{VZ} \times P_F) \leq 2\epsilon + \frac{1}{2} \sqrt{2^{\ell + \log |\mathcal V| - H_{\min}^{\epsilon}(X \mid Z)}}.    
\end{align}
\end{corollary}

\subsection{2-1 OT system setup}
In a 2-1 OT system, Alice has two messages $(M_0,M_1)\in\{0,1\}^{\ell}\times\{0,1\}^{\ell}$.
Bob has a choice bit $B\in\{0,1\}$ and aims to recover $M_B$ using a public and noiseless channel.
Let $X^n$ be the noisy channel input generated by Alice, $Y^n$ the corresponding channel output observed by Bob,
and let $\Pi$ denote the public information/discussion (all information exchanged over the public channel). Fix target parameters $\varepsilon\in[0,1]$ and $\delta_A,\delta_B\in[0,1]$. A valid OT protocol should simultaneously satisfy the following constraints:
\begin{align}
\mbox{Reliability: }&\Pr[\hat{M} \neq M_B] \le \varepsilon,\label{EQ_reliability}\\
\mbox{\gls{sfa}: }&
d_{\mathrm{var}}\!\bigl(P_{M_{\bar{B}}\,Y^n\,\Pi\,B},\;P_{M_{\bar{B}}}\times P_{Y^n\,\Pi\,B}\bigr)
\le \delta_A,\label{EQ_SFA}\\
\mbox{\gls{sfb}: }&
d_{\mathrm{var}}\!\bigl(P_{B\,M_0\,M_1\,X^n\,\Pi},\;P_B \times P_{M_0\,M_1\,X^n\,\Pi}\bigr)
\le \delta_B,\label{EQ_SFB}
\end{align}
where $\bar{B}:=1-B$.
Condition \gls{sfa} ensures that Bob's entire view $(Y^n,\Pi,B)$ reveals essentially
no information about the unchosen message $M_{\bar{B}}$.
Condition \gls{sfb} ensures that Alice's view $(M_0,M_1,X^n,\Pi)$ reveals essentially
no information about Bob's selection $B$.

The design goal is to construct valid OT protocols.
A rate $\textsf{R}$ is achievable if, for every
$0 \le \varepsilon,\delta_A,\delta_B < 1$ and sufficiently large $n$, there exists an
$(\varepsilon,\delta_A,\delta_B)$-secure OT protocol of length $\ell$ such that
$\ell/n \ge \textsf{R}$.
The OT capacity $\textsf{C}_{\mathrm{OT}}(W)$ is the supremum of all achievable OT rates.

\section{Main results}
In this section, we explain our motivation, introduce the proposed OT protocol, derive a relaxed reliability and prove the \gls{sfa} and \gls{sfb} followed by an illustrative example.

\subsection{Motivation and setup}
The asymmetric information at the transmitter Alice and the receiver Bob is essential for \gls{ot} protocol design. There is an elegant and simple \gls{ot} protocol that transmits over \gls{bec} without channel coding, because such a use of \gls{bec} efficiently provides such asymmetry. In contrast, in the classical use of polar codes, \gls{bbcs} are frozen to, e.g., zeros, and are completely useless for the receiver, while \gls{gbcs} can be successfully decoded, when the code is properly designed. In this way, the transmitter and the receiver share identical information, which cannot be used for \gls{ot}.

\begin{figure}[htbp]
    \centering
    \includegraphics[width=1\textwidth]{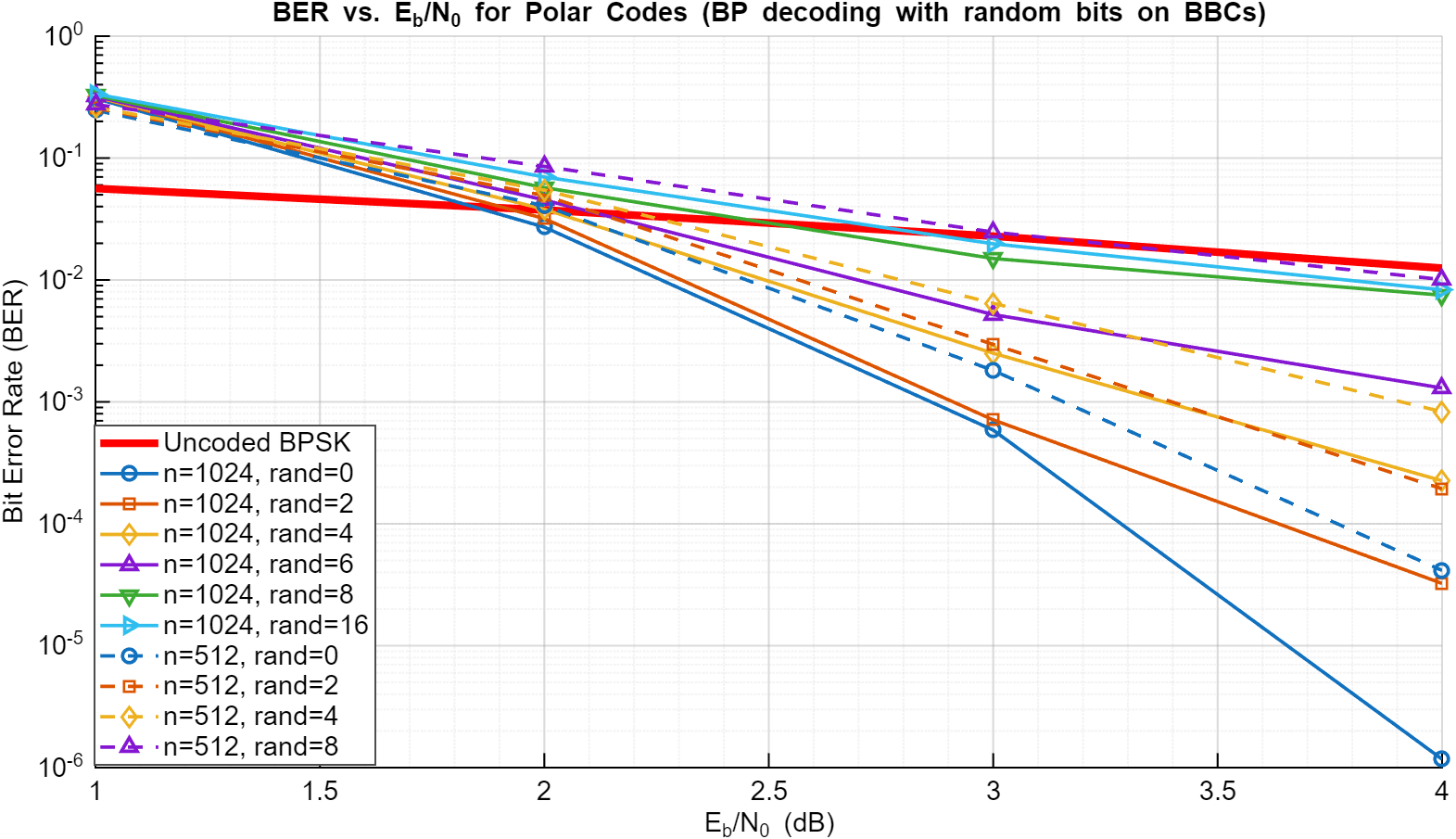}
    \caption{BER comparison of letting \gls{bbcs} carry random bits unknown to Bob.}
    \label{fig:RandomBitBBC}
\end{figure}

Our work is motivated by the presence of \gls{bbcs}, which naturally admit an erasure-like abstraction. The simulation results below support this motivation: even when a subset of \gls{bbcs} is used to carry random bits unknown to Bob, the decoder remains operational with a controlled performance degradation, providing the intended asymmetry. The setting is as follows. The polar codes decoded by \gls{bp} follow the construction in~\cite{Cammerer2018SparseGraphsBPPolar,CammererLDPCLikePolarBPCode}.
We consider blocklength $n=$512 (dashed curves) and $n=$1024 (solid curves), code rate as $1/2$, $10^4$ transmitted codewords, SNR= 1, 2, 3, and 4 dB. The cases of $n=512$ and $n=1024$ consider numbers of random bits unknown to Bob on \gls{bbcs} as rand=0, 2, 4, 8 and 0, 2, 4, 6, 8, 16, respectively. Note that rand=0 means that it is the classical way of using polar codes. The uncoded \gls{bpsk} curve serves as a baseline.
By observing Fig. \ref{fig:RandomBitBBC}, we see that even when random bits are inserted on a subset of \gls{bbcs}, the decoder can still operate reliably as long as the injected portion is chosen appropriately. In particular, the \gls{ber} increases in a controlled manner, depending on the target value. This controlled degradation provides the design degree of freedom, the \textit{asymmetric information/view} at Alice and Bob, which can be exploited to construct an \gls{ot} protocol. 
By this way, we can emulate an erasure-like abstraction reminiscent of a \gls{bec}, while keeping the \gls{gbcs} sufficiently reliable for the intended reconstruction.
In short, the idea behind Fig. \ref{fig:RandomBitBBC} illustrates a practical way that lets us trade reliability for erasures.

%\myboxb{ML decoding seems can alleviate the degradation of putting random bits on \gls{bbcs}}{Try new simulations}

Note that in the following, we will propose a protocol relying on left-permuting the polarization matrix $\bT$, i.e., permuting the rows of $\bT$.
By associativity of the linear encoding operation, encoding with a row-permuted $\bT$ is equivalent to encoding with the original $\bT$, but with the input message bits permuted.
In polar coding, for a given $\bT$ and channel, each input bit together with the channel output and the previously decoded bits (under successive decoding) defines a synthesized bit-channel.
Hence, when we view the permutation as a relabeling of the input-bit indices, permuting the input bits induces the same permutation of the bit-channels.
Therefore, if the bit-channel qualities are ordered by a performance metric, e.g., mutual information or Bhattacharyya parameter, then under the permuted labeling the quality order is permuted accordingly.
This fact will be used repeatedly in developing the proposed protocol.

\subsection{Protocol}\label{sec:III-B}
\begin{figure}[htbp]
    \centering
    \includegraphics[width=0.8\textwidth]{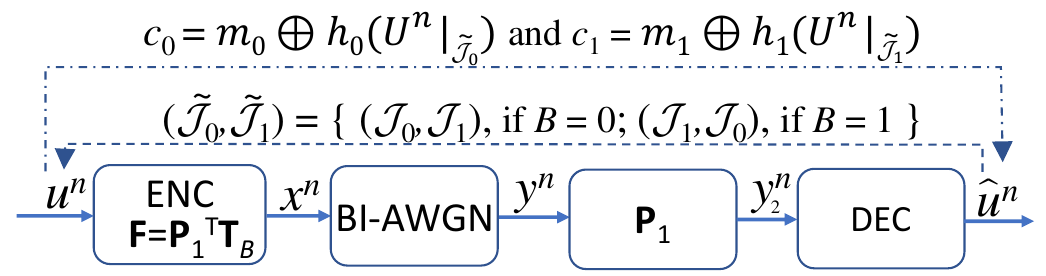}
    \caption{The proposed \gls{ot} system}
    \label{fig:sys_D}
\end{figure}

The proposed protocol is summarized in Fig. \ref{fig:sys_D}, where the non-solid arrows are public channels. In particular, the dashed arrow is Step 2 in the protocol, the solid arrows include Steps 3 and 4, the dash-dotted arrow is Step 5.  The key idea is using different viewpoints of \gls{gbcs} and \gls{bbcs} at Alice and Bob generated by an additional permutation by Bob’s hidden permutation, together with the random bits on \gls{bbcs}, to achieve \gls{sfa} and \gls{sfb}, simultaneously. 

The proposed protocol is composed of the following steps. We first introduce the setup of the proposed protocol as follows:\\
\textit{Public parameters:}
blocklength $n=2^m$; BI--AWGN channel law $W$; threshold $\gamma_n\geq 0$ and a fixed UPO rule;
a universal hash family $\mathcal{H}$.\\
\textit{Bob's private one-time setup:}
Choose $\bT_1$ as Ar{\i}kan's polarization matrix for the first view and a permutation matrix $\bA$ such that
$\bA \in \Aut(\bT_1)$
and let $\textsf{N}:=\mbox{ord}(\bA)$.
Define the second view $\bT_2 := \bA\,\bT_1,  \,  \mathcal{P} := \{\bA^k:\ 0\le k<\textsf{N}\}.$\\
\textit{Inputs:} Alice holds $m_0,\,m_1\in \{0,1\}^{\ell}$; Bob holds a choice bit $B\in\{0,1\}$.\\
We now introduce the protocol steps as follows:
\begin{enumerate}
  \item \textit{Bob selects a hidden view and publishes only the composite transform.}
  Bob samples $K\sim\Unif(\{0,1,\dots,\textsf{N}-1\})$ and sets
  \[
    \bP_1 := \bA^K \in \mathcal{P},
    \qquad
    \bT_B :=
    \begin{cases}
      \bT_1, & B=0,\\
      \bT_2, & B=1.
    \end{cases}
  \]
  Bob\! publishes\! only \!$\bF\!:=\! \bP_1^{\mathsf T}\bT_B,\!$ and keeps $(B,\bP_1,\bT_B)$ private.

  \item \textit{Bob computes and announces the index sets.}
  Using his private view $\bT_B$ and the known channel law $W$, Bob determines two sets
  $\mathcal{I}_{\mathcal G}$ and $\mathcal{I}_{\mathcal B}$ based on the bit-channel mutual information defined in \eqref{EQ_GBC_BBC_def}. Bob then selects two disjoint sets of equal size,
  $  \mathcal{J}_0 \subseteq \mathcal{I}_{\mathcal G},
  \,  \mathcal{J}_1 \subseteq \mathcal{I}_{\mathcal B}, \mathcal{J}_0\cap \mathcal{J}_1=\emptyset,\mbox{ and }
  \,  |\mathcal{J}_0|=|\mathcal{J}_1|.$
  Finally, Bob publishes the pair
  $(\tilde{\mathcal{J}}_{0},\tilde{\mathcal{J}}_{1})=(\mathcal{J}_{0},\mathcal{J}_{1})$, if $B=0$;
  $(\tilde{\mathcal{J}}_{0},\tilde{\mathcal{J}}_{1})=(\mathcal{J}_{1},\mathcal{J}_{0})$, else, over the public channel.

\item \textit{Alice encodes using only $\bF$ and transmits.}
After receiving $(\tilde{\mathcal{J}}_{0},\tilde{\mathcal{J}}_{1})$, Alice samples
$U_i\sim\mathrm{Bern}(\tfrac12)$ independently for all
$i\in \tilde{\mathcal{J}}_{0}\cup\tilde{\mathcal{J}}_{1}$,
and freezes all remaining $U_i$'s (e.g., to $0$).
She then forms the binary codeword
\(x^n := u^n\bF \in\{0,1\}^n\) and modulates it by BPSK to obtain the channel input
\(s^n := \mathbf 1 - 2x^n\), where $\mathbf 1$ is an all-one vector and the subtraction is element-wise.
She transmits \(s^n\) over the BI--\gls{awgn} channel.

\item \textit{Bob receives and permutes by $\bP_1$.}
Bob observes \(y^n = s^n + z^n\), where \(z^n\) is i.i.d.\ Gaussian noise.
He permutes the received vector by his private permutation \(\bP_1\):
$y_2^n:= y^n \bP_1\overset{(a)}=\mathbf 1 - 2u^n\bT_B + z_2^n,\nolinebreak$
where \(z_2^n:=z^n\bP_1 \overset{d}= z^n\) and (a) is due to the definitions of $s^n$, \(\bF = \bP_1^{\mathsf T}\bT_B\), and the automorphism property: $\bF\bP_1 = \bP_1^{\mathsf T}\bT_B\bP_1 = \bT_B.$
Equivalently, \(y_2^n\) is the BI--\gls{awgn} output corresponding to the BPSK-modulated codeword
\(\mathbf 1 - 2u^n\bT_B\).
Therefore, Bob performs soft decoding\footnote{The modulation step is a fixed, deterministic mapping applied independently to each coordinate. Hence, it can be absorbed into the BI--AWGN law and does not change the public channel. Therefore the \gls{sfb} argument in \cite{PHL_OT}, which considers only the distribution of the public information, remains unchanged. For \gls{sfa}, further processing at Bob is a function of \(Y^n\) and the public information, which by the data-processing inequality cannot increase Bob's knowledge about the unchosen message.}
 matched to \(\bT_B\) on \(y_2^n\) and obtains an estimate \(\hat u^n\).

\item \textit{Key generation and decryption.}
  Bob samples and publishes seeds for hash function $(h_0,h_1)\in\mathcal{H}^2$, where $h_b:\{0,1\}^{|\tilde{\mathcal{J}}_b|}\to\{0,1\}^{\ell}$ for $b\in\{0,1\}$.
  Alice computes keys
$    k_0 := h_0\!\bigl(u^n|_{\tilde{\mathcal{J}}_{0}}\bigr),\,    k_1 := h_1\!\bigl(u^n|_{\tilde{\mathcal{J}}_{1}}\bigr),$
  and sends ciphertexts $    c_0 := m_0\oplus k_0,\, c_1 := m_1\oplus k_1.$
  Bob then computes $k_B := h_B\!\bigl(\hat u^n|_{\tilde{\mathcal{J}}_{B}}\bigr)$ and outputs
  $    \hat m_B := c_B\oplus k_B.$
\end{enumerate}

The key to achieving \gls{sfa} is that we rely on the random bits injected into the \gls{bbcs} in Bob's view to make the unchosen message inaccessible to him, which is inherited from the intrinsic property of polar codes. Specifically, we ensure that even though Bob has public information, he cannot use it to gain any useful information about the unchosen message. \gls{sfb}, on the other hand, relies on the symmetry of the protocol, which is induced by the randomization of the public generator matrix and the injection of random bits into the \gls{bbcs}. Specifically, the randomization of the public generator relies on a randomly chosen permutation \(\mathbf{P}_1\) from a specific set specified by its permutation order and \(\Aut(\mathbf{T})\), which ensures that both decoder views are symmetric. The symmetry guarantees that Alice’s view of the two sets $\tilde{\mathcal{J}}_{0}$ and $\tilde{\mathcal{J}}_{1}$ is indistinguishable. Detailed proof is in \cite[Theorem 2]{PHL_OT}.

We will show that our \gls{sfb} is perfect ($\delta_B=0$ in \eqref{EQ_SFB}). Note that \cite[Remark~5]{AhlswedeCsiszar2009} shows perfect \gls{sfb} is achievable if channel input \(X\) (uncoded) is \gls{iid} uniformly distributed over \(\{0,1\}\) for a DMC under honest-but-curious Bob's behavior. In our setting, the random bits placed on the selected bit-channels are \gls{iid} uniform over \(\{0,1\}\) and we also assume an honest-but-curious Bob. Although our physical channel is BI--AWGN, our construction yields the same perfect \gls{sfb}.

\begin{table}[h]
    \centering
    \begin{normalsize}

\setlength{\arraycolsep}{4pt}
\begin{tabular}{ll}
$\sigma_{1}$: $[b_3\,b_2\,b_1\,b_0]\to[b_3\,b_2\,b_1\,b_0]$ & [16, 15, 14, 13, 12, \g{11}, 10, \g{9}, 8, \y{7}, 6, \y{5}, 4, 3, 2, 1]\\
$\sigma_{2}$: $[b_3\,b_2\,b_1\,b_0]\to[b_2\,b_3\,b_1\,b_0]$ & [16, 15, 14, 13, 8, \y{7}, 6, \y{5}, 12, \g{11}, 10, \g{9}, 4, 3, 2, 1]\\
$\sigma_{3}$: $[b_3\,b_2\,b_1\,b_0]\to[b_3\,b_1\,b_2\,b_0]$ & [16, 15, 12, \g{11}, 14, 13, 10, \g{9}, 8, \y{7}, 4, 3, 6, \y{5}, 2, 1]\\
$\sigma_{4}$: $[b_3\,b_2\,b_1\,b_0]\to[b_1\,b_3\,b_2\,b_0]$ & [16, 15, 8, \y{7}, 14, 13, 6, \y{5}, 12, \g{11}, 4, 3, 10, \g{9}, 2, 1]\\
$\sigma_{5}$: $[b_3\,b_2\,b_1\,b_0]\to[b_2\,b_1\,b_3\,b_0]$ & [16, 15, 12, 11, 8, 7, 4, 3, 14, 13, 10, 9, 6, 5, 2, 1]\\
$\sigma_{6}$: $[b_3\,b_2\,b_1\,b_0]\to[b_1\,b_2\,b_3\,b_0]$ & [16, 15, 8, 7, 12, 11, 4, 3, 14, 13, 6, 5, 10, 9, 2, 1]\\
$\sigma_{7}$: $[b_3\,b_2\,b_1\,b_0]\to[b_3\,b_2\,b_0\,b_1]$ & [16, 14, 15, 13, 12, 10, \g{11}, \g{9}, 8, 6, \y{7}, \y{5}, 4, 2, 3, 1]\\
$\sigma_{8}$: $[b_3\,b_2\,b_1\,b_0]\to[b_2\,b_3\,b_0\,b_1]$ & [16, 14, 15, 13, 8, 6, \y{7}, \y{5}, 12, 10, \g{11}, \g{9}, 4, 2, 3, 1]\\
$\sigma_{9}$: $[b_3\,b_2\,b_1\,b_0]\to[b_3\,b_0\,b_2\,b_1]$ & [16, 12, 15, \g{11}, 14, 10, 13, \g{9}, 8, 4, \y{7}, 3, 6, 2, \y{5}, 1]\\
$\sigma_{10}$: $[b_3\,b_2\,b_1\,b_0]\to[b_0\,b_3\,b_2\,b_1]$ & [16, 8, 15, \y{7}, 14, 6, 13, \y{5}, 12, 4, \g{11}, 3, 10, 2, \g{9}, 1]\\
$\sigma_{11}$: $[b_3\,b_2\,b_1\,b_0]\to[b_2\,b_0\,b_3\,b_1]$ & [16, 12, 15, 11, 8, 4, 7, 3, 14, 10, 13, 9, 6, 2, 5, 1]\\
$\sigma_{12}$: $[b_3\,b_2\,b_1\,b_0]\to[b_0\,b_2\,b_3\,b_1]$ & [16, 8, 15, 7, 12, 4, 11, 3, 14, 6, 13, 5, 10, 2, 9, 1]\\
$\sigma_{13}$: $[b_3\,b_2\,b_1\,b_0]\to[b_3\,b_1\,b_0\,b_2]$ & [16, 14, 12, 10, 15, 13, \g{11}, \g{9}, 8, 6, 4, 2, \y{7}, \y{5}, 3, 1]\\
$\sigma_{14}$: $[b_3\,b_2\,b_1\,b_0]\to[b_1\,b_3\,b_0\,b_2]$ & [16, 14, 8, 6, 15, 13, \y{7}, \y{5}, 12, 10, 4, 2, \g{11}, \g{9}, 3, 1]\\
$\sigma_{15}$: $[b_3\,b_2\,b_1\,b_0]\to[b_3\,b_0\,b_1\,b_2]$ & [16, 12, 14, 10, 15, \g{11}, 13, \g{9}, 8, 4, 6, 2, \y{7}, 3, \y{5}, 1]\\
$\sigma_{16}$: $[b_3\,b_2\,b_1\,b_0]\to[b_0\,b_3\,b_1\,b_2]$ & [16, 8, 14, 6, 15, 7, 13, 5, 12, 4, 10, 2, \g{11}, 3, \g{9}, 1]\\
$\sigma_{17}$: $[b_3\,b_2\,b_1\,b_0]\to[b_1\,b_0\,b_3\,b_2]$ & [16, 12, 8, 4, 15, 11, 7, 3, 14, 10, 6, 2, 13, 9, 5, 1]\\
$\sigma_{18}$: $[b_3\,b_2\,b_1\,b_0]\to[b_0\,b_1\,b_3\,b_2]$ & [16, 8, 12, 4, 15, 7, 11, 3, 14, 6, 10, 2, 13, 5, 9, 1]\\
$\sigma_{19}$: $[b_3\,b_2\,b_1\,b_0]\to[b_2\,b_1\,b_0\,b_3]$ & [16, 14, 12, 10, 8, 6, 4, 2, 15, 13, \g{11}, \g{9}, \y{7}, \y{5}, 3, 1]\\
$\sigma_{20}$: $[b_3\,b_2\,b_1\,b_0]\to[b_1\,b_2\,b_0\,b_3]$ & [16, 14, 8, 6, 12, 10, 4, 2, 15, 13, \y{7}, \y{5}, \g{11}, \g{9}, 3, 1]\\
$\sigma_{21}$: $[b_3\,b_2\,b_1\,b_0]\to[b_2\,b_0\,b_1\,b_3]$ & [16, 12, 14, 10, 8, 4, 6, 2, 15, \g{11}, 13, \g{9}, \y{7}, 3, \y{5}, 1]\\
$\sigma_{22}$: $[b_3\,b_2\,b_1\,b_0]\to[b_0\,b_2\,b_1\,b_3]$ & [16, 8, 14, 6, 12, 4, 10, 2, 15, \y{7}, 13, \y{5}, \g{11}, 3, \g{9}, 1]\\
$\sigma_{23}$: $[b_3\,b_2\,b_1\,b_0]\to[b_1\,b_0\,b_2\,b_3]$ & [16, 12, 8, 4, 14, 10, 6, 2, 15, \g{11}, \y{7}, 3, 13, \g{9}, \y{5}, 1]\\
$\sigma_{24}$: $[b_3\,b_2\,b_1\,b_0]\to[b_0\,b_1\,b_2\,b_3]$ & [16, 8, 12, 4, 14, 6, 10, 2, 15, \y{7}, \g{11}, 3, 13, \y{5}, \g{9}, 1]\\
\end{tabular}
\end{normalsize}
    \caption{All orders of bit channel-indices under the 24 permutations in $\Aut(\bT)$ for $n=16.$}
    \label{Table_Aut_n16}
\end{table}

\subsection{Impact of parallel decodings at Bob}
In this section, we investigate different combinations of $\bP_2$ with the type of decoders mentioned in the previous section, to verify that Bob will not break \gls{sfa} and \gls{sfb} by parallel decoding using different $\bP_2$ and decoders.

\subsubsection{Baseline decoding: DEC decodes w.r.t. the code $\bT$, and $\bP_1=\bP_2\in \Aut(\bT)$} 
In this case, $y_2^n=(u^n\cdot\bP_1^{\mathsf T}\bT+z^n)\cdot\bP_2=u^n\cdot\bT+z_2^n$, where the second equality is due to automorphism, and $z_2^n:=z^n\bP_2$, $Z_2^n\overset{d}=Z^n$. In short, $y_2^n=u^n\cdot\bT+z_2^n$, which means Bob can decode \gls{gbcs} in $u^n$ based on the \gls{gbcs} and \gls{bbcs} definitions of $\bT$, which are (11,9) and (7,5), respectively, according to the example in Step 2. The most important thing for Bob's decoding is that the random bits are uniformly allocated on the indices (11,9) and (7,5), but the genuine \gls{gbcs} or \gls{bbcs} are not distinguishable by Alice, which will be proved later. In contrast, from Bob's viewpoint, this case is equivalent to that where Alice transmits $u^n$ encoded by the polarization matrix $\bT$ and Bob decodes w.r.t. $\bT$. The only difference here compared to the normal polar code setting is that we let \gls{bbcs} carry a few random bits unknown to Bob (and also, here we do not use all \gls{gbcs} to convey messages for the symmetry of the two sets of indices). For $i$ in \gls{bbcs} we will bound the possible leakage in Lemma \ref{lem:bbc-leakage}. This leakage may invalidate the \gls{sfa}, but can be remedied by privacy amplification, and will be discussed in Sec. \ref{subsec_finite_n}. On the other hand, for any estimator $\hat U_i=\hat U_i(Y^n,U^{i-1})$, by Fano’s inequality we can simply see the lower bounded $h_2(P_{e,i})\ge H(U_i\!\mid Y^n,U^{i-1})\ge 1-\delta_n$, i.e., $P_{e,i}\ge h_2^{-1}(1-\delta_n)$, due the random bit transmitted on \gls{bbcs}.

It is clear that Alice is not allowed to allocate random bits only on \gls{gbcs} of $\bF$. Because, if $\bP_1=\bI$, Bob can get all bits transmitted over \gls{gbcs} and then there is no \gls{sfa}. In contrast, if Alice allocates random bits on both \gls{gbcs} and \gls{bbcs}, even if $\bP_1=\bI$, Bob cannot decode the bits on his \gls{bbcs}. In addition, let us have a simple check on \gls{sfb} in this case. Recall that when $\bP_{1}=\bI$, both Alice and Bob use the same polarization matrix $\bT$. In this case, they identify the \gls{gbcs} as indices $(11,9)$ and the \gls{bbcs} as indices $(7,5)$. In contrast, if $\bP_1\neq\bI$, Alice will see indices of \gls{gbcs} as (7,5) and \gls{bbcs} as (11,9), while Bob will still see indices of \gls{gbcs} as (11,9) and \gls{bbcs} as (7,5), due to the assumption $\bP_1=\bP_2\in\Aut(\bT)$. In this protocol, we can observe that Bob can only decode (11,9) for both cases: $\bP_1=\bI$ and $\bP_1\neq\bI$ and Alice can know it just based on $\bF=\bP_1^{\mathsf T}\bT$, i.e., if \gls{gbcs} are (7,5) at Alice, then Bob must use (11,9) and then no \gls{sfb} exists. 
    
\begin{figure}[htbp]
    \centering
    \includegraphics[width=1\textwidth]{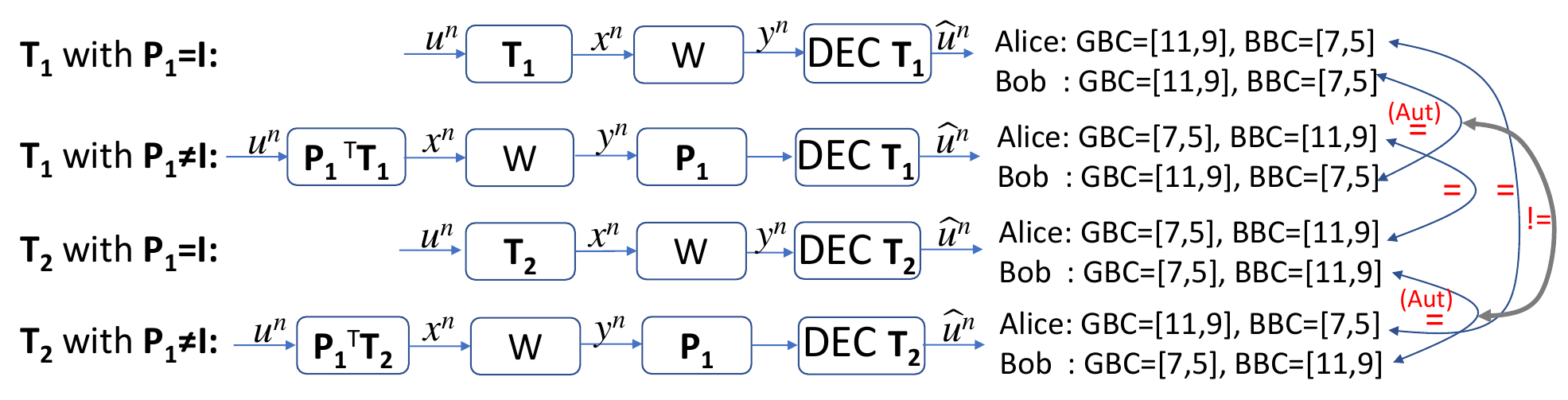}
    \caption{Let $\bT_1$-\gls{gbcs}=(11,9) and $\bT_1$-\gls{bbcs}=(7,5). Select $\bA\in\Aut(\bT_1)$, let $\bT_2:=\bA\bT_1$, such that $\bT_2$-\gls{gbcs}=(7,5) and $\bT_2$-\gls{bbcs}=(11,9). Let $\bP_1=\bP_2$.}
    \label{fig:sysD_comparison}
\end{figure}

This issue comes from the assumption that Bob uses a fixed $\bT$ to decode, which can be simply solved as follows. Let us choose $\bA\in\Aut(\bT)$ as the same the $\bP_1$ as in the previous paragraph, such that (7,5)  are \gls{gbcs} and (11,9) are \gls{bbcs}, and define $\bT_2:=\bA^T\bT_1$. Therefore, when $\bT_2$ is used to encode and decode with $\bP_1=\bP_2=\bI$, we will have \gls{gbcs} as (7,5) and \gls{bbcs} as (11,9) for both Alice and Bob. If now $\bP_1$ is selected as $\bA^{-1}$, Alice will have \gls{gbcs} as (11,9) and \gls{bbcs} as (7,5), while Bob will still have \gls{gbcs} as (7,5) and \gls{bbcs} as (11,9) when the DEC is fixed as $\bT_2$, due to automorphism. As a result, the use of $\bT_1$ and $\bT_2$ has symmetric \gls{gbcs} and \gls{bbcs}. A detailed illustration of the symmetry can be seen from Fig. \ref{fig:sysD_comparison}. Together with the multiplication of $\bP_1$, Alice can not be able to guess the $b$ selected by Bob, just based on her derived indices of \gls{gbcs} and \gls{bbcs}. More specifically, in Fig. \ref{fig:sysD_comparison}, the 1st and the 4th branches have the same/different \gls{gbcs} and \gls{bbcs} from Alice's/Bob's viewpoint, similarly to the 2nd and the 3rd branches. This is possible due to the local permutation at Bob and the automorphism, such that Bob can switch between different decoders to get different \gls{gbcs} and \gls{bbcs} from those at Alice.

\subsubsection{Case 1. DEC decodes w.r.t.\ $\mathbf T$, but $\mathbf P_1\neq \mathbf P_2$}\label{sec:III-C2}
Recall that $\Pi$ denotes the public information. Let $\Theta$ collect Bob's additional private side information, including his choice bit $B$, and define Bob's information as $T := (Y^n,\Pi,\Theta).$
In this case, Alice encodes with $\mathbf F=\mathbf P_1^{\top}\mathbf T$ (the view induced by $\mathbf P_1$),
while Bob may attempt to cheat by running multiple decoding branches indexed by $\mathbf P_2\in\Aut(\mathbf T)$ that do not
match $\mathbf P_1$. Let $\widehat{V}$ denote the collection of all outputs produced by such a stochastic
parallel-decoding strategy, which takes values in a measurable space $(\widehat{\mathcal V},\widehat{\mathscr V})$, where $\widehat{\mathcal V}$ is the alphabet of $\widehat{V}$ and $\widehat{\mathscr V}$ is a $\sigma$-algebra on
$\widehat{\mathcal V}$. Let $T_0:=(Y^n,\Pi,B)$ and let $\Theta$ collect Bob's other private information.
We model a randomized parallel-decoding strategy as a conditional distribution
$Q_{\widehat{V}\mid T_0,\Theta}$, i.e., for every measurable set $\mathcal{A}\in\widehat{\mathscr V}$,
\[
Q_{\widehat{V}\mid T_0,\Theta}(\mathcal{A}\mid t_0,\theta)
\;:=\;
\Pr\!\big[\widehat{V}\in \mathcal{A}\,\big|\,T_0=t_0,\ \Theta=\theta\big].
\]
Then we can derive the following. 

\begin{lemma}\label{cor:parallel-sfa}
If \gls{sfa} holds with leakage at most $\delta_A$, then any parallel decoding cannot increase the leakage about
$M_{\bar B}$ beyond $\delta_A$, i.e.,
\begin{align}\label{EQ_Case1_result}
d_{\mathrm{var}}\!\bigl(P_{M_{\bar B},T_0,\Theta,\widehat{V}},\,P_{M_{\bar B}}  P_{T_0,\Theta,\widehat{V}}\bigr)
\le \delta_A.
\end{align}
\end{lemma}

\begin{proof}
Since $\widehat{V}$ is the output of any randomized function of $(T_0,\Theta)$ at Bob, we have
$M_{\bar B}-(T_0,\Theta)-\widehat{V}.$
From $M_{\bar B}-T_0-\Theta$ and $M_{\bar B}-(T_0,\Theta)-\widehat{V}$, for all $(m,t_0,\theta,\hat v)$, we have
\begin{align*}
P_{\Theta,\widehat{V}\mid M_{\bar B},T_0}(\theta,\hat v\mid m,t_0)
&=P_{\Theta\mid M_{\bar B},T_0}(\theta\mid m,t_0)\,
  P_{\widehat{V}\mid M_{\bar B},T_0,\Theta}(\hat v\mid m,t_0,\theta)\\
&=P_{\Theta\mid T_0}(\theta\mid t_0)\,
  P_{\widehat{V}\mid T_0,\Theta}(\hat v\mid t_0,\theta)
= P_{\Theta,\widehat{V}\mid T_0}(\theta,\hat v\mid t_0),
\end{align*}
which means the Markov chain $M_{\bar B}-T_0-(\Theta,\widehat{V})$.
That is, there exists a conditional distribution $Q_{\Theta,\widehat{V}\mid T_0}$ such that, for all
$(m,t_0,\theta,\hat v)$,
\begin{align}
P_{M_{\bar B},T_0,\Theta,\widehat{V}}(m,t_0,\theta,\hat v)
&=P_{M_{\bar B},T_0}(m,t_0)\,Q_{\Theta,\widehat{V}\mid T_0}(\theta,\hat v\mid t_0),\nonumber\\
\bigl(P_{M_{\bar B}}  P_{T_0,\Theta,\widehat{V}}\bigr)(m,t_0,\theta,\hat v)
&=P_{M_{\bar B}}(m)\,P_{T_0}(t_0)\,Q_{\Theta,\widehat{V}\mid T_0}(\theta,\hat v\mid t_0).
\label{EQ_Markov_chain_Case1}
\end{align}

Substituting \eqref{EQ_Markov_chain_Case1} into the \gls{lhs} of \eqref{EQ_Case1_result}, we have
\begin{align*}
d_{\mathrm{var}}\!\bigl(P_{M_{\bar B},T_0,\Theta,\widehat{V}},\,P_{M_{\bar B}}  P_{T_0,\Theta,\widehat{V}}\bigr)
&=
d_{\mathrm{var}}\!\Bigl(P_{M_{\bar B},T_0}(m,t_0)\,Q_{\Theta,\widehat{V}\mid T_0}(\theta,\hat v\mid t_0),\;
P_{M_{\bar B}}(m)\,P_{T_0}(t_0)\,Q_{\Theta,\widehat{V}\mid T_0}(\theta,\hat v\mid t_0)\Bigr)\\
&=
d_{\mathrm{var}}\!\bigl(P_{M_{\bar B},T_0},\,P_{M_{\bar B}}  P_{T_0}\bigr)
\le \delta_A,
\end{align*}
where the inequality is from the \gls{sfa} constraint, which completes the proof.
\end{proof}

Lemma \ref{cor:parallel-sfa} shows that allowing Bob to exhaust all $\mathbf P_2\in\Aut(\mathbf T)$ and to collect the corresponding branch outputs
cannot further increase the unchosen message $M_{\bar B}$ leaked to him.

\subsubsection{Case 2.  DEC decodes w.r.t. the code $\bP_1^{\mathsf T}\bT\bP_2$ as a \textit{matched} decoder} Since in this case, the DEC matches the ENC for all combinations of $(\bP_1,\bP_2)$ given $\bP_1$, then it seems no matter what $\bP_2$ is selected by Bob, he should be able to decode something successfully. However, we know that baseline decoding has already decoded successfully\footnote{With a degradation due to the random bits on \gls{bbcs}} with a number of bits tied to the capacity. Then different other schemes cannot surpass the capacity law, and no further leakage is caused by this case. This statement can be confirmed in a more detailed way by checking the role of $\bP_2$ in the \textit{matched decoding} setting. The received signal after permutation at Bob in this case is expressed as:
\begin{align}
y_3^n:=y^n\bP_2=u^n\cdot\bP_1^{\mathsf T}\bT\bP_2+z_3^n:=u^n\cdot\bF_2+z_3^n,     
\end{align}
$z_3^n:=z^n\bP_2\overset{d}=z^n$. Then we can treat $u^n$ is encoded by a new encoder $\bF_2$ with new channel output $y_3^n$ under the same channel distribution. It is clear that $\bF_2:=\bP_1^{\!\top}\bT\bP_2=(\bP_1^{\!\top}\bT\bP_1)(\bP_1^{-1}\bP_2):=\bT\bS$, where the last equality is from automorphism and we define $\bS:=\bP_1^{-1}\bP_2$. Then we have the following result.

\begin{proposition}\label{prop:C3toC1-clean}
Let $\bP_1,\bP_2\in\Aut(\bT)$, $\bF_2 := \bP_1^{\!\top}\bT\bP_2$, and $ \bS := \bP_1^{-1}\bP_2.$ Then for every memoryless channel $W$, every observation $y^n$, and every $u^n\in\{0,1\}^n$,
\begin{align}
&\arg\max_{u^n} W^{\otimes n}\!\big(y_3^n \,\big|\, u^n\bF_2\big)=\arg\max_{u^n} W^{\otimes n}\!\big(y_2^n \,\big|\, u^n\bT\big).
\end{align}
\end{proposition}

Proposition \ref{prop:C3toC1-clean} shows that decoding matched to $\bF_2$ on $y_3^n$ is {identical} to decoding matched to $\bT$ on $y_2^n=u^n\bT+z_2^n$, defined in baseline decoding. In particular, for any decoder Bob may apply to $y_3^n$, there exists a decoder in the canonical system applied to 
$y_2^n$ with exactly the same performance and the same decoded sequence $u^n$. As a result, Case~2 is just the same detection problem as baseline decoding.

\begin{proof}
For any message vector $u^n$, define $c^n:= u^n\bT$. We can derive the following:
% Let σ be the permutation induced by the right-multiplication with \bS:
% for any row vector z^n, (z^n \bS)_i = z_{\sigma(i)} and ((z^n)\bS^{-1})_i = z_{\sigma^{-1}(i)}.

\begin{align}
W^{\otimes n}(y_3^n\mid u^n\bF_2)
&\overset{(a)}= W^{\otimes n}(y_3^n\mid c^n\,\bS) \overset{(b)}= \prod_{i=1}^n W\!\big( y_{3,i} \,\big|\, (c^n\bS)_i \big) \\
&\overset{(c)}= \prod_{i=1}^n W\!\big( y_{3,i} \,\big|\, c_{\sigma(i)} \big) \overset{(d)}= \prod_{j=1}^n W\!\big( y_{3,\sigma^{-1}(j)} \,\big|\, c_j \big) \\
&\overset{(e)}= \prod_{j=1}^n W\!\big( (y_3^n\bS^{-1})_j \,\big|\, c_j \big) \overset{(f)}= W^{\otimes n}\!\big(y_3^n\bS^{-1}\mid u^n\bT\big)\\
&\overset{(g)}=  W^{\otimes n}\!\big(y^n\bP_2\bS^{-1}\mid u^n\bT\big)\overset{(h)}=W^{\otimes n}\!\big(y^n\bP_1\mid u^n\bT\big)\\
&\overset{(i)}=  W^{\otimes n}\!\big(y_2^n\mid u^n\bT\big),
\end{align}
where (a) is from $u^n\bF_2=u^n\bT\bS=c^n\,\bS$, (b) is due to the memorylessness of the channel $W$, (c) is due to the index permutation operator $\sigma(i):\,\,(c^n\bS)_i=c^n_{\sigma(i)}$, (d) is due to the reindexing and bijection mapping of the function $\sigma$, (e) is due to the equivalent expression $((y_3^n)\bS^{-1})_i = y_{3,\sigma^{-1}(i)}$, (f) is due to $c^n:= u^n\bT$, (g) is due to $y_3^n=y^n\bP_2$, (h) is due to $\bS^{-1}=\bP_2^{-1}\bP_1$, (i) is due to the fact that $y^n\bP_1=u^n\bP_1^{\mathsf T}\bT\bP_1+z^n\bP_1=u^n\bT+z_2^n=y_2^n$, where the second equality is from the automorphism, and the third equality is from the definition of $y_2^n$ stated in baseline decoding, which then completes the proof.
\end{proof}

\subsection{Reliability}\label{sec:reliability}
In our OT protocol, Alice intentionally injects independent random bits on
selected \gls{bbcs} to control information leakage, where the corresponding secrecy
analysis is developed in the next section. This is a non-standard use of a polar
code: these \gls{bbcs} are not merely frozen, and their
randomness can propagate through the polar transform and influence the decoding
result of the bits that Bob uses to generate the key for recovering \(M_b\). To the best of our knowledge, there is no error analysis in
the literature for this setting. Therefore, we first estimate the resulting
hash-input decoding error probability
\(\textsf{P}_{\mathrm{e,hin}}:=\Pr(\widehat U_{\mathcal{S}_b}\neq U_{\mathcal{S}_b})\), where \(\mathcal{S}_b\subseteq[n]\) denotes the index set of hash-input bits
used to generate the key \(K_b\), and then use the empirical estimate of $\textsf{P}_{\mathrm{e,hin}}$ as a tool to upper bound the OT reliability error \(\textsf{P}_{\mathrm{e,OT}}\) through
\(\textsf{P}_{\mathrm{e,OT}}:=\Pr(\widehat M_b\neq M_b)\le \textsf{P}_{\mathrm{e,hin}}\).

\begin{lemma}\label{lem:rel-bridge}
Fix \(b\in\{0,1\}\). Let \(g\) be drawn from a \gls{uhf} family and then fixed, and define
\(K_b:=g(U_{\mathcal{S}_b})\) and \(\widehat K_b:=g(\widehat U_{\mathcal{S}_b})\).
Define \(\textsf{P}_{\mathrm{e,key}}:=\Pr(\widehat K_b\neq K_b)\).
Assume \(\widehat M_b\) is decoded by one-time-pad using \(\widehat K_b\) at Bob. Then the following holds:
\begin{equation}
  \textsf{P}_{\mathrm{e,OT}}=\textsf{P}_{\mathrm{e,key}}
  \;\le\;
  \textsf{P}_{\mathrm{e,hin}}.
\end{equation}
\end{lemma}

\begin{proof}
Define the events
\(\mathcal{E}_{\mathrm{hin}}:=\{\widehat U_{\mathcal{S}_b}\neq U_{\mathcal{S}_b}\}\)
and \(\mathcal{E}_{\mathrm{key}}:=\{\widehat K_b\neq K_b\}\).
If \(\mathcal{E}_{\mathrm{hin}}^c\) occurs, i.e., \(\widehat U_{\mathcal{S}_b}=U_{\mathcal{S}_b}\), then
\(\widehat K_b=g(\widehat U_{\mathcal{S}_b})=g(U_{\mathcal{S}_b})=K_b\), so
\(\mathcal{E}_{\mathrm{key}}\) cannot occur, i.e.,
\(\mathcal{E}_{\mathrm{key}}\cap \mathcal{E}_{\mathrm{hin}}^c=\emptyset\), and thus
\(\Pr(\mathcal{E}_{\mathrm{key}}\cap \mathcal{E}_{\mathrm{hin}}^c)=0\).
Therefore,
\(\Pr(\mathcal{E}_{\mathrm{key}})
=\Pr(\mathcal{E}_{\mathrm{key}}\cap \mathcal{E}_{\mathrm{hin}})
+\Pr(\mathcal{E}_{\mathrm{key}}\cap \mathcal{E}_{\mathrm{hin}}^c)
=\Pr(\mathcal{E}_{\mathrm{key}}\cap \mathcal{E}_{\mathrm{hin}})
\le \Pr(\mathcal{E}_{\mathrm{hin}})\),
i.e., \(\textsf{P}_{\mathrm{e,key}}\le \textsf{P}_{\mathrm{e,hin}}\).
Finally, since \(\widehat M_b=(M_b\oplus K_b)\oplus \widehat K_b\), we have
\(\widehat M_b\neq M_b\) if and only if \(\widehat K_b\neq K_b\), and therefore
\(\textsf{P}_{\mathrm{e,OT}}=\textsf{P}_{\mathrm{e,key}}\).
\end{proof}

Let $\mathcal{A}\subseteq[n]$ denote the set of bit-channel indices $i$ for which the random $U_i$ is unknown to Bob a priori. In our OT construction, $\mathcal{A}=\mathcal{S}_b\cup\mathcal{R}_b$ includes the set of indices of random bits placed on \gls{gbcs} $U_{\mathcal{S}_b}=(U_i)_{i\in\mathcal{S}_b}$ , i.e., $\mathcal{S}_b$, and the set of indices of the random bits placed on \gls{bbcs}, i.e., $\mathcal{R}_b$. In addition, bit-channel indices belonging to $\mathcal{A}^c$ are frozen to zero. Let $i^\star:=\max(\mathcal{S}_b)$. Since \gls{scd} proceeds sequentially in the index order $1,2,\dots,n$, the decisions on the hash-input bits in $\mathcal{S}_b$ depend only on the decoder's past decisions up to time $i^\star$. Therefore, only unfrozen/unknown indices before $i^\star$ can influence the correctness of $\widehat U_{\mathcal{S}_b}$ through error propagation. This motivates the prefix set
$\mathcal{A}_{\le i^\star}:=\mathcal{A}\cap\{1,2,\dots,i^\star\}$, which collects exactly those indices that are decided by the SCD and occur no later than the last hash-input index.

\begin{lemma}\label{lem:prefix-bound}
Using SCD, the decoding error of the hash input satisfies
\begin{align}
    \textsf{P}_{\mathrm{e,hin}}    \le \Pr(\exists\,j\in\mathcal{A}_{\le i^\star}:\ \widehat U_j\neq U_j)
    \le \sum_{j\in\mathcal{A}_{\le i^\star}} Z(W_n^{(j)})    
\end{align}

\end{lemma}

\begin{proof}
Recall $\mathcal{E}_{\mathrm{hin}}:=\{\widehat U_{\mathcal{S}_b}\neq U_{\mathcal{S}_b}\}$ and define
$\mathcal{E}_{\mathrm{pre}}:=\{\exists\,j\in\mathcal{A}_{\le i^\star}:\ \widehat U_j\neq U_j\}$.
If $\mathcal{E}_{\mathrm{pre}}^c$ occurs, then all indices in $\mathcal{A}_{\le i^\star}$ are decoded correctly. Since $\mathcal{S}_b\subseteq\{1,\dots,i^\star\}$, this implies $\widehat U_{\mathcal{S}_b}=U_{\mathcal{S}_b}$, i.e., $\mathcal{E}_{\mathrm{hin}}$ cannot occur. Equivalently, $\mathcal{E}_{\mathrm{hin}}\cap \mathcal{E}_{\mathrm{pre}}^c=\emptyset$. Hence $\mathcal{E}_{\mathrm{hin}}  =
  \bigl(\mathcal{E}_{\mathrm{hin}}\cap \mathcal{E}_{\mathrm{pre}}\bigr)
  \,\cup\,  \bigl(\mathcal{E}_{\mathrm{hin}}\cap \mathcal{E}_{\mathrm{pre}}^c\bigr)  =
  \mathcal{E}_{\mathrm{hin}}\cap \mathcal{E}_{\mathrm{pre}},$ which implies that $
  \Pr(\mathcal{E}_{\mathrm{hin}})  =  \Pr(\mathcal{E}_{\mathrm{hin}}\cap \mathcal{E}_{\mathrm{pre}})
  \le  \Pr(\mathcal{E}_{\mathrm{pre}})$. For the second inequality, we can apply the standard bound from polar codes analysis \cite{arikan2009} to the set $\mathcal{A}_{\le i^\star}$ to obtain $\Pr(\mathcal{E}_{\mathrm{pre}})\le \sum_{j\in\mathcal{A}_{\le i^\star}} Z(W_n^{(j)})$.
\end{proof}

Combine Lemma~\ref{lem:rel-bridge} and Lemma~\ref{lem:prefix-bound}, we can have a simple upper bound of $\textsf{P}_{\mathrm{e,OT}}$ as follows:
\begin{align}\label{EQ_Pe_UB}
  \textsf{P}_{\mathrm{e,OT}} = \textsf{P}_{\mathrm{e,key}}
  \le \sum_{j\in\mathcal{A}_{\le i^\star}} Z(W_n^{(j)}).  
\end{align}
  
However, the rightmost term takes into account the error probabilities of \gls{bbcs}, which implies that it is loose.
In the following, we develop an upper bound on $\textsf{P}_{\mathrm{e,OT}}$ with a prescribed confidence level via
Monte--Carlo simulation.
In each trial, we declare an error if the polar-decoding is wrong, i.e., Bob fails to reconstruct the hash input for key generation. This error event implies an OT error, and thus its probability upper-bounds
$\textsf{P}_{\mathrm{e,OT}}$. Consequently, it suffices to upper-bound this error probability from Monte--Carlo
simulation, which yields a relaxed version of the reliability requirement in~\eqref{EQ_reliability}.

\begin{lemma}\label{lem:cp-ot}
Fix the protocol parameters and run $\textsf{M}$ independent Monte--Carlo trials. Let $k\in\{0,\dots,\textsf{M}\}$
be the number of errors.
For any $\delta\in(0,1)$, define $\overline p_{\mathrm{CP}}(k;\textsf{M},\delta)$ as the unique $u\in(0,1]$ satisfying
$\Pr(X\le k)=\delta$, where $X\sim\mathrm{Bin}(\textsf{M},u)$. Then
\begin{align}\label{EQ_reliability_to_PeGBC}
\Pr\!\Big(\textsf{P}_{\mathrm{e,OT}}\le \overline p_{\mathrm{CP}}(k;\textsf{M},\delta)\Big)\ \ge\ 1-\delta .
\end{align}
\end{lemma}

The proof is relegated to Appendix \ref{APP_proof_lem:cp-ot}.

\begin{remark}
In principle, we want a deterministic reliability constraint
$\textsf{P}_{\mathrm{e,OT}}\le \varepsilon$ as in \eqref{EQ_Pe_UB}. However, obtaining a tractable analytic upper bound on $\textsf{P}_{\mathrm{e,OT}}$ with sufficient tightness is difficult.
We therefore relax this deterministic constraint by a statistical one
based on Monte--Carlo experiments. In particular, we run the polar code decoding $\textsf{M}$ times and let
$K:=\sum_{t=1}^  \textsf{M} E_t$ be the number of errors, where
$E_t:=\mathds{1}\{\text{the $t$-th trial fails}\}$ as defined in Lemma~\ref{lem:cp-ot}.
Let $p:=\textsf{P}_{\mathrm{e,hin}}$, we have $K\sim\mathrm{Bin}(  \textsf{M},p)$.
Given the observation $K=k$, we can compute the upper confidence bound
$\overline p_{\mathrm{CP}}(k;  \textsf{M},\delta)$, which is a deterministic number.
Then we can guarantee \eqref{EQ_reliability_to_PeGBC}. Therefore, enforcing the verifiable condition $\overline p_{\mathrm{CP}}(k;  \textsf{M},\delta)\le \varepsilon$
implies that $\textsf{P}_{\mathrm{e,OT}}\le \varepsilon$
with confidence at least $1-\delta$.
\end{remark}

\begin{example}
Assume we use $\textsf{M}=10^6$ i.i.d.\ trials to conduct the Monte-Carlo simulation, which measures an empirical error probability $\widehat p=10^{-3}$. Let $\delta=10^{-6}$. Then, we have $K=\widehat p\cdot   \textsf{M}=1000$ errors. Since $K\sim\mathrm{Bin}(  \textsf{M},\textsf{P}_{\mathrm{e,hin}})$, Lemma~\ref{lem:cp-ot} yields
$\Pr(\textsf{P}_{\mathrm{e,OT}}\le \overline p_{\mathrm{CP}}(K;  \textsf{M},\delta))\ge 1-\delta$, where for $K<  \textsf{M}$,
$\overline p_{\mathrm{CP}}(K;  \textsf{M},\delta)=I^{-1}_{1-\delta}(K+1,  \textsf{M}-K)$, where \(I_x(a,b):=\frac{1}{B(a,b)}\int_0^x t^{a-1}(1-t)^{b-1}\,dt,\, B(a,b):=\;\int_{0}^{1} t^{a-1}(1-t)^{b-1}\,dt\) is the beta function.
Substituting in $(K,  \textsf{M},\delta)=(1000,10^6,10^{-6})$ gives
$\overline p_{\mathrm{CP}}(1000;10^6,10^{-6}) = 1.16\times 10^{-3}$. Therefore, with probability at least $1-10^{-6}$, $\textsf{P}_{\mathrm{e,OT}}\le 1.1586\times 10^{-3}$.
\end{example}

\subsection{\gls{sfa} and \gls{sfb}}
% Requires: amsmath, amssymb, amsthm (or your theorem envs)

In this section, we prove \gls{sfa} and \gls{sfb}. We first do the following setup. Over a BI--AWGN channel with \gls{snr} $\rho$, polarization splits the bit-channels into \gls{gbcs} and \gls{bbcs} indices. In particular, in the asymptotic case, \gls{gbcs} behave as noiseless (non-erasures) and \gls{bbcs} as useless
(erasures). In the asymptotic regime, no privacy-amplification is needed. In contrast, at finite $n$ the impact of imperfections and the proposed solutions are discussed in Sec.~\ref{subsec_finite_n}. Fix $\mathbf A\in\Aut(\mathbf T)$ with order $\textsf{N}$ (cf. Definition \ref{def:cycle-decomp-type}) and let permutation matrices act on left multiplication, $\mathbf P:\mathbf T\mapsto \mathbf P\mathbf T$. Consider a vanishing sequence $\gamma_n\downarrow 0,\,n\in\mathds{N}$, and we parameterize the definition from \eqref{EQ_GBC_BBC_def} as follows:
\begin{align}
\mathcal{I}_{\mathcal G}(\mathbf T)&:=\{i:\ I_i^{(n)}(\mathbf T;\rho)\ge 1-\gamma_n\},\label{EQ_Def_GBC}\\
\mathcal{I}_{\mathcal B}(\mathbf T)&:=\{i:\ I_i^{(n)}(\mathbf T;\rho)\le \gamma_n\},\label{EQ_Def_BBC}
\end{align}
where we additionally parameterize $I_i^{(n)}$ as $I_i^{(n)}(\mathbf T;\rho)$, to emphasize that the bit-channel capacity is a function of the polarization matrix $\bT$ and the channel, i.e., here, via the \gls{snr} $\rho$ for the \gls{awgn} channel. Recall that the bit--channel mutual information
$I_i = I(U_i;Y^n,U^{i-1}),\,i\in [n]$ is defined by the joint
distribution $(U^n,Y^n)$ and do not depend on a particular decoder.
In particular, for any $i$ and any decoding strategy applied to $Y^n$, we have $I(U_i;\hat{U}_i)\le I(U_i;Y^n)\le I(U_i;Y^n,U^{i-1}) = I_i$. Thus, if an index is
classified into \gls{bbcs} in the sense $I_i\le\gamma_n$, no decoder can
extract more than $\gamma_n$ bits of information about that random bit.

In our OT protocol, Alice injects independent uniform random bits on selected
indices. Although these bits are random a priori, at finite blocklength the
corresponding bit--channels are not perfectly polarized, so mutual
information of \gls{bbcs} may be small but not close to zero. As a result, Bob's channel output
(and any decoder side information, including previously decoded bits) can still
be statistically correlated with these injected bits, leading to nonzero leakage.
Moreover, the public permutation/automorphism used in the protocol may map some
of these indices to positions that are comparatively more reliable for Bob,
which can further increase this finite-\(n\) leakage. Therefore, we require an
explicit upper bound on how much information Bob can obtain about the injected
random bits to design the privacy-amplification that removes the
residual leakage. The following lemma quantifies this leakage in terms of the
bit--channel mutual information.

\begin{lemma}\label{lem:bbc-leakage}
Let \(\gamma_n\in[0,1]\) and fix any subset \(\mathcal{S}\subseteq\mathcal{I}_{\mathcal B}(\gamma_n)\) and list its elements as
\(i_1<\dots<i_{|\mathcal{S}|}\).
Assume \(U_i\sim\mathrm{Bern}(\tfrac12)\) independently for all \(i\in\mathcal{S}\), and set \(U_i:=0\) for all
\(i\in \mathcal{I}_{\mathcal B}(\gamma_n)\setminus\mathcal{S}\).
Then, we have
\[
  I\bigl(U_{\mathcal{S}};\,Y^n\bigr)  \;\le\;  |\mathcal{S}|\,\gamma_n.
\]
\end{lemma}

The proof is relegated to Appendix~\ref{APP_proof_lem:bbc-leakage}. In particular,
if Alice places independent random bits on a subset
$\mathcal{S}\subseteq\mathcal{I}_{\mathcal B}(\gamma_n)$, even though these bits cannot be
reliably decoded by Bob, the total information leakage  is at most $|\mathcal{S}|\,\gamma_n$, which can be removed
by privacy amplification. Detailed discussion will be given in Section \ref{subsec_finite_n}.

\begin{remark}\label{rem:S_tradeoff}
The choice of the selected set $\mathcal{S}$ induces a tradeoff among
(leakage, reliability, OT rate), as follows.
\begin{itemize}
\item \emph{Leakage impact:}
Lemma~\ref{lem:bbc-leakage} upper bounds the leakage due to the injected random bits on
selected \gls{bbcs} by $|\mathcal{S}|\,\gamma_n.$
Hence, enlarging $|\mathcal{S}|$ increases the leakage bound linearly.
Moreover, polar code analysis yields $\gamma_n=2^{-n^{\beta}}$ for any
$\beta\in(0,\tfrac12)$. Therefore, if we select \(|\mathcal S|=\Theta(n)\), then the OT bit-rate contributed by these
\(|\mathcal S|\) bits is as
$\textsf{R}_{\mathcal S}\;:=\;\frac{|\mathcal S|}{n}\;=\;\Theta(1)$. Moreover, the leakage vanishes for the usual polar choice \(\gamma_n = 2^{-n^{\beta}}\) as $
I(U_{\mathcal S};Y^n)\ \le\ |\mathcal S|\,\gamma_n\ \le\ n\,2^{-n^{\beta}}\ \to\ 0.$

\item \emph{Reliability impact:}
Lemma~\ref{lem:cp-ot} can be stated as follows: after running \(\textsf{M}\) trials and observing
\(K=k\) errors, one can certify
$  \textsf{P}_{\mathrm{e,OT}}\le \overline p_{\mathrm{CP}}(k;\textsf{M},\delta)
  \text{ with confidence at least } 1-\delta .$ The confidence parameter \(1-\delta\) is chosen independently of \(|\mathcal S|\).
What \emph{does} depend on \(|\mathcal S|\) is the underlying error probability
\(p(|\mathcal S|):=\textsf{P}_{\mathrm{e,hin}}(|\mathcal S|)\), and hence the typical realization of
\(K\sim\mathrm{Bin}(\textsf{M},p(|\mathcal S|))\). Lemma~\ref{lem:prefix-bound} provides a loose estimate for this dependence: $  p(|\mathcal S|)\le \overline p_{\mathrm{hin}}(|\mathcal S|)
  := \sum_{j\in\mathcal{A}_{\le i^\star}} Z\!\bigl(W_n^{(j)}\bigr).$ Enlarging \(|\mathcal S|\) may increase \(i^\star\) and thus enlarge \(\mathcal{A}_{\le i^\star}\). Since
\(Z(W_n^{(j)})\ge 0\), this implies that \(\overline p_{\mathrm{hin}}(|\mathcal S|)\) can only increase.
Note that this only says that
our \emph{available upper bound} on \(p(|\mathcal S|)\) may become looser when \(|\mathcal S|\) grows.

For fixed \((\textsf{M},\delta)\), the map \(k\mapsto \overline p_{\mathrm{CP}}(k;\textsf{M},\delta)\) is
nondecreasing\footnote{For fixed \((\textsf{M},\delta)\), \(\overline p_{\mathrm{CP}}(k;\textsf{M},\delta)\) is obtained by inverting the binomial CDF: it is the largest \(p\) such that \(\Pr(\mathrm{Bin}(\textsf{M},p)\le k)\ge \delta\).
Since \(\Pr(\mathrm{Bin}(\textsf{M},p)\le k)\) is nondecreasing in \(k\) and
nonincreasing in \(p\) (the binomial distribution shifts to the right as \(p\) increases), the inverted
bound \(k\mapsto \overline p_{\mathrm{CP}}(k;\textsf{M},\delta)\) is nondecreasing.
}. Therefore, larger observed error counts \(k\) lead to larger certified bounds
\(\overline p_{\mathrm{CP}}(k;\textsf{M},\delta)\).
To express a target reliability requirement \(\textsf{P}_{\mathrm{e,OT}}\le \varepsilon\) in terms of the
observable \(K\), define the maximal admissible number of errors
$  k_{\max}(\varepsilon)  := \max\bigl\{k:\overline p_{\mathrm{CP}}(k;\textsf{M},\delta)\le \varepsilon\bigr\}.$ Then, by monotonicity in \(k\), if we want $
  \overline p_{\mathrm{CP}}(k;\textsf{M},\delta)\le \varepsilon$, then we must follow $  k\le k_{\max}(\varepsilon).$ Hence, meeting the target is equivalent to observing at most \(k_{\max}(\varepsilon)\) errors. In
particular, for \(K\sim\mathrm{Bin}(\textsf{M},p)\), the success probability
\(\Pr[K\le k_{\max}(\varepsilon)]\) is nonincreasing in \(p\). Thus, whenever the underlying error
probability becomes larger, the target becomes harder to reach.
%A more rigorous derivation is shown in Appendix \ref{APP_proof_tradeoff}.

\item \emph{OT-rate impact:}
The OT rate is determined by the maximum key length
that simultaneously satisfies (i) \gls{sfa} and \gls{sfb}, whose finite-blocklength bounds are affected by the
leakage contribution \(I(U_{\mathcal S};Y^n)\) and (ii) a reliability constraint certified from the
observed test outcome $K=k$ via $\overline p_{\mathrm{CP}}(k;\textsf{M},\delta)$.
Thus, enlarging $\mathcal{S}$ can improve the nominal OT payload, but it also reduces the output length of privacy amplification through the leakage bound and may violate the reliability constraint by increasing the typical
observed $k$. Therefore, $\mathcal{S}$ should be properly designed and a more detailed investigation of this issue is derived in Section \ref{subsec_finite_n}.
\end{itemize}
\end{remark}

To prove \gls{sfa} and \gls{sfb}, we first establish two auxiliary lemmas. Recall in the proposed protocol, Bob selects a private bit $B\in\{0,1\}$ and chooses the polarization matrix and a decoder corresponding to it as follows:
\[
\bT_B = \left\{
\begin{array}{ll}
  \bT_1, & \text{if } B=0,\\
  \bT_2, & \text{if } B=1.
\end{array}
\right.
\]
Bob also selects a private permutation $\mathbf P_1\in\mathcal{P}\subseteq\Aut(\bT_1)$ with induced index
permutation $\pi_{\mathbf P_1}$, where $\mathcal{P}$ is defined in the proposed protocol. Then Bob shares
\begin{equation}\label{eq:F-published}
\ {\ \mathbf F \ :=\ \mathbf P_1^{\mathsf T}\,\bT_B\ }
\end{equation}
with Alice together with the two disjoint index sets
\begin{equation}\label{eq:I0I1-def}
\ {\;
\mathcal J_0  \!:= \!\big(\pi_{\mathbf P_1}\big(\mathcal{I}_{\mathcal G}(\bT_B)\big)\big)_{\downarrow \ell},\,
\mathcal J_1  \!:=\! \big(\pi_{\mathbf P_1}\big(\mathcal{I}_{\mathcal B}(\bT_B)\big)\big)_{\downarrow \ell},
\;}
\end{equation}
where $\ell$ is a parameter to be designed and will be shown in Theorem \ref{thm:polarOT-AWGN}, and recall a truncation by taking $\ell$ entries is denoted by
$(\cdot)_{\downarrow \ell}$. Finally, Bob publishes the pair
  $(\tilde{\mathcal{J}}_{0},\tilde{\mathcal{J}}_{1})=(\mathcal{J}_{0},\mathcal{J}_{1})$ if $B=0$;
  $(\tilde{\mathcal{J}}_{0},\tilde{\mathcal{J}}_{1})=(\mathcal{J}_{1},\mathcal{J}_{0})$, else, over the public channel and defines
\begin{equation}\label{eq:Pi-def}
\Pi_{\mathrm{sel}} \;:=\; (\mathbf F,\tilde{\mathcal{J}}_{0},\tilde{\mathcal{J}}_{1}),
\qquad
\Pi_{\mathrm{pub}} \;:=\; \bigl(\Pi_{\mathrm{sel}},\,S,C_0,\,C_1 \bigr),
\end{equation}
where \(\Pi_{\mathrm{pub}}\) denotes the complete public information, including $\Pi_{\mathrm{sel}}$ and the one-time-padded messages \( C_b := M_b \oplus K_b\), and $S$, the seed to select hash functions from a \gls{uhf}, which is independent of all other random variables. By construction, \(K_b\) is a deterministic function of \((Y^n,\Pi_{\mathrm{sel}},S)\), i.e.,
\(K_b = f_b(Y^n,\Pi_{\mathrm{sel}})\), while \((M_0,M_1)\) are chosen independently of
\((U^n,Y^n,\Pi_{\mathrm{sel}})\). 

% ================= PRE-LEMMA SETUP =================
% ===================== SETUP =====================
Recall if $\pi_{\mathbf A}=C_1\cdots C_r$ has cycle lengths $\ell_1,\dots,\ell_r$ and $  \textsf{N}:=\mathrm{lcm}(\ell_1,\dots,\ell_r)$, then $  \textsf{N}$ is the order of $\pi_{\mathbf A}$ (and of the permutation matrix $\mathbf A$). Consequently, $\mathbf A^{  \textsf{N}}=\mathbf I$ and $\mathbf A^{t}\ne \mathbf I$ for $1\le t<  \textsf{N}$, so the powers $\mathbf A^{0},\dots,\mathbf A^{  \textsf{N}-1}$ are pairwise distinct. Intuitively, we want Bob's choice bit \(B\) to be hidden in the public
matrix \(\mathbf F\). Our construction randomizes the polarization matrix by cycling
through all powers (within the order) of a fixed automorphism \(\mathbf A\).
The next lemma shows that the distribution of \(\mathbf F\) does not depend on \(B\) as long as the
power \(K\) is chosen uniformly from $\mathrm{Unif}\bigl(\{0,1,\dots,\textsf{N}-1\}\bigr)$.

%============= Two-option lemma (combined) =============
\begin{lemma}\label{lem:cyclic-hideB}
Let \(\mathbf A\in\Aut(\mathbf T)\) and let \(\pi_{\mathbf A}\) be its induced
permutation. Let \(\pi_{\mathbf A}=C_1\cdots C_r\) as disjoint cycles with lengths
\(\ell_1,\dots,\ell_r\), and set \(  \textsf{N}:=\mathrm{lcm}(\ell_1,\dots,\ell_r)\).
Define \(\mathbf T_2:=\mathbf A\mathbf T_1\) and the set
$\mathcal P:=\{\mathbf A^{k}:\ 0\le k<  \textsf{N}\}.$
Let \(K\sim\Unif(\{0,\ldots,  \textsf{N}-1\})\) be a local randomness and is independent of \(B\), set \(\mathbf P_1:=\mathbf A^{K}\), and define
\(\mathbf F:=\mathbf P_1^{\mathsf T}\mathbf T_B\). Then \(\mathbf F\pperp B\).
\end{lemma}
The proof is relegated to Appendix \ref{APP_proof_cyclic-hideB}.

\begin{remark}\label{rem:cycle-period}
 In Sec.~III-A and III-B, $\pi_{\mathbf A}$ swaps two disjoint pairs and fixes the
  others, e.g. $\pi_{\mathbf A}=(11\ 8)(9\ 6)\times(\text{fixed indices})$. Hence $  \textsf{N}=\mathrm{lcm}(2,2,1,\ldots)=2$.
  Selecting $\mathbf P_1$ uniformly from the two options $\mathcal P=\{\mathbf I,\mathbf A\}$ (independent of $B$) is exactly the $  \textsf{N}=2$ case and yields $\mathbf F\pperp B$.
 If $  \textsf{N}>2$, two options are in general insufficient, which can be easily seen from the proof of Lemma \ref{lem:cyclic-hideB}. 
\end{remark}

Combining Lemma~\ref{lem:cyclic-hideB} and the leftover-hash lemma, we obtain the following secrecy guarantees.

\begin{thm}\label{thm:polarOT-AWGN}
With the construction above, assume the chosen $\bA\in\Aut(\bT_1)$ satisfies the cross-cut swap
\begin{align}
    \mathcal{I}_{\mathcal G}(\bT_2)=\pi_{\bA}(\mathcal{I}_{\mathcal B}(\bT_1)),\qquad
    \mathcal{I}_{\mathcal B}(\bT_2)=\pi_{\bA}(\mathcal{I}_{\mathcal G}(\bT_1)),
    \quad\text{where }\bT_2:=\bA\bT_1.    \label{EQ_cross_cut_swap}
\end{align}

Define Bob's total information as
\(
T := (Y^n,\Pi_{\mathrm{pub}},B)
\)
and let \(\ell = |\mathcal J_{\bar b}|=|\mathcal J_{b}|\) for each \(b\in\{0,1\}\).
Let \(V_{\bar b}\) denote the hash-input random variable used to generate the
unchosen key \(K_{\bar b}\), and let
\(K_{\bar b} := h_{S}(V_{\bar b})\), where \(h_{\sigma}\) is drawn uniformly at random
from a \gls{uhf} family with public seed \(S\), independent of \((Y^n,\Pi_{\mathrm{sel}},B)\).
Fix any smoothing parameter \(\varepsilon_{\mathrm{sm}}\in(0,1)\), and define, for each \(b\),
\begin{equation}\label{eq:eps-def-new}
  \varepsilon_b
  :=
  2\varepsilon_{\mathrm{sm}}
  \;+\;
  \tfrac12 \sqrt{2^{\ell - H_{\min}^{\varepsilon_{\mathrm{sm}}}(V_{\bar b}\mid Y^n,\Pi_{\mathrm{sel}},S,B=b)}},
\end{equation}
then the following hold:
\begin{align}
\text{\gls{sfa}: }\quad d_{\mathrm{var}}\!\bigl(P_{M_{\bar B}\,T},\;P_{M_{\bar B}}\times P_{T}\bigr) \;\le\; 2\max\{\varepsilon_1,\varepsilon_2\},\label{eq:SfA_new}
\\
\text{\gls{sfb}: }\quad
d_{\mathrm{var}}\!\big(P_{B\,M_0\,M_1\,X^n\,\Pi_{\mathrm{pub}}},\ P_B\times P_{M_0\,M_1\,X^n\,\Pi_{\mathrm{pub}}}\big)
&=0.
\label{eq:SfB_new}
\end{align}
\end{thm}
The proof is relegated to Appendix \ref{APP_proof_thm:polarOT-AWGN}.

\begin{remark}\label{rem:RB_screening_SfA}
In this remark we discuss whether including Bob's local randomness \(R_{\mathrm B}\) in the side
information affects \gls{sfa}. Under the honest protocol, there exists a
deterministic map \(\eta\) such that $\Pi_{\mathrm{sel}}=\eta(B,R_{\mathrm B})$
from Protocol Steps~1--2. In Protocol Step~3, Alice generates \(U^n\) using only
\(\Pi_{\mathrm{sel}}\) and her local randomness \(R_{\mathrm A}\), and the
channel output is then formed using only the AWGN noise \(N^n\). Hence there
exist deterministic maps \(\varphi\) and \(\psi_b\) such that, for each
\(b\in\{0,1\}\), $U^n=\varphi(\Pi_{\mathrm{sel}},R_{\mathrm A}),\,
(V_{\bar b},Y^n)=\psi_b(\Pi_{\mathrm{sel}},R_{\mathrm A},N^n),$
where \(N^n\) denotes the AWGN. Since \((R_{\mathrm A},N^n)\pperp
R_{\mathrm B}\) and \(\Pi_{\mathrm{sel}}\) is part of \(\Pi_{\mathrm{pub}}\),
we obtain the Markov chain
$R_{\mathrm B}\;-\;(\Pi_{\mathrm{pub}},B)\;-\;(V_{\bar B},Y^n).$
In particular, for each \(b\), $P_{V_{\bar b}\mid Y^n,\Pi_{\mathrm{pub}},B=b,R_{\mathrm B}}
=P_{V_{\bar b}\mid Y^n,\Pi_{\mathrm{pub}},B=b}.$
Now consider the privacy amplification step in SfA. Recall that \(S\) is the published
seed contained in \(\Pi_{\mathrm{pub}}\), and define the unchosen key
\(K_{\bar b}:=h_{\bar b}(V_{\bar b};S)\), which is a deterministic function of
\((V_{\bar b},\Pi_{\mathrm{pub}},B=b)\). Let \(E_b:=(Y^n,\Pi_{\mathrm{pub}},B=b)\), the above implies $
P_{K_{\bar b}\mid E_b,R_{\mathrm B}}
=P_{K_{\bar b}\mid E_b}.$
Extract $E_b$ and $R_B$ from the variation distance, we have
\begin{align}
 d_{\mathrm{var}}\!\bigl(P_{K_{\bar b},E_b,R_{\mathrm B}},\;\Unif\times P_{E_b,R_{\mathrm B}}\bigr)
  &= \mathbb{E}_{E_b,R_{\mathrm B}}\!\Big[
    d_{\mathrm{var}}\!\bigl(P_{K_{\bar b}\mid E_b,R_{\mathrm B}},\Unif\bigr)\Big]\\
  &=
\mathbb{E}_{E_b}\!\Big[
d_{\mathrm{var}}\!\bigl(P_{K_{\bar b}\mid E_b},\Unif\bigr)
\Big]
=d_{\mathrm{var}}\!\bigl(P_{K_{\bar b},E_b},\Unif\times P_{E_b}\bigr),   
\end{align}
where the second equality is due to $P_{K_{\bar b}\mid E_b,R_{\mathrm B}}=P_{K_{\bar b}\mid E_b}.$
Hence, additionally considering \(R_{\mathrm B}\) as part of the side information when chekcing \gls{sfa} does not change the \gls{sfa} bound.
\end{remark}

\subsection{Characterization of $\Aut(\mathbf T)$}\label{sec_Aut_characterization}
In this section we show that every automorphism of \(\mathbf T=\mathbf F^{\otimes m}\) is induced by permuting the \(m\) bit positions, which translates the matrix condition into an equivalent poset viewpoint. To achieve the goal, we first show that the entries of \(\mathbf T\) are exactly the indicators of the bit-wise partial order on \(\mathcal X=\{0,1\}^m\), namely \(\mathbf T_{x,y}=1\) if and only if \(y\le x\) (Lemma~\ref{lem:polar-incidence}).
Next, we show that a permutation matrix \(\mathbf P_\pi\) satisfies \(\mathbf P_\pi^{\!\top}\mathbf T\mathbf P_\pi=\mathbf T\) if and only if the underlying relabeling \(\pi\) preserves this order,
\(x\le y \mbox{ iff } \pi(x)\le \pi(y)\) (Lemma~\ref{lem:T-poset-auto-merged}),
thereby identifying \(\Aut(\mathbf T)\) with the automorphism group of the poset \((\mathcal X,\le)\).
Finally, we use the combinatorial fact that every order-automorphism of \((\mathcal X,\le)\) must be a coordinate permutation (a bit-permutation), and that this permutation is unique (Lemma~\ref{lem:bool-aut}).
Combining these equivalences yields that all matrix automorphisms of \(\mathbf T\) come from permuting bit positions (Theorem~\ref{thm:AutT}), hence \(\Aut(\mathbf T)\cong\mathcal S_m\) and \(|\Aut(\mathbf T)|=m!\), rather than exhausting all permutations.

Recall that permutation matrices are defined in
Section~\ref{Sec_Permute_Aut} and recall $  \Aut(\mathbf T)
  \;:=\;  \bigl\{\mathbf P\in\{0,1\}^{n\times n}:\ \mathbf P^{\!\top}\mathbf T\,\mathbf P=\mathbf T\bigr\}$ for the automorphism group of $\mathbf T$. Let $\mathcal{X}:=\{0,1\}^m$ and recall the definition of partial order set in Definition \ref{Def_POSET} with $\leq$ defined bit-wise, i.e., $x\leq_b y\mbox{ iff } x_i\leq y_i,\,\forall i\in[m]$. We index the rows and columns of $\mathbf T=\mathbf F^{\otimes m}$ by the
elements of $\mathcal{X}$ and let $\mathbf T_{x,y}$ be the entry of $\bT$ in row $x$ and column $y$.

\begin{lemma}\label{lem:polar-incidence}
For every $m\ge 1$ and every $x,y\in\mathcal{X}$,
\begin{equation}\label{eq:T-encodes-order-compact}
  \mathbf T_{x,y}
  \;=\;  \mathds{1}\{\,y \le_b x\,\}.
\end{equation}
\end{lemma}

The proof is relegated to Appendix~\ref{APP_proof_lem:polar-incidence}.

Next we identify matrix automorphisms of \(\mathbf T\) with automorphisms of the
poset \((\mathcal X,\le)\).
For a bijection \(\pi:\mathcal X\to\mathcal X\), let \(\mathbf P_\pi\) denote the
corresponding permutation matrix whose \(x\)-th column is \(e_{\pi(x)}\), i.e.,
\begin{align}\label{EQ_P_pi-compact}
  (\mathbf P_\pi)_{u,x}
  \;=\;
  \begin{cases}
    1, & u = \pi(x),\\
    0, & \text{otherwise},
  \end{cases}
  \qquad u,x\in\mathcal X.
\end{align}
The next lemma shows exactly \(\mathbf T_{x,y}=\mathbf T_{\pi(x),\pi(y)}\) for all \(x,y\), i.e., when \(\pi\) preserves the underlying order relation \(y\le_b x\) represented by \(\mathbf T\), which is analogous to \cite[Thm.~2.1.6]{Knauer2011AGT}.

\begin{lemma}\label{lem:T-poset-auto-merged}
Let \(\pi:\mathcal{X}\to\mathcal{X}\) be a bijection and let \(\mathbf P_\pi\)
be defined by~\eqref{EQ_P_pi-compact}. Then the following are equivalent:
\begin{enumerate}
  \item \(\mathbf P_\pi\in\Aut(\mathbf T)\);
  \item   $x\le_b y \mbox{ iff } \pi(x)\le_b \pi(y),\text{for all }x,y\in\mathcal{X}.$
\end{enumerate}
In particular, the correspondence \(\pi \mapsto \mathbf P_\pi\) is one-to-one and onto, and it
identifies \(\Aut(\mathbf T)\) with the set of all poset automorphisms of \((\mathcal{X},\le_b)\).
\end{lemma}

For completeness and to keep the paper self-contained, we provide a full proof in Appendix~\ref{APP_proof_lem:T-poset-auto-merged}.

Note that Lemma~\ref{lem:T-poset-auto-merged} is stated for a general alphabet set \(\mathcal X\)
with the bitwise partial order \(\le_b\). To achieve our setting \(\mathcal X=\{0,1\}^m\) with \(\le_b\), we invoke the following known result: every automorphism of the poset \((\{0,1\}^m,\le_b)\) is induced by a unique permutation \(\sigma\in S_m\), i.e., a unique way to shuffle the \(m\) bit
positions of the index label (a coordinate permutation)~\cite[p.~44]{StanleyAlgComb}.

%we include a proof in Appendix~\ref{APP_proof_lem:bool-aut} for completeness.

\begin{lemma}\label{lem:bool-aut}
Let \(m\in\mathbb N\), \(\mathcal X:=\{0,1\}^m\) with the bit-wise partial order
\(x\le y \mbox{ iff } x_i\le y_i\) for all \(i\in[m]\).
Let \(\pi:\mathcal X\to\mathcal X\) be a bijection that preserves this order, i.e.,
$  x\le y \mbox{ iff } \pi(x)\le \pi(y)\,\text{for all }x,y\in\mathcal X.$ Then, for each \(i\in[m]\) there exists a unique \(j\in[m]\) such that
\(\pi(e_i)=e_j\), so \(\pi\) induces a unique \(\sigma\in\mathcal S_m\). Order preservation then forces \(\pi\) to act on every \(x\in\mathcal X\) by permuting coordinates according to \(\sigma\), i.e., \((\pi(x))_j = x_{\sigma^{-1}(j)}\) for all \(x\in\mathcal X\) and \(j\in[m]\).
\end{lemma}

% The proof is relegated to Appendix~\ref{APP_proof_lem:bool-aut}.

\medskip
Combining Lemma~\ref{lem:T-poset-auto-merged} with Lemma~\ref{lem:bool-aut}
gives us the following characterization of $\Aut(\mathbf T)$. Recall $\mathcal{S}_m$ is the symmetric group defined in Definition \ref{def:symmetric-group}.
\begin{thm}\label{thm:AutT}
Let $\mathbf T := \mathbf F^{\otimes m}$, with rows and columns indexed by
$\mathcal{X}:=\{0,1\}^m$. Then
\begin{align}
&\hspace{-0.3cm}\Aut(\mathbf T)
    =\Bigl\{\mathbf P_\sigma: \sigma\in\mathcal{S}_m,\,\mathbf P_\sigma e_{(x_1,\dots,x_m)}=e_{(x_{\sigma(1)},\dots,x_{\sigma(m)})}, \forall\,(x_1,\dots,x_m)\in\mathcal{X}
    \,\Bigr\}.    
\end{align}
\end{thm}

The proof is relegated to Appendix~\ref{APP_proof_thm:AutT}. By Theorem \ref{thm:AutT},
it is clear that \(\Aut(\mathbf{T})\cong \mathcal{S}_m\) and $|\Aut(\mathbf T)|=m!$.

\medskip
In our OT protocol, Bob may use a polarization matrix
$\mathbf T_{\mathbf P}:=\mathbf P\,\mathbf T$ obtained from $\mathbf T$
by a fixed row permutation $\mathbf P\in\Aut(\bT)$. The next corollary shows how
$\Aut(\mathbf T_{\mathbf P})$ is related to $\Aut(\mathbf T)$.

\begin{corollary}\label{cor:one-sided-iso-clean}
Let \(\mathbf P\in\Aut(\mathbf T)\) and define \(\mathbf T_{\mathbf P}:=\mathbf P\,\mathbf T\).
Then
\[
  \Aut(\mathbf T_{\mathbf P})
  \;=\;
  \bigl\{\,\mathbf Q\in\Aut(\mathbf T):\ \mathbf Q\,\mathbf P=\mathbf P\,\mathbf Q\,\bigr\}.
\]
Moreover, by Theorem~\ref{thm:AutT} there exists \(\sigma\in\mathcal S_m\) such that
\(\mathbf P=\mathbf P_\sigma\), and for \(\mathbf Q=\mathbf P_\tau\) we have
\(\mathbf Q\mathbf P=\mathbf P\mathbf Q\) if and only if \(\tau\sigma=\sigma\tau\).
Equivalently, $  \Aut(\mathbf T_{\mathbf P})  \;=\;  \bigl\{\,\mathbf P_\tau:\ \tau\in\mathcal S_m,\ \tau\sigma=\sigma\tau\,\bigr\}.$
\end{corollary}

% \begin{corollary}\label{cor:one-sided-iso-clean}
% Let $\mathbf P\in\Aut(\mathbf T)$ and define $\mathbf T_{\mathbf P}:=\mathbf P\,\mathbf T$.
% Then
% \[
%   \Aut(\mathbf T_{\mathbf P})
%   \;=\;
%   \bigl\{\,\mathbf Q\in\Aut(\mathbf T):\ \mathbf Q\,\mathbf P=\mathbf P\,\mathbf Q\,\bigr\}.
% \]
% % Moreover, identifying $\Aut(\mathbf T)\cong\mathcal{S}_m$ via
% % $\mathbf Q=\mathbf P_\tau$ and $\mathbf P=\mathbf P_\sigma$, we have
% % \[
% %   \Aut(\mathbf T_{\mathbf P})
% %   \;\cong\;
% %   \bigl\{\tau\in\mathcal{S}_m:\ \tau\sigma=\sigma\tau\bigr\}.
% % \]
% \end{corollary}

The proof is relegated to Appendix~\ref{APP_proof_cor:one-sided-iso-clean}.

% \myboxr{Connect to (the proof of) Lemma \ref{lem:cyclic-hideB}}{Check how it will affect Lemma \ref{lem:cyclic-hideB} and \gls{sfb} proof. In addition, how it will affect the OT rate.}

%=============================
% (Optional) References mentioned in text
%=============================
% Arıkan entry test:
%  E. Arıkan, “Channel polarization,” IEEE Trans. Inf. Theory, 2009.
% For the zeta-matrix viewpoint and conjugation = poset automorphisms:
%  R. P. Stanley, Enumerative Combinatorics, Vol. 1, 2nd ed., §3.6–3.7.

In the following, we use a simple example with $n=16$ to show that the existence of a solution, feasible to the conditions of selecting permutations uniformly from $\mathcal{P}$, required in Lemma \ref{lem:cyclic-hideB} and Theorem \ref{thm:polarOT-AWGN}, to guarantee the \gls{sfb}.

\begin{example}\label{ex:n16-119-75}
Let $n=16$ ($m=4$) and let $  \bT_1 = \bT = \mathbf F^{\otimes 4}.$ Let $\sigma_2$ be the bit-permutation
\[
  \sigma_2 : [b_3\,b_2\,b_1\,b_0] \mapsto [b_2\,b_3\,b_1\,b_0],
\]
and let $\bA = \bP_{\sigma_2}\in\Aut(\bT)$ be the corresponding
permutation matrix.  The induced permutation
$\pi_{\bA}:[16]\to[16]$ is defined by
$\bA e_i = e_{\pi_{\bA}(i)}$.  A direct check using the binary
representations\footnote{This example uses the $i\mapsto (i-1)$ binary indexing convention consistent with the definition of $\pi_{\bA}$ used earlier.}
\[
  11-1=(1010)_2,\;7-1=(0110)_2,\;9-1=(1000)_2,\;5-1=(0100)_2
\]
gives
\[
  \sigma_2(1010)=0110,\quad
  \sigma_2(0110)=1010,\quad
  \sigma_2(1000)=0100,\quad
  \sigma_2(0100)=1000,
\]
hence $  \pi_{\bA}(11)=7,\;  \pi_{\bA}(7)=11,\;  \pi_{\bA}(9)=5,\;  \pi_{\bA}(5)=9$.  Thus $\pi_{\bA}$ contains the two 2-cycles $(11\ 7)$ and $(9\ 5)$.
The remaining cycles are not
relevant for this construction since the corresponding bit-channels are frozen.

Recall that $I^{(n)}_i(\bT_1;\rho)$ denotes the mutual information of the $i$-th
bit-channel of $\bT_1$ at some fixed SNR~$\rho$.  For a threshold
$\gamma>0$, recall from \eqref{EQ_Def_GBC} and \eqref{EQ_Def_BBC}:
\[
  \mathcal{I}_{\mathcal G}(\bT_1)
  :=\bigl\{ i : I^{(n)}_i(\bT_1;\rho)\ge 1-\gamma\bigr\},
  \qquad
  \mathcal{I}_{\mathcal B}(\bT_1)
  :=\bigl\{ i : I^{(n)}_i(\bT_1;\rho)\le \gamma\bigr\}.
\]
For rate $\textsf{R}=1/2$ we choose the cut as in Table~\ref{Table_Aut_n16}, i.e.,
\[
  \mathcal{I}_{\mathcal G}(\bT_1) = \{16,15,14,13,12,11,10,9\},\qquad
  \mathcal{I}_{\mathcal B}(\bT_1) = \{8,7,6,5,4,3,2,1\}.
\]
We only use the four indices $\mathcal I_{\rm OT} := \{11,9,7,5\},$
with $\{11,9\}\subseteq\mathcal{I}_{\mathcal G}(\bT_1)\cap\mathcal I_{\rm OT}$ and
$\{7,5\}\subseteq\mathcal{I}_{\mathcal B}(\bT_1)\cap\mathcal I_{\rm OT}$.

Now let $  \bT_2 := \bA\bT_1.$ Row~$\sigma_2$ of Table~\ref{Table_Aut_n16} shows that the order of bit-channel
qualities under $\bT_2$ is
\[
  (16,15,14,13,8,7,6,5,12,11,10,9,4,3,2,1),
\]
so for the same rate $\textsf{R}=1/2$ we have $ \mathcal{I}_{\mathcal G}(\bT_2)
   = \{16,15,14,13,8,7,6,5\},\;  \mathcal{I}_{\mathcal B}(\bT_2)   = \{12,11,10,9,4,3,2,1\}.$ Masking by $\mathcal I_{\rm OT}$, we have
\begin{align}
  \mathcal{I}_{\mathcal G}(\bT_2)\cap\mathcal I_{\rm OT}
    &= \{7,5\}
     = \pi_{\bA}\bigl(\mathcal{I}_{\mathcal G}(\bT_1)\cap\mathcal I_{\rm OT}\bigr),\notag\\
  \mathcal{I}_{\mathcal B}(\bT_2)\cap\mathcal I_{\rm OT}
    &= \{11,9\}
     = \pi_{\bA}\bigl(\mathcal{I}_{\mathcal B}(\bT_1)\cap\mathcal I_{\rm OT}\bigr).
\end{align}
Thus the two 2-cycles $(11\ 7)$ and $(9\ 5)$ of $\pi_{\bA}$ each connect
a \gls{gbcs} of $\bT_1$ with a \gls{bbcs} of $\bT_1$, while
the roles are reversed under $\bT_2$. This realizes exactly the
cross-cut pairing required by \eqref{EQ_cross_cut_swap}.

In Lemma \ref{lem:cyclic-hideB} we assume $\bA\in\Aut(\bT_1)$ with cycle lengths
$\ell_1,\dots,\ell_r$, define \(\textsf{N}:=\mathrm{lcm}(\ell_1,\dots,\ell_r)\) and
\[
\mathcal P=\{\bA^k:0\le k<\textsf{N}\},\qquad \bT_2=\bA\bT_1,
\]
and introduce a local random exponent $K\sim\Unif(\{0,\dots,\textsf{N}-1\})$
independent of Bob's bit $B$. In our instance we have $\textsf{N}=2$ and
$\mathcal P=\{\bI,\bA\}$. Recall that $B\in\{0,1\}$ is Bob's choice and let
$  \bT_B :=     \bT_1, \, B=0; \bT_B :=  \bT_2,  B=1,$ and we select $K\sim\Unif(\{0,1\})$ independent of $B$, set
$\bP_1 := \bA^K\in\mathcal P$, and define $\bF := \bP_1^{\mathsf T}\bT_B$.
Hence this construction is exactly the $\textsf{N}=2$ specialization of
Lemma \ref{lem:cyclic-hideB} with the symmetric \gls{gbcs} and \gls{bbcs} pairing
$  \{11,9\}\longleftrightarrow\{7,5\}$ induced by $\pi_{\bA}$.
\end{example}

\subsection{Unify the orders from real channel and Table \ref{Table_Aut_n16}} \label{Sec_unify_order}
In our OT construction, the design based on $\Aut(\bT)$ is carried out under a canonical reliability order $ \mathcal{O}_{can}$, e.g., $\mathcal{O}_{can}=(16,15,14,13,12,11,10,9,8,7,6,5,4,3,2,1)$, when $n=16$, with a fixed \gls{gbcs} and \gls{bbcs} partition at rate $\textsf{R}=\tfrac12$, e.g., $\mathcal{I}_{\mathcal{G}} = \{16,15,14,13,12,11,10,9\}$ and $\mathcal{I}_{\mathcal{B}} = \{8,7,6,5,4,3,2,1\}.$ However, for the physical BI--AWGN channel at a fixed SNR, the {true} bit-channel reliability order is $\mathcal{O}_{\rm real} \;:=\; (i_1,i_2,\dots,i_{16})$ by polar code analysis, which in general does not coincide with $\mathcal{O}_{\rm can}$. For example, numerically we may obtain $\mathcal{O}_{\rm real}=(16,15,14,12,8,13,11,10,7,6,4,9,5,3,2,1)$. Let $\mathcal{O}_{\rm real}=(i_1,i_2,\dots,i_n)$ list indices from most to least reliable at the operating SNR. Define the relabeling permutation
$\pi_{\rm rel}\in\mathcal{S}_n$ by $ \pi_{\rm rel}(i_t)=n+1-t,\, t=1,\dots,n,$ so that under the new labels, the physical order $\mathcal{O}_{\rm real}$ is
mapped to the canonical order $\mathcal{O}_{\rm can}=(n,n-1,\dots,1)$. Let \(\bP_{\rm rel}\) be the permutation matrix of \(\pi_{\rm rel}\), and represent
the same physical vectors under the new labels by
\[
\widetilde{u}^n := u^n\bP_{\rm rel}^{-1},
\qquad
\widetilde{y}^n := y^n\bP_{\rm rel}^{-1}.
\]
Then, using \(y^n=u^n\bF\) and \(u^n=\widetilde{u}^n\bP_{\rm rel}\), we obtain
\[
\widetilde{y}^n
= y^n\bP_{\rm rel}^{-1}
= u^n\bF\bP_{\rm rel}^{-1}
= \widetilde{u}^n \bigl(\bP_{\rm rel}\bF\bP_{\rm rel}^{-1}\bigr).
\]
Hence the matrix that represents the same physical coordinate permutation under the
new labels is
\(
\widetilde{\bF} := \bP_{\rm rel}\bF\bP_{\rm rel}^{-1}.
\) Accordingly, for every protocol-relevant permutation $\bP$ we write its relabeled
version as 
\begin{align}\label{EQ_conjugation}
  \widetilde{\bP} \;:=\; \bP_{\rm rel}\,\bP\,\bP_{\rm rel}^{-1}.
\end{align}
Likewise, for any index set $\mathcal{S}\subseteq[n]$ we relabel it by
$  \widetilde{\mathcal{S}} \;:=\; \pi_{\rm rel}(\mathcal{S})  \;=\;\{\pi_{\rm rel}(i): i\in\mathcal{S}\}.$

% This mismatch matters because the proofs of Lemma~\ref{lem:cyclic-hideB},
% Corollary~\ref{lem:markov}, and Theorem~\ref{thm:polarOT-AWGN} rely crucially
% on the cycle structure of $\bA\in\Aut(\bT)$ relative to
% $(\mathcal{G}_{\rm can},\mathcal{B}_{\rm can})$ and on the commutation
% condition $\bQ\bP=\bP\bQ$ for $\bP,\bQ\in\Aut(\bT)$.

We emphasize that the relabeling $\pi_{\rm rel}\in\mathcal{S}_n$ is not a new
protocol operation but a purely notational tool. Its only purpose is to
resolve the indexing mismatch between the physical reliability order
$\mathcal{O}_{\rm real}$ induced by the BI--AWGN channel at the operating SNR
and the canonical order $\mathcal{O}_{\rm can}$ used to tabulate and reason about
$\Aut(\bT)$ (e.g., Table~\ref{Table_Aut_n16}). Concretely, after renaming indices
via $\pi_{\rm rel}$, we can describe the same physical bit-channels using the
canonical labels, while the underlying channel and the actual coordinate
permutations applied to codewords remain unchanged. This step matters because the proofs of Lemma~\ref{lem:cyclic-hideB} and Theorem~\ref{thm:polarOT-AWGN} are formulated in terms
of (i) the cycle structure of $\bA\in\Aut(\bT)$ relative to a fixed
\gls{gbcs}/\gls{bbcs} partition and (ii) the commutation relation $\bQ\bP=\bP\bQ$
for $\bP,\bQ\in\Aut(\bT)$. The key point is that a consistent relabeling preserves
both properties: it does not change the underlying permutation pattern, but only
renames the indices.

To see the effect of \eqref{EQ_conjugation} formally, recall the cycle
decomposition (cf.\ Definition~\ref{Def_POSET}). Let $\alpha,\pi\in\mathcal{S}_n$ and define
the relabeled permutation $\widetilde{\alpha}:=\pi\alpha\pi^{-1}$.
If $\alpha$ contains a cycle $(a_1\,a_2\,\dots\,a_k)$, i.e.,
$\alpha(a_j)=a_{j+1}$ for $j=1,\dots,k-1$ and $\alpha(a_k)=a_1$, then
$\widetilde{\alpha}$ contains the cycle \cite[Proposition~10, p.~125]{DummitFoote2004}
\begin{align}\label{EQ_conjugation2}
(\pi(a_1)\,\pi(a_2)\,\dots\,\pi(a_k)).
\end{align}
Let \(b_j:=\pi(a_j)\). Then the computation
$\widetilde{\alpha}(\pi(a_j))=\pi(\alpha(a_j))=\pi(a_{j+1})$
is exactly $\widetilde{\alpha}(b_j)=b_{j+1}\; (j=1,\dots,k-1),$ and similarly \(\widetilde{\alpha}(b_k)=b_1\).
\footnote{Example ($n=6$).
Let $\alpha=(1\,4\,2)(3\,6)\in\mathcal{S}_6$ and $\pi=(1\,5\,3)(2\,4)\in\mathcal{S}_6$.
First see how $\pi$ relabels the indices:
$\pi:\ 1\mapsto 5,\ 5\mapsto 3,\ 3\mapsto 1,\;2\mapsto 4,\ 4\mapsto 2,\;6\mapsto 6.$
Now consider the $3$-cycle $(1\,4\,2)$ of $\alpha$. Under relabeling, the elements
$1,4,2$ become $\pi(1)=5$, $\pi(4)=2$, $\pi(2)=4$, so we expect a $3$-cycle
$(5\,2\,4)$ in $\widetilde{\alpha}=\pi\alpha\pi^{-1}$.
This can be verified directly:
$\widetilde{\alpha}(5)=\pi(\alpha(1))=\pi(4)=2,\;
\widetilde{\alpha}(2)=\pi(\alpha(4))=\pi(2)=4,\;
\widetilde{\alpha}(4)=\pi(\alpha(2))=\pi(1)=5,$
hence $(5\,2\,4)$ is indeed a cycle of $\widetilde{\alpha}$.
Next, for the $2$-cycle $(3\,6)$ of $\alpha$, relabeling gives $\pi(3)=1$ and
$\pi(6)=6$, so it becomes $(1\,6)$. Again,
$\widetilde{\alpha}(1)=\pi(\alpha(3))=\pi(6)=6,\;\widetilde{\alpha}(6)=\pi(\alpha(6))=\pi(3)=1,$
confirming the cycle $(1\,6)$. Finally, $\alpha$ fixes $5$, hence
$\widetilde{\alpha}$ fixes $\pi(5)=3$:
$\widetilde{\alpha}(3)=\pi(\alpha(5))=\pi(5)=3$.
Therefore,
$\widetilde{\alpha}=(5\,2\,4)(1\,6)$ has the same cycle type as $\alpha$
(one $3$-cycle, one $2$-cycle, and one fixed point).} That is, relabeling cannot create or destroy cycles, nor can it change
their lengths; it only renames the elements inside each cycle. Consequently, once the
\gls{gbcs}/\gls{bbcs} sets and all protocol permutations are re-indexed
consistently via $\pi_{\rm rel}$ (equivalently, via conjugation as in
\eqref{EQ_conjugation}), the cycle-based constraints and commutation-based
conditions used in Lemma~\ref{lem:cyclic-hideB} and
Corollary~\ref{cor:one-sided-iso-clean} remain valid under the new labeling. The result is summarized in the following lemma.

\begin{lemma}\label{lem:relabel-invariance}
Let $\pi_{\rm rel}\in\mathcal{S}_n$ be any relabeling permutation and let
$\bP_{\rm rel}$ be its permutation matrix. Define the relabeled polarization
matrix by $\widetilde{\bT} := \bP_{\rm rel}\,\bT\,\bP_{\rm rel}^{-1}.$
For any $\bU\in\Aut(\bT)$ define its relabeled version by
$\widetilde{\bU} := \bP_{\rm rel}\,\bU\,\bP_{\rm rel}^{-1},$ 
and for any index set $\mathcal{S}\subseteq[n]$ define
$\widetilde{\mathcal{S}} := \pi_{\rm rel}(\mathcal{S})$.
Then 
\begin{align}
    \bU\in\Aut(\bT) \mbox{ if and only if } \widetilde{\bU}\in\Aut(\widetilde{\bT}).    
\end{align}
Moreover, for any $\bA,\bP,\bQ\in\Aut(\bT)$ and any \gls{gbcs}/\gls{bbcs} partition
$(\mathcal{G},\mathcal{B})$, the results of
Lemma~\ref{lem:cyclic-hideB} and Corollary~\ref{cor:one-sided-iso-clean} hold for
$(\bA,\bP,\bQ,\mathcal{G},\mathcal{B},\bT)$ if and only if they hold for the relabeled tuple
$(\widetilde{\bA},\widetilde{\bP},\widetilde{\bQ}, \widetilde{\mathcal{G}},\widetilde{\mathcal{B}},\widetilde{\bT}).$
In particular, the OT construction and its security proof are invariant under
the relabeling $\pi_{\rm rel}$.
\end{lemma}

% \begin{lemma}\label{lem:relabel-invariance}
% Let $\pi_{\rm rel}\in\mathcal{S}_n$ be any relabeling permutation and let
% $\bP_{\rm rel}$ be its permutation matrix. For every $\bR\in\Aut(\bT)$, define
% \[
% \widetilde{\bR} \;:=\; \bP_{\rm rel}\,\bR\,\bP_{\rm rel}^{-1},\qquad
% \widetilde{\mathcal{G}} \;:=\; \bP_{\rm rel}(\mathcal{G}),\quad
% \widetilde{\mathcal{B}} \;:=\; \bP_{\rm rel}(\mathcal{B}).
% \]
% Then Lemma \ref{lem:cyclic-hideB} and Corollary~\ref{cor:one-sided-iso-clean} hold for the relabeled objects
% \[
% (\widetilde{\bA},\widetilde{\bP},\widetilde{\bQ},
%  \widetilde{\mathcal{G}},\widetilde{\mathcal{B}})
% =
% (\bP_{\rm rel}\bA\bP_{\rm rel}^{-1},
%  \bP_{\rm rel}\bP\bP_{\rm rel}^{-1},
%  \bP_{\rm rel}\bQ\bP_{\rm rel}^{-1},
%  \bP_{\rm rel}(\mathcal{G}),
%  \bP_{\rm rel}(\mathcal{B}))
% \]
% if and only if they hold for $(\bA,\bP,\bQ,\mathcal{G},\mathcal{B})$. In particular, the OT construction and its security proof are invariant under the relabeling $\bP_{\rm rel}$.
% \end{lemma}
The proof is relegated to Appendix \ref{APP_proof_lem:relabel-invariance}.

\subsection{OT rate optimization}\label{subsec_finite_n}
The OT analysis in Sec.~III-C to Sec.~III-F relies on polarization:
$\mathcal{G}$ becomes almost noiseless and $\mathcal{B}$ becomes almost useless as the blocklength $n\to\infty$.
At finite $n$, issues occur due to the following two major reasons: (i) \gls{bbcs} still carry nonzero information, which causes leakage and hinders \gls{sfa}, and (ii) \gls{gbcs} are not perfectly reliable and hinder reliability. Both effects directly reduce the payload length $\ell$ that can satisfy
\eqref{EQ_reliability}, \eqref{EQ_SFA}, and \eqref{EQ_SFB}.
Moreover, the leakage caused by placing random bits on \gls{bbcs} can be upper bounded by Lemma~\ref{lem:bbc-leakage} and incorporated into the general leftover-hash lemma. However, in the previous discussion, we assume \gls{gbcs} and \gls{bbcs} are already selected, which is a missing step in practical design. Therefore, in this section, we aim to develop a systematic way to select (a) the paired index sets of \gls{gbcs} and \gls{bbcs} and (b) the automorphism in $\Aut(\bT)$ by explicitly optimizing the finite-$n$ bit-channel \gls{mi}, so that the designed OT payload $\ell$ is feasible to satisfy required reliability, \gls{sfa}, and \gls{sfb} constraints, while it is maximized at the operating SNR and blocklength $n$.

Fix $\bP\in\Aut(\bT)$ and let $\pi_\bP:[n]\to[n]$ be the induced index permutation.
We choose one reference transform $\bT_0$ and define the paired sets
$\mathcal{J}_1 := \pi_{\bP}(\mathcal{J}_0),\, \mathcal{J}_0\cap\mathcal{J}_1=\varnothing,$
so that in the proposed protocol, e.g., Fig. \ref{fig:sysD_comparison}, one branch uses $\mathcal{J}_0$ as \gls{gbcs} and $\mathcal{J}_1$ as \gls{bbcs},
while in the other branch they swap the roles. Let $I_i^{(n)}(\bT_0)\in[0,1]$ denote the polarized bit-channel mutual informations under $\bT_0$.
The total leakage under finite-$n$, contributed by the non-ideal \gls{bbcs} can be described as follows:
\begin{equation}\label{eq:leakage-def}
\textsf{L}
\;\coloneqq\;
\sum_{j\in\mathcal{J}_1} I_j^{(n)}(\bT_0)
\;=\;
\sum_{i\in\mathcal{J}_0} I_{\pi_{\bP}{(i)}}^{(n)}(\bT_0).
\end{equation}

% --- drop-in replacement for your current PA paragraph + (eq:lhl-design-rule) ---
% --- revised PA paragraph (keeps your structure, but makes the “why only (Pi_sel,S)” point rigorous) ---
To remove this leakage due to finite-$n$ which threatens \gls{sfa}, we design privacy amplification based on the \gls{lhl}.
Recall that $\Pi_{\mathrm{sel}}:=(\bF,\mathcal J_0,\mathcal J_1)$ denotes the selection-related public information,
and let $S$ be the public hash seed. Define the pre-transfer side information at Bob as
\[
E_{\mathrm{pre}} \;:=\; (Y^n,\Pi_{\mathrm{sel}},S,B).
\]
Hence privacy amplification must ensure that $K_{\bar B}$ is almost uniform and independent of Bob's entire view
available at the key-extraction stage, namely $E_{\mathrm{pre}}$. Concretely, for a suitable $\varepsilon\in(0,1)$,
the \gls{lhl} yields
\begin{equation}\label{eq:LHL-pre-TV}
  d_{\mathrm{TV}}\!\Big(P_{K_{\bar B},E_{\mathrm{pre}}},\; P_{\Unif(\{0,1\}^\ell)}\times P_{E_{\mathrm{pre}}}\Big)
  \;\le\; \varepsilon,
\end{equation}
i.e., the unchosen key $K_{\bar B}$ is $\varepsilon$-close to uniform and (approximately) independent of $E_{\mathrm{pre}}$.

Note that in \eqref{eq:LHL-pre-TV} we do not consider ciphertexts $C_0,C_1$ as side information when invoking the
\gls{lhl}, even though they are public information. The reason is that the ciphertexts are formed as
$C_b=M_b\oplus K_b$ and thus are deterministic functions of the messages and keys. If we include $C_0,C_1$
directly in the \gls{lhl} side information, we would introduce a dependence on the very key we aim to prove
uniform.
Instead, we first establish \eqref{eq:LHL-pre-TV} for the pre-transfer view $E_{\mathrm{pre}}$, and then use a
separate one-time-pad  step to extend this bound to the full public information including ciphertexts,
which is exactly what \gls{sfa} requires. The validity of considering only $\Pi_{\mathrm{sel}}\mbox{ and }\,S$ instead of
$\Pi_{\mathrm{pub}}$ is proved in the following lemma.

\begin{lemma}\label{lem:lhl-lift-coupling}
Let $K\in\{0,1\}^\ell$ and let
$U_\ell\sim\Unif(\{0,1\}^\ell)$ be independent of $E_{\mathrm{pre}}$.
Assume $d_{\mathrm{TV}}\!\Big(P_{K,E_{\mathrm{pre}}},\; P_{U_\ell}\times P_{E_{\mathrm{pre}}}\Big)\le\varepsilon.$
Let $M$ be any random variable independent of $(K,E_{\mathrm{pre}})$, and define $C:=M\oplus K$.
Then
\begin{equation}\label{eq:MCE-close}
d_{\mathrm{TV}}\!\Big(P_{M,C,E_{\mathrm{pre}}},\; P_M\times P_{U_\ell}\times P_{E_{\mathrm{pre}}}\Big)\le\varepsilon .
\end{equation}
\end{lemma}

\begin{proof}
Let $\delta \;:=\; d_{\mathrm{TV}}\!\Big(P_{K,E_{\mathrm{pre}}},\;P_{U_\ell}\times P_{E_{\mathrm{pre}}}\Big).$
By the maximal coupling theorem \cite{shaked2007}, there exist random variables
\((\widetilde K,\widetilde E_{\mathrm{pre}},\widetilde U,\widetilde E_{\mathrm{pre}}')\) on a common probability space such that
\[
(\widetilde K,\widetilde E_{\mathrm{pre}})\sim P_{K,E_{\mathrm{pre}}},\qquad
(\widetilde U,\widetilde E_{\mathrm{pre}}')\sim P_{U_\ell}\times P_{E_{\mathrm{pre}}},
\qquad\text{and}\qquad
\Pr\!\big((\widetilde K,\widetilde E_{\mathrm{pre}})\neq(\widetilde U,\widetilde E_{\mathrm{pre}}')\big)=\delta.
\]
Generate \(\widetilde M\sim P_M\) independently of
\((\widetilde K,\widetilde E_{\mathrm{pre}},\widetilde U,\widetilde E_{\mathrm{pre}}')\)
and define
\(\widetilde C:=\widetilde M\oplus \widetilde K\) and
\(\widetilde C':=\widetilde M\oplus \widetilde U\).
Then \((\widetilde M,\widetilde C,\widetilde E_{\mathrm{pre}})\sim P_{M,C,E_{\mathrm{pre}}}\).
Moreover, since \(\widetilde U\sim \Unif(\{0,1\}^\ell)\) and \(\widetilde U\pperp(\widetilde M,\widetilde E_{\mathrm{pre}}')\),
the Crypto Lemma implies that \(\widetilde C'\sim \Unif(\{0,1\}^\ell)\) and
\(\widetilde C'\pperp(\widetilde M,\widetilde E_{\mathrm{pre}}')\), hence
\((\widetilde M,\widetilde C',\widetilde E_{\mathrm{pre}}')\sim P_M\times P_{U_\ell}\times P_{E_{\mathrm{pre}}}\).

Now fix any measurable set \(\mathcal A\) in the alphabet of \((M,C,E_{\mathrm{pre}})\) and let \(\mathds 1_{\mathcal A}(\cdot)\)
be its indicator. Using the above coupling, we can derive
\begin{align}
\big|\Pr\big((\widetilde M,\widetilde C,\widetilde E_{\mathrm{pre}})\in\mathcal A\big)
-\Pr\big((\widetilde M,\widetilde C',\widetilde E_{\mathrm{pre}}')\in\mathcal A\big)\big|
&=\big|\mathbb E\big[\mathds 1_{\mathcal A}(\widetilde M,\widetilde C,\widetilde E_{\mathrm{pre}})
-\mathds 1_{\mathcal A}(\widetilde M,\widetilde C',\widetilde E_{\mathrm{pre}}')\big]\big|\notag\\
&\le \mathbb E\big[\big|\mathds 1_{\mathcal A}(\widetilde M,\widetilde C,\widetilde E_{\mathrm{pre}})
-\mathds 1_{\mathcal A}(\widetilde M,\widetilde C',\widetilde E_{\mathrm{pre}}')\big|\big]\notag\\
&\le \Pr\!\big((\widetilde K,\widetilde E_{\mathrm{pre}})\neq(\widetilde U,\widetilde E_{\mathrm{pre}}')\big)\notag\\
&=\delta,\label{eq:tv-via-indicator}
\end{align}
where the second inequality holds because the difference of indicator functions can be nonzero only on outcomes where
\((\widetilde K,\widetilde E_{\mathrm{pre}})\) and \((\widetilde U,\widetilde E_{\mathrm{pre}}')\) are different.
Taking the supremum of \eqref{eq:tv-via-indicator} over all measurable \(\mathcal A\) yields
\(d_{\mathrm{TV}}\!\big(P_{M,C,E_{\mathrm{pre}}},\;P_M\times P_{U_\ell}\times P_{E_{\mathrm{pre}}}\big)\le \delta \le \varepsilon\),
which completes the proof.
\end{proof}

Define $\mathsf V_{\mathrm{sfa}}:=(C_{\bar B},E_{\mathrm{pre}})$. The bound \eqref{eq:MCE-close} directly implies \gls{sfa}, which can be shown as follows:
\begin{align}
d_{\mathrm{TV}}\!\Big(P_{M_{\bar B},\mathsf V_{\mathrm{sfa}}},\; P_{M_{\bar B}}\times P_{\mathsf V_{\mathrm{sfa}}}\Big)
&\le
d_{\mathrm{TV}}\!\Big(P_{M_{\bar B},\mathsf V_{\mathrm{sfa}}},\; P_{M_{\bar B}}\times P_{U_\ell}\times P_{E_{\mathrm{pre}}}\Big)
+
d_{\mathrm{TV}}\!\Big(P_{M_{\bar B}}\times P_{U_\ell}\times P_{E_{\mathrm{pre}}},\; P_{M_{\bar B}}\times P_{\mathsf V_{\mathrm{sfa}}}\Big)\notag\\
&=
d_{\mathrm{TV}}\!\Big(P_{M_{\bar B},C_{\bar B},E_{\mathrm{pre}}},\; P_{M_{\bar B}}\times P_{U_\ell}\times P_{E_{\mathrm{pre}}}\Big)
+
d_{\mathrm{TV}}\!\Big(P_{U_\ell}\times P_{E_{\mathrm{pre}}},\; P_{\mathsf V_{\mathrm{sfa}}}\Big)\notag\\
&\le \varepsilon+\varepsilon
\;=\;2\varepsilon,
\label{eq:SFA-from-LHL}
\end{align}
where the first inequality is the triangle inequality, the second inequality upper-bounds the two terms separately: the first term is bounded by Lemma~\ref{lem:lhl-lift-coupling}, and the second term can be derived by marginalizing the first term with respect to $M_{\bar B}$.

% --- Replace the whole block by the following ---

Specifically, in the proposed protocol Alice draws i.i.d.\ $\mathrm{Bern}(\tfrac12)$ bits on the two published index sets
$\tilde{\mathcal J}_0$ and $\tilde{\mathcal J}_1$, and freezes all remaining positions. Hence the key for Bob is extracted from
$U_{\tilde{\mathcal J}_B}$, while $U_{\tilde{\mathcal J}_{1-B}}$ is the potential source of leakage.
For example, if Bob selects $B=0$, he extracts the key from $U_{\tilde{\mathcal J}_0}$ and the leakage comes from $U_{\tilde{\mathcal J}_1}$;
for $B=1$ the roles are swapped. Recall in the Step 2 of the proposed protocol, Bob publishes the pair $(\tilde{\mathcal{J}}_{0},\tilde{\mathcal{J}}_{1})=(\mathcal{J}_{0},\mathcal{J}_{1})$, if $B=0$; $(\tilde{\mathcal{J}}_{0},\tilde{\mathcal{J}}_{1})=(\mathcal{J}_{1},\mathcal{J}_{0})$, else, over the public channel.

Therefore, by the \gls{lhl} in Corollary~\ref{Corollary_LHL}, the extracted key length $\ell$ must satisfy
\begin{equation}\label{eq:lhl-design-rule}
\ell
\;\le\;
H_{\min}^{\varepsilon_s}\!\bigl(U_{\tilde{\mathcal J}_{\bar{B}}}\mid Y^n,\Pi_{\mathrm{sel}},S,B\bigr)
\;-\;
2\log_2(1/\varepsilon_p),
\end{equation}
where $\varepsilon_s$ is the smoothing parameter and $\varepsilon_p$ is the target upper bound of \eqref{EQ_LHL}.

To efficiently calculate $\ell$, we lower-bound the smooth min-entropy by the conditional Shannon entropy with an explicit correction term.
We use \cite[Lemma~1]{heerklotz2025neuralestimationinformationleakage} as shown below.
\begin{lemma}\label{Lemma_Hmin_H_gap}
Let $X$ be a discrete random variable, $Z\in\mathcal Z$ a continuous random variable, and fix $\varepsilon\in(0,1)$.
Select a measurable set $\mathcal E\subseteq\mathcal Z$ such that $P_Z(\mathcal E)=1-\varepsilon$.
Assume that for every $z\in\mathcal E$ the conditional \gls{pmf} $p_{X\mid Z}(\cdot\mid z)$ satisfies
$v_{z}:=\bigl|\supp_x p_{X\mid Z=z}\bigr|<\infty$ and $t_{z}:=\max_{x}p_{X\mid Z}(x\mid z)<\infty$.
Define the random variables $V:=v_{Z}\mathds{1}_{\mathcal E}$ and $T:=t_{Z}\mathds{1}_{\mathcal E}$.
Define $\psi_{v}(t):=H_{\mathrm b}(t)+(1-t)\log_{2}(v-1)+\log_{2}t.$ Then
{\small
\begin{align}\label{EQ_Hmin_H_gap}
-H_{\min}^{\varepsilon}(X\mid Z)
 \;\le\;&
   -H(X\mid Z)
   +\mathbb{E}_Z\!\bigl[\psi_{V}(T)\bigr]-\log_{2}(1-\varepsilon)
   +\frac{\varepsilon}{1-\varepsilon}\,H_{\max}(X).
\end{align}}
\end{lemma}

To be self-contained, we restate the proof in Appendix~\ref{APP_proof_LHL_cond_entropy}.
Rearranging \eqref{EQ_Hmin_H_gap} gives
\begin{equation}\label{eq:Hmin-lower-via-H}
H_{\min}^{\varepsilon_s}(X\mid Z)
\;\ge\;
H(X\mid Z)
\;-\;
\Delta_{\varepsilon_s}(X\mid Z),
\end{equation}
where $\Delta_{\varepsilon_s}(X\mid Z):=\mathbb{E}\!\big[\psi_V(T)\big]
-\log_2(1-\varepsilon_s)+\frac{\varepsilon_s}{1-\varepsilon_s}H_{\max}(X).$

To apply Lemma~\ref{Lemma_Hmin_H_gap} to our problem, we substitute
$(X,Z):=\bigl(U_{\tilde{\mathcal J}_{\bar{B}}},(Y^n,\Pi_{\mathrm{sel}},S,B)\bigr)$
into \eqref{eq:Hmin-lower-via-H} and then into \eqref{eq:lhl-design-rule}, yielding
\begin{align}
\ell
\;&\le\;
H\!\bigl(U_{\tilde{\mathcal J}_{\bar{B}}}\mid Y^n,\Pi_{\mathrm{sel}},S,B\bigr)
\;-\;
\Delta_{\varepsilon_s}\!\bigl(U_{\tilde{\mathcal J}_{\bar{B}}}\mid Y^n,\Pi_{\mathrm{sel}},S,B\bigr)
\;-\;
2\log_2(1/\varepsilon_p)\notag\\
&:=\;
H\!\bigl(U_{\tilde{\mathcal J}_{\bar{B}}}\mid Y^n,\Pi_{\mathrm{sel}},S,B\bigr)
\;-\;
c_{\varepsilon},\label{EQ_ell}
\end{align}
where we define
$c_{\varepsilon}:=\Delta_{\varepsilon_s}\!\bigl(U_{\tilde{\mathcal J}_{\bar{B}}}\mid Y^n,\Pi_{\mathrm{sel}},S,B\bigr)
+2\log_2(1/\varepsilon_p).$

We can equivalently express $H(U_{\tilde{\mathcal J}_{\bar{B}}}\mid Y^n,\Pi_{\mathrm{sel}},S,B)$
as Lemma~\ref{lem:hidden-set-ell-vs-leakage} shows, due to the proposed protocol.
Before that, we introduce a tool lemma.

\begin{lemma}\label{lem:markov-B-F-UY}
In the proposed OT protocol, we have the Markov chain
$ B\;-\;\mathbf F\;-\;(U^n,Y^n).$
\end{lemma}

\begin{proof}
Fix any realization $\mathbf f$ of $\mathbf F$ and any $b\in\{0,1\}$.
In our protocol, Bob generates $(B,\mathbf F)$ using only local randomness.
After $\mathbf F$ (and the selection-related public information) are published, Alice forms $U^n$ as follows:
she sets $U_i=0$ on frozen indices, and samples $U_i\sim\mathrm{Bern}(\tfrac12)$ independently on the randomized indices
specified by the published index sets. In particular, conditioned on $\mathbf F=\mathbf f$, 
$P_{U^n\mid \mathbf F=\mathbf f}$ is fully determined by Alice's local randomness and does not depend on $B$. Hence, we have
\begin{equation}\label{eq:U-indep-B-given-F}
P_{U^n\mid \mathbf F,B}(u\mid \mathbf f,b)=P_{U^n\mid \mathbf F}(u\mid \mathbf f),
\qquad \forall\,(u,\mathbf f,b).
\end{equation}
Moreover, the physical channel noise is independent of $(B,\mathbf F,U^n)$, and
the channel input and output are $X^n:=U^n\mathbf F$ and $Y^n = X^n + N^n = U^n\mathbf F + N^n,$ respectively.
Now fix $(u,y)$ and condition on $\{U^n=u,\mathbf F=\mathbf f,B=b\}$. Then
\begin{align}
P_{Y^n\mid U^n,\mathbf F,B}(y\mid u,\mathbf f,b)
&=\Pr\!\bigl(U^n\mathbf F+N^n=y \,\big|\, U^n=u,\mathbf F=\mathbf f,B=b\bigr)\notag\\
&=\Pr\!\bigl(N^n=y-u\mathbf f \,\big|\, U^n=u,\mathbf F=\mathbf f,B=b\bigr)\notag\\
&\overset{(a)}{=}\Pr\!\bigl(N^n=y-u\mathbf f\bigr)
=\Pr\!\bigl(u\mathbf f+N^n=y\bigr)\notag\\
&=\Pr\!\bigl(U^n\mathbf F+N^n=y \,\big|\, U^n=u,\mathbf F=\mathbf f\bigr)
\;=\;P_{Y^n\mid U^n,\mathbf F}(y\mid u,\mathbf f),
\label{EQ_YUFB_YUF}
\end{align}
where (a) uses the fact that the AWGN noise $N^n$ is independent of $(U^n,\mathbf F,B)$. Therefore, for all $(u,y)$, we have
\begin{align}
P_{U^n,Y^n\mid \mathbf F,B}(u,y\mid \mathbf f,b)
&=P_{U^n\mid \mathbf F,B}(u\mid \mathbf f,b)\,P_{Y^n\mid U^n,\mathbf F,B}(y\mid u,\mathbf f,b)\notag\\
&\overset{(b)}{=}P_{U^n\mid \mathbf F}(u\mid \mathbf f)\,P_{Y^n\mid U^n,\mathbf F}(y\mid u,\mathbf f)
\;=\;P_{U^n,Y^n\mid \mathbf F}(u,y\mid \mathbf f),
\label{EQ_UY_FB_UY_F}
\end{align}
where (b) uses \eqref{eq:U-indep-B-given-F} and \eqref{EQ_YUFB_YUF}, which completes the proof.
\end{proof}

To proceed, we derive a lower bound of $H(U_{\tilde{\mathcal J}_{\bar{B}}}\mid Y^n,\Pi_{\mathrm{sel}},B)$ in terms of bit-channel capacities as the following lemma. Define $I_i^{(n)}(\bF)\;:=\;I\!\bigl(U_i;Y^n,U^{i-1}\,\big|\,\mathbf F\bigr)$. 

\begin{lemma}\label{lem:hidden-set-ell-vs-leakage}
Assume that the bits $\{U_i:\ i\in\mathcal J_0\cup\mathcal J_1\}$ are i.i.d.\ $\mathrm{Bern}(\tfrac12)$, while the remaining bits $U_{([n]\setminus(\mathcal J_0\cup\mathcal J_1))}$
are frozen to zeros. Assume $|\mathcal J_0|=|\mathcal J_1|$.
Then the design rule \eqref{EQ_ell} of $\ell$ can be expressed as follows
\begin{equation}\label{eq:ell-vs-L}
\ell
\;\le\;
|\mathcal J_0|-\sum_{i\in\mathcal J_1} I_i^{(n)}(\mathbf F)
\;-\;
c_{\varepsilon}.
\end{equation}
\end{lemma}

The proof is relegated to Appendix \ref{APP_proof_hidden-set-ell-vs-leakage}.

% --- Replace \mathcal{J}_0 by the selected key index set \mathcal J_B (or its reconciled subset),
%     keeping your structure/steps unchanged. ---

When $\mathcal J_0$ contains non-ideal \gls{gbcs}, Alice can send a public reconciliation
message with length $\ell_{\rm SWC}$ to help Bob reconstruct the required raw
bits reliably from the side information $Y^n$. For any $\mathcal{S}\subseteq[n]$, define the subvector
$U_{\mathcal{S}} := (U_i)_{i\in\mathcal{S}}$. Then, we can use Slepian--Wolf coding via universal hashing, where
$\ell_{\rm SWC}$ must satisfy \cite[Proposition~6.8]{SudaWatanabe25}
\begin{equation}\label{eq:swc-one-shot}
\ell_{\rm SWC} \;\ge\; H_{\max}^{\varepsilon_{\rm sw}}(X\mid Z),
\end{equation}
where $X$ denotes the reconciliation target and $Z$ denotes the RX side information.
Concretely, choose $\mathcal A_{\rm SI}\subseteq \mathcal J_0$ as the side-information index set, i.e., indices within $\mathcal J_0$
that are decoded with negligible error and thus can be provided to the RX as side information without
reconciliation. Define the reconciliation index set as the remaining indices in $\mathcal J_0$:
\[
\mathcal A_{\rm SI}^c \;:=\; \mathcal J_0\setminus \mathcal A_{\rm SI}.
\]
Accordingly, the reconciliation target is $X:=U_{\mathcal A_{\rm SI}^c}$ and the RX side information can be taken as
$Z:=(Y^n,U_{\mathcal A_{\rm SI}})$.

\begin{lemma}\label{lem:sw-hhat-discreteX-contY}
Fix an index set $\mathcal J_0\subseteq[n]$.
Fix a side-information index set $\mathcal A_{\rm SI}\subseteq\mathcal J_0$ and define the corresponding
reconciliation index set
$\mathcal A_{\rm SI}^c.$ Define the RX side information $Z := \bigl(Y^n,\,U_{\mathcal A_{\rm SI}}\bigr).$
Then for any target Slepian--Wolf decoding error probability
$\varepsilon_{\rm sw}\in(0,1)$, there exists a
reconciliation encoder that sends a public message $M_{p}$ of length
$\ell_{\rm SWC}$ bits such that the RX can reconstruct
$U_{\mathcal A_{\rm SI}^c}$ from $(Z,M_{p})$ with error probability at
most $\varepsilon_{\rm sw}$, provided
\begin{equation}\label{eq:hhat-lemma-b}
  \ell_{\rm SWC}\;\ge\;
  \sum_{i\in\mathcal A_{\rm SI}^c}\bigl(1-I_i^{(n)}(\bF)\bigr)
  \;+\;
  \beta_n(\varepsilon_{\rm sw}),
\end{equation}
where $\beta_n(\varepsilon_{\rm sw})=O(\sqrt{n})$.
\end{lemma}

The proof is delegated in Appendix
\ref{APP_proof_sw-hhat-discreteX-contY}, where the finite blocklength result of \gls{swc} with continuous side information at the decoder is derived in Appendix \ref{APP_proof_sw_continuous_SI_second_order} following the explanation in
\cite[Remark~1]{Hayashi_SWC_Entropy20}.
The term $\sum_{i\in\mathcal{A}_{\rm SI}^c}(1-I_i^{(n)}(\bF))$ in
\eqref{eq:hhat-lemma-b} quantifies the residual decoding uncertainty of the indices
that must be reconciled. We take a conservative design by setting $\mathcal A_{\rm SI}=\varnothing$, such that
$\mathcal A_{\rm SI}\cap\mathcal J_0=\varnothing$, i.e., $\mathcal{A}_{\rm SI}^c=\mathcal J_0$.
A sufficient reconciliation length that guarantees Slepian--Wolf decoding error at most
$\varepsilon_{\rm sw}$ is
\begin{equation}\label{eq:hhat}
\widehat{\ell}_{\rm SWC}(\mathcal J_0)
\;\coloneqq\;
\sum_{i\in\mathcal J_0}\bigl(1-I_i^{(n)}(\bF)\bigr)\;+\;\beta_n(\varepsilon_{\rm sw}).
\end{equation}
Accordingly, in the protocol we choose the public reconciliation message length as
$\ell_{\rm SWC}:=\widehat{\ell}_{\rm SWC}(\mathcal J_0)$.

Since the reconciliation message $M_p$ is public and has length $\ell_{\rm SWC}$, from Corollary~\ref{Corollary_LHL} we know that revealing $M_p$ can reduce the extractable key length by at least $\ell_{\rm SWC}$ bits.
Therefore, a sufficient condition for an achievable OT key length is
\begin{equation}\label{eq:ell-net-derivation-1}
\ell_{\rm net}(\mathcal J_0,\mathcal J_1)
\;\le\;
\ell \;-\; \widehat{\ell}_{\rm SWC}(\mathcal J_0).
\end{equation}

To obtain an explicit MI-based achievable net key length, we substitute the
design rule \eqref{eq:ell-vs-L} into \eqref{eq:ell-net-derivation-1} and use the
choice $\ell_{\rm SWC}:=\widehat{\ell}_{\rm SWC}(\mathcal J_0)$. This yields
\begin{align}
\ell_{\rm net}(\mathcal J_0,\mathcal J_1)
\;&\le\;
\Bigl(|\mathcal J_0|-\sum_{i\in\mathcal J_1} I_i^{(n)}(\mathbf F)-c_{\varepsilon}\Bigr)
\;-\;
\widehat{\ell}_{\rm SWC}(\mathcal J_0)\notag\\
&=
|\mathcal J_0|
-\sum_{i\in\mathcal J_1} I_i^{(n)}(\mathbf F)
-\sum_{i\in\mathcal J_0}\bigl(1-I_i^{(n)}(\bF)\bigr)
-\beta_n(\varepsilon_{\rm sw})
-c_{\varepsilon}\notag\\
&=
\sum_{i\in\mathcal J_0} I_i^{(n)}(\bF)
\;-\;
\sum_{i\in\mathcal J_1} I_i^{(n)}(\mathbf F)
\;-\;
\beta_n(\varepsilon_{\rm sw})
\;-\;
c_{\varepsilon}.\label{eq:ell-net-lb-step}
\end{align}

To reduce complexity relative to exact density evolution, we adopt the Gaussian
approximation (GA) \cite{Trifonov2012EfficientPolarGA}, which approximates the \gls{llr} of each intermediate
synthetic channel by a symmetric Gaussian distribution parameterized by a
single scalar (equivalently, its \gls{mi}). In particular, GA models the \gls{llr} as
$  L \sim \mathcal{N}\!\Bigl(\tfrac{\sigma^2}{2},\,\sigma^2\Bigr),$
and tracks the corresponding mutual information via the standard $J$-function
\cite{Trifonov2012EfficientPolarGA}: $  J(\sigma)
  \;:=\;  1-\mathbb{E}\Bigl[\log_2(1+e^{-L})\Bigr],\;  L\sim\mathcal{N}\!\Bigl(\tfrac{\sigma^2}{2},\sigma^2\Bigr),$
together with its inverse $J^{-1}(\cdot)$ on $[0,1]$. For the BI-AWGN model, GA is
initialized with $I_0(\textsf{SNR})=J(\sigma_0:=2\sqrt{\textsf{SNR}})$. Starting from $I_0(\textsf{SNR})$,
the GA recursion propagates \gls{mi} through the polarization tree: at each stage,
the ``$+$'' child uses $f_{+}$ and the ``$-$'' child uses $f_{-}$, where
\begin{align}
  f_{+}(I) &:= J\!\Bigl(\sqrt{2}\,J^{-1}(I)\Bigr),\qquad
  f_{-}(I) := 1 - J\!\Bigl(\sqrt{2}\,J^{-1}(1-I)\Bigr),
\end{align}
set $f_{1}\equiv f_{+}$, $f_{0}\equiv f_{-}$.
Equivalently, for $n=2^{m}$ and index $i\in[n]$, write the binary expansion
$i-1=(b_m\cdots b_1)_2$ with $b_j\in\{0,1\}$. Then GA yields the computable approximation
\begin{equation}\label{eq:Ii-explicit-composition}
  I_i^{(n)}(\bF;\textsf{SNR})
  \;\approx\;
  \bigl(f_{b_{m}}\circ f_{b_{m-1}}\circ\cdots\circ f_{b_1}\bigr)
  \bigl(I_0(\textsf{SNR})\bigr).
\end{equation}
A simple example can be seen below\footnote{Example ($n=4$): $i-1=(b_2b_1)_2$ gives
$I_1^{(4)}\approx f_{-}(f_{-}(I_0))$,
$I_2^{(4)}\approx f_{+}(f_{-}(I_0))$,
$I_3^{(4)}\approx f_{-}(f_{+}(I_0))$,
$I_4^{(4)}\approx f_{+}(f_{+}(I_0))$.}.

For each $\sigma\in\Aut(\bT_1)$ with induced permutation $\pi_\sigma$, define
\[
  w_i(\sigma;\textsf{SNR})
  \;\coloneqq\;
  I_i^{(n)}(\bF;\textsf{SNR})
  -
  I_{\pi_\sigma(i)}^{(n)}(\bF;\textsf{SNR}),
  \qquad i\in[n].
\]

For each $i\in[n]$, let $x_i\in\{0,1\}$ indicate whether index $i$ is selected as a key index on the \gls{gbcs} side or not by $x_i=1$ and $x_i=0$, respectively.
Define the feasible set $\mathcal J_0 := \{\, i\in[n] : x_i=1 \,\},$ which will be used in the proposed OT protocol to generate the keys.
Let $\tilde{\mathcal{G}}(\bT_0;\textsf{SNR})$ be the set of indices corresponding to the largest $n/2$
values among $\{I_i^{(n)}(\bF;\textsf{SNR})\}_{i=1}^{n}$, and define $\tilde{\mathcal{B}}(\bF;\textsf{SNR}) := [n]\setminus \tilde{\mathcal{G}}(\bF;\textsf{SNR}).$ Given a desired error probability for the reliability, which can be derived from Lemma \ref{lem:cp-ot}, we select $k$ indices by solving the following optimization problem:
\begin{subequations}\label{prob:ot-rate-opt}
\begin{align}
\max_{\sigma\in\Aut(\bT_1)}\ \max_{x\in\{0,1\}^{n}}\quad
& \sum_{i=1}^{n} w_i(\sigma;\textsf{SNR})\,x_i
\label{prob:ot-rate-opt-obj}
\\
\text{s.t.}\quad
& x_i = 0,\quad \forall\, i\in[n]\setminus \tilde{\mathcal{G}}(\bF;\textsf{SNR}),
\label{prob:ot-rate-opt-goodside}
\\
& x_i = 0,\quad \forall\, i\in[n]\ \text{with }\pi_\sigma(i)\notin \tilde{\mathcal{B}}(\bF;\textsf{SNR}),
\label{prob:ot-rate-opt-crosscut}
\\
& \sum_{i=1}^{n} x_i = k,\label{prob:ot-rate-opt-card}
\end{align}
\end{subequations}
where constraint~\eqref{prob:ot-rate-opt-goodside} enforces that the selected set $\mathcal J_0 \;:=\; \{\, i\in[n] : x_i=1 \,\}$
is contained in the candidate \gls{gbcs} set, i.e., $\mathcal J_0\subseteq \tilde{\mathcal{G}}(\bF;\textsf{SNR})$.
Constraint~\eqref{prob:ot-rate-opt-crosscut} enforces that if $i$-th bit-channel is selected, its paired index $\pi_\sigma(i)$ must lie
in the candidate \gls{bbcs} set $\tilde{\mathcal{B}}(\bF;\textsf{SNR})$.
Constraint~\eqref{prob:ot-rate-opt-card} fixes the number of selected indices to
$k$, which is chosen according to the target reliability when random bits are
placed on the paired \gls{bbcs}
\footnote{This constraint may be relaxed or removed if information reconciliation
(e.g., \gls{swc}) is employed to tolerate a larger number of random bits on
\gls{bbcs}.}.

Problem~\eqref{prob:ot-rate-opt} is not convex because $x\in\{0,1\}^{n}$ is discrete and the outer maximization
over $\sigma\in\Aut(\bT_1)$ is also discrete. Nevertheless, in our present setting the inner problem for a fixed
$\sigma$ admits a direct closed-form solution (largest-$k$ selection), so generic \gls{milp} solvers are unnecessary.

We first fix a bit-permutation $\sigma\in\mathcal{S}_m$
and focus on the inner problem. Constraint~\eqref{prob:ot-rate-opt-goodside} forces $x_i=0$ for all
$i\notin \tilde{\mathcal{G}}(\bF;\textsf{SNR})$, and constraint~\eqref{prob:ot-rate-opt-crosscut} forces $x_i=0$
whenever $\pi_\sigma(i)\notin \tilde{\mathcal{B}}(\bF;\textsf{SNR})$. Hence only indices in the following set are eligible
\begin{align}\label{eq:Vsig-def}
\mathcal{V}_\sigma
\;:=\;
\Bigl\{\, i\in \tilde{\mathcal{G}}(\bF;\textsf{SNR}) :
\pi_\sigma(i)\in \tilde{\mathcal{B}}(\bF;\textsf{SNR}) \Bigr\}
\end{align}
to satisfy $x_i=1$, and we set $x_i=0$ for all $i\notin\mathcal{V}_\sigma$.

Moreover, since $\tilde{\mathcal{B}}(\bF;\textsf{SNR})=[n]\setminus\tilde{\mathcal{G}}(\bF;\textsf{SNR})$,
every $i\in\mathcal{V}_\sigma$ satisfies $\pi_\sigma(i)\in\tilde{\mathcal{B}}(\bF;\textsf{SNR})$ and thus
$\pi_\sigma(i)\notin\tilde{\mathcal{G}}(\bF;\textsf{SNR})$. By \eqref{prob:ot-rate-opt-goodside},
this implies $x_{\pi_\sigma(i)}=0$ for all $i\in\mathcal V_\sigma$. Therefore, once we restrict to indices $i\in\mathcal{V}_\sigma$, the only constraint that couples these variables is the cardinality constraint~\eqref{prob:ot-rate-opt-card}. Hence, the fixed-$\sigma$ inner problem reduces to
\begin{align}\label{eq:inner-topk-prob}
\max\Bigl\{\sum_{i\in\mathcal{V}_\sigma} w_i(\sigma;\textsf{SNR})\,x_i :
x_i\in\{0,1\}\ \forall i\in\mathcal{V}_\sigma,\ \sum_{i\in\mathcal{V}_\sigma} x_i = k\Bigr\}.
\end{align}
If $|\mathcal{V}_\sigma|<k$, then \eqref{eq:inner-topk-prob} is infeasible and we discard such $\sigma$ in the outer
maximization. In contrast, when $|\mathcal{V}_\sigma|\ge k$, the optimizer of \eqref{eq:inner-topk-prob} is obtained by selecting the $k$ largest
weights among $\{w_i(\sigma;\textsf{SNR})\}_{i\in\mathcal{V}_\sigma}$. If the values are the same, we choose the smaller index.
Equivalently, letting $\mathcal{J}_0^\star(\sigma)\subseteq\mathcal{V}_\sigma$ be the indices of the largest-$k$ weights,
the optimal inner value is
\[
s(\sigma)\;:=\;\sum_{i\in\mathcal{J}_0^\star(\sigma)} w_i(\sigma;\textsf{SNR}),
\]
and the paired set and OT index set follow deterministically as
\[
\mathcal{J}_1^\star(\sigma):=\pi_\sigma\bigl(\mathcal{J}_0^\star(\sigma)\bigr),
\qquad
\mathcal{I}_{\rm OT}^\star(\sigma):=\mathcal{J}_0^\star(\sigma)\cup\mathcal{J}_1^\star(\sigma).
\]
% Computationally, this requires only sorting (or partial selection), i.e., $\mathcal{O}(|\mathcal{V}_\sigma|\log|\mathcal{V}_\sigma|)$ time per $\sigma$.

For the implementation, we define the following variables and functions.
Assume $n=2^m$ with $m:=\log_2 n$.
Define
\begin{align}
\pi_\sigma(i)
&:= 1+\mathrm{bin2int}\!\Bigl(\sigma\bigl(\mathrm{int2bin}(i-1)\bigr)\Bigr),
\qquad i\in[n],
\tag{63}\label{eq:pi-sigma}
\\
\tilde{\mathcal{I}_{\mathcal G}}(\mathbf F;\textsf{SNR})
&:= \text{indices of the largest $n/2$ values in }
\bigl\{I_i^{(n)}(\mathbf F;\textsf{SNR})\bigr\}_{i=1}^n,
\tag{64}\label{eq:G-tilde}
\\
\tilde{\mathcal{I}_{\mathcal B}}(\mathbf F;\textsf{SNR})
&:= [n]\setminus \tilde{\mathcal{I}_{\mathcal G}}(\mathbf F;\textsf{SNR}),
\tag{65}\label{eq:B-tilde}
\\
w_i(\sigma;\textsf{SNR})
&:= \Bigl(
I_i^{(n)}(\mathbf F;\textsf{SNR})
- I_{\pi_\sigma(i)}^{(n)}(\mathbf F;\textsf{SNR})
\Bigr),
\qquad i\in[n].
\tag{66}\label{eq:w-def}
\end{align}
Here $\mathrm{int2bin}(i-1)\in\{0,1\}^m$ is the length-$m$ binary expansion of $i-1$
and $\mathrm{bin2int}(\cdot)$ maps a binary vector back to an integer in
$\{0,\dots,n-1\}$.

\begin{algorithm}[t]
\caption{Inner solver for fixed $\sigma$}
\label{alg:inner-topk}
\begin{algorithmic}[1]
\Input $\sigma$, $\textsf{SNR}$, $\{I_i^{(n)}(\bF;\textsf{SNR})\}_{i=1}^n$, target size $k$
\Solve $\mathcal{J}_0^\star,\mathcal{J}_1^\star,\mathcal{I}_{\rm OT}^\star$ for \eqref{eq:inner-topk-prob}

\State Compute $\pi_\sigma(\cdot)$ using \eqref{eq:pi-sigma}
\State Compute $\tilde{\mathcal{G}}(\bF;\textsf{SNR})$ using \eqref{eq:G-tilde}
\State Compute $\tilde{\mathcal{B}}(\bF;\textsf{SNR})$ using \eqref{eq:B-tilde}
\State Compute weights $\{w_i(\sigma;\textsf{SNR})\}_{i=1}^n$ using \eqref{eq:w-def}
\State Compute eligible set $\mathcal{V}_\sigma$ using \eqref{eq:Vsig-def}

\State \textbf{If} $|\mathcal{V}_\sigma|<k$ \textbf{then return} ``infeasible for this $\sigma$''

\State Form list $\mathcal{L} := \{(i,w_i(\sigma;\textsf{SNR})) : i\in\mathcal{V}_\sigma\}$
\State Sort $\mathcal{L}$ by decreasing weight (break ties by smaller index)

\State Set $\mathcal{J}_0^\star := \{\, i : (i,\cdot)\ \text{is among the first $k$ pairs in }\mathcal{L}\,\}$
\State Set $\mathcal{J}_1^\star := \pi_\sigma(\mathcal{J}_0^\star)$
\State Set $\mathcal{I}_{\rm OT}^\star := \mathcal{J}_0^\star \cup \mathcal{J}_1^\star$
\State \Return $(\mathcal{J}_0^\star,\mathcal{J}_1^\star,\mathcal{I}_{\rm OT}^\star)$
\end{algorithmic}
\end{algorithm}

\begin{example}
Let $n=16$ and select
$\sigma=\sigma_2:[b_3\,b_2\,b_1\,b_0]\mapsto[b_2\,b_3\,b_1\,b_0]$
with induced index permutation $\pi_{\sigma_2}$, and GA at $I_0=\tfrac12$.
Assume $\tilde{\mathcal{G}}(\bF;\textsf{SNR})$ is formed by the largest $n/2$ values of
$\{I_i^{(n)}(\bF;\textsf{SNR})\}_{i=1}^n$. Under GA at $I_0=\tfrac12$ this yields
\[
\tilde{\mathcal{G}}(\bF;\textsf{SNR})=\{16,15,14,13,12,11,10,8\},\qquad
\tilde{\mathcal{B}}(\bF;\textsf{SNR})=\{9,7,6,5,4,3,2,1\}.
\]
From Table~\ref{Table_Aut_n16}, $\pi_{\sigma_2}=(12\ 8)(11\ 7)(10\ 6)(9\ 5)$.
Hence \eqref{eq:Vsig-def} gives
\[
\mathcal{V}_{\sigma_2}=\{10,11\},
\qquad
\pi_{\sigma_2}(\mathcal{V}_{\sigma_2})=\{6,7\}\subseteq\tilde{\mathcal{B}}(\bF;\textsf{SNR}).
\]

The inner problem \eqref{eq:inner-topk-prob} is solved by choosing the largest-$k$ weights among
$\{w_i(\sigma_2;\textsf{SNR})\}_{i\in\mathcal{V}_{\sigma_2}}$.
Given $I_0(\textsf{SNR})=\tfrac12$, which corresponds to $\textsf{SNR}= 1.044$ (or $0.187$ dB),
GA yields
$I_{12}^{(16)}= 0.946,\ I_{8}^{(16)}= 0.888,\;
I_{11}^{(16)}= 0.623,\ I_{7}^{(16)}= 0.477,\;
I_{10}^{(16)}= 0.523,\ I_{6}^{(16)}= 0.377,\;
I_{9}^{(16)}= 0.112,\ I_{5}^{(16)}= 0.054,$
where $I_i^{(16)}$ abbreviates $I_i^{(16)}(\bF;\textsf{SNR})$.
Hence the weights
$w_i(\sigma_2;\textsf{SNR}) := I_i^{(16)}-I_{\pi_{\sigma_2}(i)}^{(16)}$ are
$w_{12}= 0.029,\;w_{11}= 0.073,\;w_{10}= 0.073,\;w_{9}= 0.029.$
For $k=2$, the largest-$k$ rule yields $\mathcal{J}_0^\star=\{10,11\}$ and thus
$\mathcal{J}_1^\star := \pi_{\sigma_2}(\mathcal{J}_0^\star)=\{6,7\}$, with
$\mathcal{I}_{\rm OT}^\star := \mathcal{J}_0^\star\cup\mathcal{J}_1^\star=\{10,11,6,7\}$.

We now evaluate the \emph{net} achievable key length using \eqref{eq:ell-net-lb-step}.
For this choice,
\[
\sum_{i\in\mathcal{J}_0^\star} I_i^{(16)}-\sum_{i\in\mathcal{J}_1^\star} I_i^{(16)}
=
(I_{10}^{(16)}+I_{11}^{(16)})-(I_{6}^{(16)}+I_{7}^{(16)})
=
(0.523+0.623)-(0.377+0.477)
=
0.292.
\]
Therefore, an achievable \emph{net} OT key length satisfies
\[
\ell_{\rm net}(\sigma_2)
\;\le\;
0.292\;-\;\beta_{16}(\varepsilon_{\rm sw})\;-\;c_\varepsilon,
\]
and the corresponding \emph{net} OT rate is
\[
\textsf{R}_{{\rm OT,net}}(\sigma_2)
\;:=\;
\frac{\ell_{\rm net}(\sigma_2)}{n}
\;\le\;
\frac{0.292-\beta_{16}(\varepsilon_{\rm sw})-c_\varepsilon}{16}.
\]
In particular, ignoring the second-order reconciliation term and the security slack
(i.e., setting $\beta_{16}(\varepsilon_{\rm sw})=c_\varepsilon=0$ for illustration),
we obtain the nominal value
$\textsf{R}_{{\rm OT,net}}(\sigma_2)=\frac{0.292}{16}=1.825\times 10^{-2}$ bits/channel use.
\end{example}

Now we consider the outer maximization. For each $\sigma\in\Aut(\bT_1)$, we evaluate the outer objective by running Algorithm~\ref{alg:inner-topk} to obtain
$s(\sigma)$, and then select
\[
\sigma^\star \in \arg\max_{\sigma\in\Aut(\bT_1):\,|\mathcal{V}_\sigma|\ge k} s(\sigma).
\]
In practice, the simple inner evaluation (sorting on $\mathcal{V}_\sigma$) enables efficient pruning in the outer search,
e.g., by terminating early whenever $\mathcal{V}_\sigma=\varnothing$ or $|\mathcal{V}_\sigma|<k$.

\section{Conclusion}
We developed an explicit polar-code-based one-out-of-two OT protocol over binary-input memoryless
channels, and we quantified its performance over the BI--AWGN channel.
The construction views polarization through a virtual binary-erasure interface: reliable bit-channels carry Bob's
chosen message, while a paired set of unreliable bit-channels is arranged to convey essentially no information about
the unchosen message.
To mitigate leakage caused by revealing the polar transform, we restricted to permutations that preserve the
polar transform and used them to create different views of the good/bad index structure at Alice and Bob.
We further strengthened the erasure-like behavior at finite blocklength by injecting random bits on a carefully
selected subset of unreliable bit-channels and applying privacy amplification via universal hashing.
One of the contributions was a complete characterization of the transform-preserving permutations:
we showed that they are exactly those induced by 
permuting the bit positions of the binary index representation, yielding a concrete and fully implementable family of
admissible permutations.
Building on this structure, we introduced a finite-blocklength OT-rate optimization method that selects both the
permutation and the paired index sets using standard polar-code reliability estimates. All information-theoretic claims were proved within the paper.

\appendices 
\label{sec:appendix} % so you can refer to it

\section{Proof of Lemma \ref{lem:cp-ot}}\label{APP_proof_lem:cp-ot}
% The result follows from the standard one-sided Clopper--Pearson (exact) upper confidence limit
% obtained by inverting the binomial \gls{cdf} \cite{BrownCaiDasGupta2001}. To be self-contained, the proof is provided below.
\begin{proof}
Let $p:=\textsf{P}_{\mathrm{e,hin}}$ denote the true error probability of decoding a polar code where some of the \gls{bbcs} transmit random bits unknown to the decoder. A closed-form expression of $p$ is not available, so we resort to \gls{mcm} with $\textsf{M}$ i.i.d.\ trials, which produces the random number of error events
$K\sim\mathrm{Bin}(\textsf{M},p)$. Given the observation $K=k$, we want to derive an upper bound of $p$, namely $u_k$, which is data-dependent. Thus, if $u_k\leq \epsilon$, then the original reliability holds. Due to the randomness of $K$, it is possible that $u_k<p$, for some $k$, which cannot guarantee reliability, even if $u_k\leq \epsilon.$ Therefore, we want to avoid such an event by design, i.e., by minimizing Pr$\{u_K<p\}$. The proof is sketched as follows:
(i) We define $u_k$ by applying a binomial lower-tail probability so that $F(k;u_k)=\delta$;
(ii) we show that for fixed $k$, the map $q\mapsto F(k;q)$ is strictly decreasing;
(iii) we use this monotonicity to show that if $p>u_k$ then $F(k;p)<\delta$;
(iv) we show that $k\mapsto u_k$ is non-decreasing, hence the bad event $\{u_K<p\}$ is equivalent to the
event $\{K\le k^\star\}$ for some threshold $k^\star$;
(v) we bound $\Pr_p(K\le k^\star)=F(k^\star;p)<\delta$, which yields the desired $\Pr_p(p\le u_K)\ge 1-\delta$.

It suffices to prove 
\begin{equation}\label{eq:goal_pk_clean}
\Pr\bigl(p\le u_K\bigr)\ge 1-\delta,
\qquad p:=\textsf{P}_{\mathrm{e,hin}},
\end{equation}
when $K\sim\mathrm{Bin}(\textsf{M},p).$ Assume we do the \gls{mcm} for decoding \gls{gbcs} $\textsf{M}$-time under identical conditions and define the error indicator $E_t :=  \mathds{1}\{\mbox{the $t$-th trial fails}\};\text{ otherwise}, \; t\in[\textsf{M}].$
Assume $E_t\sim\mathrm{Bern}(p)$ i.i.d. Then the total error count is $
K:=\sum_{t=1}^{\textsf{M}} E_t \sim \mathrm{Bin}(\textsf{M},p),
\, K\in\{0,1,\dots,\textsf{M}\}.$ For each observation $k$, we define a deterministic upper bound $u_k$ of $p$. For $q\in[0,1]$ and integers $0\le k\le \textsf{M}$, define
\begin{equation}\label{eq:def_Fkq_again}
F(k;q)
:=\Pr_q(K\le k)
=\sum_{j=0}^k \binom{\textsf{M}}{j} q^j(1-q)^{\textsf{M}-j},
\end{equation}
where $\Pr_q(\cdot)$ denotes the probability calculated based on $E_t\sim\mathrm{Bern}(q)$.
In particular, for the true system, we have $\Pr_p(K\le k)=F(k;p)$.

For a fixed $k<\textsf{M}$, an increasing $q$ makes errors more likely. Hence, $K$ has higher probability to have larger values. Therefore, $\Pr_q(K\le k)$ decreases with $q$.
Formally, we know that $F(k;q)=1-I_q(k+1,\textsf{M}-k)$, where $I_q(a,b)$ is the regularized incomplete
beta function. By \cite[(8.17.18)]{NIST:DLMF},
we know that ${\frac{\mathrm{d}}{\mathrm{d}q} I_q(a,b)
=\frac{1}{B(a,b)}\,q^{a-1}(1-q)^{b-1}>0},\,q\in(0,1),\ a>0,\ b>0.$
Taking $a=k+1$ and $b=\textsf{M}-k$ yields $\frac{\mathrm{d}}{\mathrm{d}q}I_q(k+1,\textsf{M}-k)>0$, and hence
$\frac{\mathrm{d}}{\mathrm{d}q}F(k;q)=-\frac{\mathrm{d}}{\mathrm{d}q}I_q(k+1,\textsf{M}-k)<0$ for $q\in(0,1)$,
so $q\mapsto F(k;q)$ is strictly decreasing on $(0,1)$.

Recall that we want Pr$(p\leq u_K)\geq 1-\delta$. To achieve this goal, for each $k<\textsf{M}$ we define $u_k$ as the unique value such that, under
$K\sim\mathrm{Bin}(\textsf{M},u_k)$, the following equality is fulfilled
\begin{equation}\label{eq:def_uk_again}
F(k;u_k)=\delta,
\end{equation}
which is equivalent to $\Pr_{u_k}(K\le k)=\delta,$
i.e., under the candidate parameter $u_k$, the lower-tail event $\{K\le k\}$ has probability $\delta$.
We also set $u_{\textsf{M}}:=1$ since $F(\textsf{M};q)=\Pr_q(K\le \textsf{M})=1$ for all $q\in[0,1]$ and thus
\eqref{eq:def_uk_again} cannot be satisfied when $k=\textsf{M}$.
For any fixed $k<\textsf{M}$, the map $q\mapsto F(k;q)$ is continuous and strictly decreasing on $(0,1)$, with
$F(k;0)=1$ and $F(k;1)=0$, hence the solution to \eqref{eq:def_uk_again} exists and is unique.
Moreover, since $q\mapsto F(k;q)$ is strictly decreasing, $p>u_k$ implies
$F(k;p)<F(k;u_k)=\delta,$ which means that if the true error probability is worse than $u_k$, then the probability of observing as few as $k$ errors, is at most $\delta$.

In the following we show that $k\mapsto u_k$ is nondecreasing. Fix $k_1<k_2<\textsf{M}$.
For any $q$, since \gls{cdf} is nondecreasing in $k$, we have $F(k_2;q)\ge F(k_1;q)$.
In particular, when $q=u_{k_1}$, we have
\begin{align}\label{EQ_FFF}
    F(k_2;u_{k_1})\ge F(k_1;u_{k_1})=F(k_2;u_{k_2})=\delta,    
\end{align}
where the second and third equalities use \eqref{eq:def_uk_again}.
Since $q\mapsto F(k_2;q)$ is strictly decreasing, by comparing the first and third term in \eqref{EQ_FFF} we know that $u_k$ is nondecreasing in $k$.

We now prove \eqref{eq:goal_pk_clean} by upper-bounding \(\Pr(u_K<p)\). Define the deterministic function $g:\{0,1,\dots,\textsf{M}\}\to[0,1]$ by $g(k):=u_k$, so that
$u_K=g(K)$. Fix $p\in(0,1)$. Define $\mathcal{A}(p):=\{k\in\{0,1,\dots,\textsf{M}\}: g(k)<p\}.$
Then we have the equivalent events
\begin{align}\label{EQ_eq_events}
    \{u_K<p\}=\{g(K)<p\}=\{K\in\mathcal{A}(p)\}.    
\end{align}

Since $u_k$ is nondecreasing in $k$, for any $k'\le k$ we have $u_{k'}\le u_k$. 
Thus if $k\in\mathcal{A}(p)$ and $u_{k'}\le u_k<p$, then $k'\in\mathcal{A}(p)$,
which means that $k\in\mathcal{A}(p)$ implies $\{0,1,\dots,k\}\subseteq \mathcal{A}(p)$.
Therefore, $\mathcal{A}(p)$ must be of the form $\mathcal{A}(p)=\{0,1,\dots,k^\star\}$ for
$k^\star:=\max\mathcal{A}(p)$ with the convention $k^\star=-1$ if $\mathcal{A}(p)=\emptyset$. Hence, we have the following equivalent events
\begin{equation}\label{eq:event_equiv_again}
\{u_K<p\}=\{K\in\mathcal{A}(p)\}=\{K\le k^\star\}.
\end{equation}
Using $K\sim\mathrm{Bin}(\textsf{M},p)$ we obtain
\[
\Pr(u_K<p)=\Pr(K\le k^\star)=F(k^\star;p).
\]
By definition of $k^\star$ we know $p>u_{k^\star}$, and since $q\mapsto F(k^\star;q)$ is strictly decreasing,
\[
F(k^\star;p)<F(k^\star;u_{k^\star})=\delta,
\]
where the equality follows from \eqref{eq:def_uk_again} with $k=k^\star$.
Thus $\Pr(u_K<p)<\delta$, i.e., $\Pr(p\le u_K)\ge 1-\delta$, which completes the proof.
\end{proof}

\section{Proof of Lemma~\ref{lem:bbc-leakage}}\label{APP_proof_lem:bbc-leakage}
\begin{proof}
List elements of $\mathcal{I}_{\mathcal B}(\gamma_n)$ as
\(i_1<\dots<i_{|\mathcal{S}|}\) and let \(U_{\mathcal{S}} := (U_{i_1},\dots,U_{i_{|\mathcal{S}|}})\).
Define \(I_i:=I(U_i;Y^n,U^{i-1})\). Then
\begin{align}
  I\bigl(U_{\mathcal S};\,Y^n\bigr)
  &= \sum_{k=1}^{|\mathcal S|}
     I\bigl(U_{i_k};\,Y^n \,\big|\,U_{i_1},\dots,U_{i_{k-1}}\bigr)\label{eq:chain-mi}\\
  &\overset{(a)}\le \sum_{k=1}^{|\mathcal S|}
     I\bigl(U_{i_k};\,Y^n \,\big|\,U^{i_k-1}\bigr)\label{eq:step-a}\\
  &\overset{(b)}= \sum_{k=1}^{|\mathcal S|}
     I\bigl(U_{i_k};\,Y^n,U^{i_k-1}\bigr)\label{eq:step-b}\\
  &= \sum_{k=1}^{|\mathcal S|} I_{i_k}
  \;\overset{(c)}{\le}\;|\mathcal S|\,\gamma_n,\label{eq:step-c}
\end{align}

where in (a) we use the following
\begin{align*}
I\bigl(U_{i_k};Y^n \,\big|\,U_{i_1},\dots,U_{i_{k-1}}\bigr)
&= H\bigl(U_{i_k}\,\big|\,U_{i_1},\dots,U_{i_{k-1}}\bigr)
   -H\bigl(U_{i_k}\,\big|\,Y^n,U_{i_1},\dots,U_{i_{k-1}}\bigr)\\
&= H(U_{i_k})
   -H\bigl(U_{i_k}\,\big|\,Y^n,U_{i_1},\dots,U_{i_{k-1}}\bigr)\\
&\le H(U_{i_k})
   -H\bigl(U_{i_k}\,\big|\,Y^n,U^{i_k-1}\bigr)\\
&= I\bigl(U_{i_k};Y^n \,\big|\,U^{i_k-1}\bigr),
\end{align*}
where the second equality uses the fact $U_{i_k}\pperp U^{i_k-1}$ and the inequality uses that $U^{i_k-1}$ contains $(U_{i_1},\dots,U_{i_{k-1}})$, so conditioning on the larger set cannot increase conditional entropy, in (b), we use the fact 
$I(U_{i_k};Y^n,U^{i_k-1})=I(U_{i_k};U^{i_k-1})+I(U_{i_k};Y^n\mid U^{i_k-1})
=I(U_{i_k};Y^n\mid U^{i_k-1})$ due to the fact $U_{i_k}\pperp U^{i_k-1}$, in (c), because $\mathcal S\subseteq\mathcal I_{\mathcal B}(\gamma_n)$, we have $I_i\le \gamma_n$ for all
$i\in\mathcal S$, hence $\sum_{k=1}^{|\mathcal S|} I_{i_k}\le |\mathcal S|\gamma_n$.
This completes the proof.
\end{proof}

\section{Proof of Lemma \ref{lem:cyclic-hideB}}\label{APP_proof_cyclic-hideB}
\begin{proof}
Fix any realization $\mathbf f$ of $\bF$. Recall that $\bP_1=\bA^{K}$ with
$K\sim\Unif(\{0,\dots,\textsf{N}-1\})$, $\bT_2=\bA\bT_1$, and
$\bF=\bP_1^{\mathsf T}\bT_B=(\bA^{K})^{\mathsf T}\bT_B=(\bA^{\mathsf T})^{K}\bT_B=\bA^{-K}\bT_B$,
where we use $(\bA^{K})^{\mathsf T}=(\bA^{\mathsf T})^{K}=\bA^{-K}$ since $\bA$ is a permutation matrix.
Since $\bA^{\textsf{N}}=\bI$, we have $\bA^{-K}=\bA^{K'}$ with
$K':=(\textsf{N}-K)\ (\mathrm{mod}\ \textsf{N})$. Moreover, $K'\sim\Unif(\{0,\dots,\textsf{N}-1\})$
because the map $k\mapsto (\textsf{N}-k)\ (\mathrm{mod}\ \textsf{N})$ is a bijection on
$\{0,\dots,\textsf{N}-1\}$. Hence $\bF=\bA^{K'}\bT_B$.

Therefore, conditioned on $B=b$, the support of $\bF$ is
$\mathrm{supp}(\bF\mid B=b)=\{\bA^{k}\bT_b:\ 0\le k<\textsf{N}\}$.
Since $\bT_b$ is invertible, the map $k\mapsto \bA^{k}\bT_b$ is injective on
$\{0,\dots,\textsf{N}-1\}$. Therefore, $\mathrm{supp}(\bF\mid B=b)$ has cardinality $\textsf{N}$.

For $B=0$, we have
\begin{align}
\Pr(\bF=\mathbf f\mid B=0)
&=\sum_{k=0}^{\textsf{N}-1}\Pr(\bF=\mathbf f\mid B=0,K'=k)\Pr(K'=k)\notag\\
&\stackrel{(a)}{=}\frac{1}{\textsf{N}}\sum_{k=0}^{\textsf{N}-1}\mathds{1}\{\bA^{k}\bT_1=\mathbf f\},
\label{eq:cyclic-hideB-b0}
\end{align}
while for $B=1$, using $\bT_2=\bA\bT_1$, we obtain
\begin{align}
\Pr(\bF=\mathbf f\mid B=1)
&=\sum_{k=0}^{\textsf{N}-1}\Pr(\bF=\mathbf f\mid B=1,K'=k)\Pr(K'=k)\notag\\
&\stackrel{(b)}{=}\frac{1}{\textsf{N}}\sum_{k=0}^{\textsf{N}-1}\mathds{1}\{\bA^{k}\bT_2=\mathbf f\}\notag\\
&\stackrel{(c)}{=}\frac{1}{\textsf{N}}\sum_{k=0}^{\textsf{N}-1}\mathds{1}\{\bA^{k}\bA\bT_1=\mathbf f\}
=\frac{1}{\textsf{N}}\sum_{k=0}^{\textsf{N}-1}\mathds{1}\{\bA^{k+1}\bT_1=\mathbf f\}\notag\\
&\stackrel{(d)}{=}\frac{1}{\textsf{N}}\sum_{k=0}^{\textsf{N}-1}\mathds{1}\{\bA^{k}\bT_1=\mathbf f\}
=\Pr(\bF=\mathbf f\mid B=0),
\label{eq:cyclic-hideB-b1}
\end{align}
where (a) and (b) use $\bF=\bA^{K'}\bT_b$ given $(B=b,K'=k)$ and $\Pr(K'=k)=1/\textsf{N}$,
(c) uses $\bT_2=\bA\bT_1$, and (d) uses that the shift $k\mapsto k+1\ (\mathrm{mod}\ \textsf{N})$
is a bijection on $\{0,\dots,\textsf{N}-1\}$.
Since $\Pr(\bF=\mathbf f\mid B=0)=\Pr(\bF=\mathbf f\mid B=1)$ for all $\mathbf f$,
we conclude $\bF\pperp B$.
\end{proof}

\section{Proof of Theorem \ref{thm:polarOT-AWGN}}\label{APP_proof_thm:polarOT-AWGN}
\begin{proof}

We first prove \gls{sfb}. Let $Z := (M_0,M_1,X^n)$. Assume $\delta_B=0$, then \gls{sfb} requires
\begin{align}
I(B;Z,\Pi_{\mathrm{pub}})
&= I(B;Z)+I(B;\Pi_{\mathrm{pub}}\mid Z)=0.
\label{eq:SfB_topdown}
\end{align}
Hence it suffices to show $I(B;Z)=0$ and $I(B;\Pi_{\mathrm{pub}}\mid Z)=0$. We first check $I(B;Z)$. Recall $\bP_1=\bA^{K}$ with $K\sim\Unif(\{0,\dots,\textsf{N}-1\})$,
$\bT_2=\bA\bT_1$, and $\bF=\bP_1^{\mathsf T}\bT_B=\bA^{-K}\bT_B$.
Given $b\in\{0,1\}$ and $k\in\{0,\dots,\textsf{N}-1\}$, recall
\begin{align}\label{EQ_J0J1_gen}
    (\mathcal J_0,\mathcal J_1)=
\Bigl(
  (\pi_{\bA^{-k}}\mathcal G(\bT_{b+1}))_{\downarrow \ell},\,
  (\pi_{\bA^{-k}}\mathcal B(\bT_{b+1}))_{\downarrow \ell}
\Bigr).    
\end{align}

Moreover, $(\tilde{\mathcal J}_0,\tilde{\mathcal J}_1)=(\mathcal J_0,\mathcal J_1)$ if $B=0$;
$(\tilde{\mathcal J}_0,\tilde{\mathcal J}_1)=(\mathcal J_1,\mathcal J_0)$ if $B=1$, and
$\Pi_{\mathrm{sel}}=(\bF,\tilde{\mathcal J}_0,\tilde{\mathcal J}_1)$.

Fix an arbitrary event $\mathcal E\subseteq \mathrm{range}(\Pi_{\mathrm{sel}})$.
Write elements of $\mathcal E$ as triples $(\mathbf f,j_0,j_1)$.
Then under $B=0$, we have $\Pi_{\mathrm{sel}}=(\bF,\mathcal J_0,\mathcal J_1)$ and
\begin{align}
\Pr(\Pi_{\mathrm{sel}}\in\mathcal E\mid B=0)
&=\sum_{(\mathbf f,j_0,j_1)\in\mathcal E}
   \Pr(\bF=\mathbf f,\mathcal J_0=j_0,\mathcal J_1=j_1\mid B=0)\notag\\
&=\sum_{(\mathbf f,j_0,j_1)\in\mathcal E}
   \Pr(\bF=\mathbf f\mid B=0)\Pr(\mathcal J_0=j_0,\mathcal J_1=j_1\mid \bF=\mathbf f,B=0)\notag\\
&\stackrel{(a)}{=}\sum_{(\mathbf f,j_0,j_1)\in\mathcal E}
   \Pr(\bF=\mathbf f)\Pr(\mathcal J_0=j_0,\mathcal J_1=j_1\mid \bF=\mathbf f,B=0)\notag\\
&\stackrel{(b)}{=}\sum_{(\mathbf f,j_0,j_1)\in\mathcal E}
   \Pr(\bF=\mathbf f)\Pr(\mathcal J_1=j_0, \mathcal J_0=j_1\mid \bF=\mathbf f,B=1)\notag\\
&\stackrel{(c)}=\Pr(\Pi_{\mathrm{sel}}\in\mathcal E\mid B=1),
\label{eq:Pi_sel_indep_B_optionA}
\end{align}
where (a) uses $\bF\pperp B$ from Lemma~\ref{lem:cyclic-hideB}, (c) comes from the following fact: under $B=1$, Bob publishes $(\tilde{\mathcal J}_0,\tilde{\mathcal J}_1)=(\mathcal J_1,\mathcal J_0)$,
hence $\Pi_{\mathrm{sel}}=(\bF,\mathcal J_1,\mathcal J_0)$.
Therefore, expanding $\Pr(\Pi_{\mathrm{sel}}\in\mathcal E\mid B=1)$ in the same way as above yields
the right-hand side of step (b) and gives (c). We now derive (b) as follows:

\begin{align}
\Pr(\mathcal J_0=j_0,\mathcal J_1=j_1\mid \bF=\mathbf f,B=0)
&\!=\!\sum_{k=0}^{\textsf{N}-1}\Pr(K=k\mid \bF=\mathbf f,B=0)
        \Pr(\mathcal J_0=j_0,\mathcal J_1=j_1\mid K=k,\bF=\mathbf f,B=0)\notag\\
&\!=\!\sum_{k=0}^{\textsf{N}-1}\!\Pr(K=k\mid \bF=\mathbf f,B=0)
        \mathds{1}\!\left\{(\pi_{\bA^{-k}}\mathcal G(\bT_1))_{\downarrow\ell}=j_0,
                           (\pi_{\bA^{-k}}\mathcal B(\bT_1))_{\downarrow\ell}=j_1\right\}\notag\\
&\overset{(d)}{=}\frac{1}{c(\mathbf f)}\sum_{k=0}^{\textsf{N}-1}
   \mathds{1}\{\bA^{-k}\bT_1=\mathbf f\}\,
   \mathds{1}\!\left\{(\pi_{\bA^{-k}}\mathcal G(\bT_1))_{\downarrow\ell}=j_0,\ 
                      (\pi_{\bA^{-k}}\mathcal B(\bT_1))_{\downarrow\ell}=j_1\right\}\notag\\
&\overset{(e)}{=}\!\frac{1}{c(\mathbf f)}\!\!\sum_{k=0}^{\textsf{N}-1}\!\!
   \mathds{1}\{\!\bA^{-(k+1)}\bT_2\!=\!\mathbf f\}
   \!\mathds{1}\!\left\{(\pi_{\bA^{-(k+1)}}\mathcal G(\bT_2))_{\downarrow\ell}=j_1, 
                      (\pi_{\bA^{-(k+1)}}\mathcal B(\bT_2))_{\downarrow\ell}=j_0\right\}\notag\\
&\overset{(f)}{=}\frac{1}{c(\mathbf f)}\sum_{k=0}^{\textsf{N}-1}
   \mathds{1}\{\bA^{-k}\bT_2=\mathbf f\}\,
   \mathds{1}\!\left\{(\pi_{\bA^{-k}}\mathcal G(\bT_2))_{\downarrow\ell}=j_1,\ 
                      (\pi_{\bA^{-k}}\mathcal B(\bT_2))_{\downarrow\ell}=j_0\right\}\notag\\
&\overset{(g)}{=}\Pr(\mathcal J_1=j_0,\mathcal J_0=j_1\mid \bF=\mathbf f,B=1),
\end{align}
where the second equality comes from the fact that conditioned on $K=k,B=0$, we have $\bT_B=\bT_1$ and $\bP_1=\bA^k$, hence $\bF=\bP_1^{\mathsf T}\bT_B=\bA^{-k}\bT_1$ deterministically. Therefore, from \eqref{EQ_J0J1_gen} we know that $(\mathcal J_0,\mathcal J_1)$ is also deterministic, then
$\Pr(\mathcal J_0=j_0,\mathcal J_1=j_1\mid K=k,\bF=\mathbf f,B=0)$ equals $1$ iff
$(\pi_{\bA^{-k}}\mathcal G(\bT_1))_{\downarrow\ell}=j_0$ and
$(\pi_{\bA^{-k}}\mathcal B(\bT_1))_{\downarrow\ell}=j_1$, and equals $0$, otherwise,
which yields the indicator term, (d) uses the Bayes' rule
\begin{align}
\Pr(K=k\mid \bF=\mathbf f,B=0)
&=\frac{\Pr(\bF=\mathbf f\mid K=k,B=0)\Pr(K=k\mid B=0)}
        {\sum_{k'=0}^{\textsf{N}-1}\Pr(\bF=\mathbf f\mid K=k',B=0)\Pr(K=k')}.\label{EQ_SfB_bayes}
\end{align}
Since conditioned on $K=k,B=0$, $\bF=\bA^{-K}\bT_1=\bA^{-k}\bT_1$ is
deterministic, so $\Pr(\bF=\mathbf f\mid K=k,B=0)=\mathds{1}\{\bA^{-k}\bT_1=\mathbf f\}.$ In addition, $K\pperp B$ implies $\Pr(K=k\mid B=0)=\Pr(K=k)$, and since $K\sim\Unif(\{0,\dots,\textsf{N}-1\})$ we have $\Pr(K=k)=1/\textsf{N}$.
Therefore,
\[
\Pr(K=k\mid \bF=\mathbf f,B=0)
=\frac{\mathds{1}\{\bA^{-k}\bT_1=\mathbf f\}}
        {\sum_{k'=0}^{\textsf{N}-1}\mathds{1}\{\bA^{-k'}\bT_1=\mathbf f\}}
=\frac{\mathds{1}\{\bA^{-k}\bT_1=\mathbf f\}}{c(\mathbf f)},
\]
where $c(\mathbf f):=\sum_{k=0}^{\textsf{N}-1}\mathds{1}\{\bA^{-k}\bT_1=\mathbf f\}$,
 (e) uses $\bT_2=\bA\bT_1$ and the assumed swap
$\mathcal G(\bT_2)=\pi_{\bA}(\mathcal B(\bT_1))$ and $\mathcal B(\bT_2)=\pi_{\bA}(\mathcal G(\bT_1))$,
so that $\bA^{-(k+1)}\bT_2=\bA^{-k}\bT_1$ and
$\pi_{\bA^{-(k+1)}}\mathcal G(\bT_2)=\pi_{\bA^{-k}}\mathcal B(\bT_1)$ and
$\pi_{\bA^{-(k+1)}}\mathcal B(\bT_2)=\pi_{\bA^{-k}}\mathcal G(\bT_1)$.
Step (f) is the reindexing $k\mapsto k+1\ (\mathrm{mod}\ \textsf{N})$, a bijection on $\{0,\dots,\textsf{N}-1\}$.
Finally, (g) follows by applying the same expansions in a reverse order over $K$ under $B=1$ and
using Bayes' rule with the deterministic relation $\bF=\bA^{-K}\bT_2$, i.e.,
$\Pr(K=k\mid \bF=\mathbf f,B=1)=\frac{\mathds{1}\{\bA^{-k}\bT_2=\mathbf f\}}{c(\mathbf f)},$
where $\sum_{k=0}^{\textsf{N}-1}\mathds{1}\{\bA^{-k}\bT_2=\mathbf f\}=c(\mathbf f)$ by the same reindexing induced by $\bT_2=\bA\bT_1$.

Let $R_{\mathrm A}$ be generated independently of all Bob's random variables, i.e., $R_{\mathrm A} \ \pperp\ (B,\Pi_{\mathrm{sel}})$, and denote all of Alice’s randomness used to generate $M_0,M_1$ and the random components of $U^n$.
After observing $\Pi_{\mathrm{sel}}=(\bF,\tilde{\mathcal J}_0,\tilde{\mathcal J}_1)$ and
$\tilde{\mathcal J}:=\tilde{\mathcal J}_0\cup \tilde{\mathcal J}_1$, Alice sets
$U_i := 
(R_{\mathrm A})_i, \mbox{ if } i\in \tilde{\mathcal J};
0, \mbox{ if } i\notin \tilde{\mathcal J}$. So there exists a deterministic map $\psi_U$ such that $U^n:=\psi_U(\Pi_{\mathrm{sel}},R_{\mathrm A})$. Since $X^n=U^n\bF$ and $Z=(M_0,M_1,X^n)$, there exists a deterministic map $\psi_Z$ such that $Z=\psi_Z(\Pi_{\mathrm{sel}},R_{\mathrm A})$. Therefore, for any $b\in\{0,1\}$ and any $z$,
\begin{align}
P_{Z\mid B}(z\mid b)
&=\sum_{\pi} P_{\Pi_{\mathrm{sel}}\mid B}(\pi\mid b)\,P_{Z\mid \Pi_{\mathrm{sel}},B}(z\mid \pi,b)\notag\\
&\overset{(h)}{=}\sum_{\pi} P_{\Pi_{\mathrm{sel}}\mid B}(\pi\mid b)\,P_{Z\mid \Pi_{\mathrm{sel}}}(z\mid \pi)\notag\\
&\overset{(i)}{=}\sum_{\pi} P_{\Pi_{\mathrm{sel}}}(\pi)\,P_{Z\mid \Pi_{\mathrm{sel}}}(z\mid \pi)
= P_Z(z),
\label{eq:B_indep_Z}
\end{align}
where (h) is due to the following: since $Z=\psi_Z(\Pi_{\mathrm{sel}},R_{\mathrm A})$, conditioning on
$\Pi_{\mathrm{sel}}=\pi$, $Z$ is a deterministic function of $R_{\mathrm A}$.
Moreover, $R_{\mathrm A}\pperp (B,\Pi_{\mathrm{sel}})$ implies
$P_{R_{\mathrm A}\mid \Pi_{\mathrm{sel}},B}(r\mid \pi,b)=P_{R_{\mathrm A}}(r)
= P_{R_{\mathrm A}\mid \Pi_{\mathrm{sel}}}(r\mid \pi)$.
Therefore,
\begin{align}
P_{Z\mid \Pi_{\mathrm{sel}},B}(z\mid \pi,b)
&=\sum_{r} P_{Z\mid R_{\mathrm A},\Pi_{\mathrm{sel}},B}(z\mid r,\pi,b)\,
          P_{R_{\mathrm A}\mid \Pi_{\mathrm{sel}},B}(r\mid \pi,b)\notag\\
&\overset{(j)}=\sum_{r} \mathds{1}\{\psi_Z(\pi,r)=z\}\,
          P_{R_{\mathrm A}\mid \Pi_{\mathrm{sel}},B}(r\mid \pi,b)\notag\\
&\overset{(k)}{=}\sum_{r} \mathds{1}\{\psi_Z(\pi,r)=z\}\,
          P_{R_{\mathrm A}\mid \Pi_{\mathrm{sel}}}(r\mid \pi)\notag\\
&\overset{(\ell)}{=}\sum_{r} P_{Z\mid R_{\mathrm A},\Pi_{\mathrm{sel}}}(z\mid r,\pi)
          P_{R_{\mathrm A}\mid \Pi_{\mathrm{sel}}}(r\mid \pi),\notag\\
&\overset{}{=}P_{Z\mid \Pi_{\mathrm{sel}}}(z\mid \pi),\label{eq:pushforward_Z_given_Pi}
\end{align}
where (j) uses $Z=\psi_Z(\Pi_{\mathrm{sel}},R_{\mathrm A})$, (k) uses  $R_{\mathrm A} \pperp (B,\Pi_{\mathrm{sel}})$, ($\ell$) uses  $Z=\psi_Z(\Pi_{\mathrm{sel}},R_{\mathrm A})$ again. Step (i) in \eqref{eq:B_indep_Z} uses $\Pi_{\mathrm{sel}}\pperp B$ from \eqref{eq:Pi_sel_indep_B_optionA}.
Hence, \eqref{eq:B_indep_Z} shows $B\pperp Z$, and therefore, $I(B;Z)=0$. 

Since \eqref{eq:Pi_sel_indep_B_optionA} gives \(\Pi_{\mathrm{sel}}\pperp B\) and \eqref{eq:pushforward_Z_given_Pi} gives \(Z\pperp B\mid \Pi_{\mathrm{sel}}\),
for any \(b\in\{0,1\}\) and any realizations \((\pi,z)\) we have
\begin{align}
P_{\Pi_{\mathrm{sel}},Z\mid B}(\pi,z|b)
&\overset{(m)}{=}P_{\Pi_{\mathrm{sel}}\mid B}(\pi|b)\,P_{Z\mid \Pi_{\mathrm{sel}},B}(z|\pi,b)\notag\\
&\overset{(n)}{=}P_{\Pi_{\mathrm{sel}}}(\pi)\,P_{Z\mid \Pi_{\mathrm{sel}}}(z|\pi),\label{eq:PiZ-given-B}\\
&=P_{Z \Pi_{\mathrm{sel}}}(z,\pi)\label{eq:PiZ-given-B2}
\end{align}
where (m) is by Bayes' rule and (n) uses \(\Pi_{\mathrm{sel}}\pperp B\) and \(Z\pperp B\mid \Pi_{\mathrm{sel}}\).
Hence, for any \(z\) with \(P_{Z\mid B}(z|b)>0\), we have
\begin{align}
P_{\Pi_{\mathrm{sel}}\mid Z,B}(\pi|z,b)
&\overset{(o)}{=}\frac{P_{\Pi_{\mathrm{sel}},Z\mid B}(\pi,z|b)}{P_{Z\mid B}(z|b)}\notag\\
&\overset{(p)}{=}\frac{P_{Z \Pi_{\mathrm{sel}}}(z,\pi)}
{\sum_{\pi'} P_{Z \Pi_{\mathrm{sel}}}(z,\pi')}\notag\\
&\overset{(q)}{=}P_{\Pi_{\mathrm{sel}}\mid Z}(\pi|z),\label{eq:Pi-indep-B-given-Z}
\end{align}
where (o) is Bayes' rule and (p) uses \eqref{eq:PiZ-given-B} on both numerator and denominator; (q) follows Bayes' rule again.
Therefore \(\Pi_{\mathrm{sel}}\pperp B\mid Z\), i.e., \(\Pi_{\mathrm{sel}}-(X^n,M_0,M_1)-B\) is a Markov chain.

Recall that \(S_1,S_2\) is the public UHF seed and define
\(\Pi_{\mathrm{pub}}:=(\Pi_{\mathrm{sel}},S_1,S_2,C_0,C_1)\), where
\(C_b:=M_b\oplus K_b,\,b=0,1\).
Since $\bF$ is invertible over $\mathds{F}_2$, $U^n$ is uniquely determined by $(X^n,\bF)$ via
$U^n = X^n\bF^{-1}$. Hence, the hash inputs $U^n|_{\tilde{\mathcal J}_0}$ and $U^n|_{\tilde{\mathcal J}_1}$
are deterministic functions of $(Z,\Pi_{\mathrm{sel}})$. In addition, the seed \(S\) fixes the chosen hash function, then the resulting keys \((K_0,K_1)\) are deterministic
functions of \((Z,\Pi_{\mathrm{sel}},S)\). Therefore, there exists a deterministic map \(\varphi\) such that
$(C_0,C_1)=\varphi(Z,\Pi_{\mathrm{sel}},S).$ Consequently,
$P(C_0,C_1\mid Z,\Pi_{\mathrm{sel}},S_1,S_2,B)=
P(C_0,C_1\mid Z,\Pi_{\mathrm{sel}},S_1,S_2),$ i.e., \(B-(Z,\Pi_{\mathrm{sel}},S)-(C_0,C_1)\). Using this chain rule, we have
\begin{align}
I(B;\Pi_{\mathrm{pub}}\mid Z)
&= I(B;\Pi_{\mathrm{sel}},S_1,S_2,C_0,C_1\mid Z)\notag\\
&= I(B;\Pi_{\mathrm{sel}}\mid Z)+I(B;S\mid Z,\Pi_{\mathrm{sel}})
   +I(B;C_0,C_1\mid Z,\Pi_{\mathrm{sel}},S)\notag\\
&= 0,
\label{EQ_I_zero_pub}
\end{align}
where the last equality uses \(\Pi_{\mathrm{sel}}-Z-B\) from \eqref{eq:Pi-indep-B-given-Z}, \(S\pperp (B,Z,\Pi_{\mathrm{sel}})\), and
\(B-(Z,\Pi_{\mathrm{sel}},S)-(C_0,C_1)\) for each of the three terms in the second equality. Hence we have the Markov chain \(\Pi_{\mathrm{pub}}-Z-B\), i.e., $I(B;\Pi_{\mathrm{pub}}\mid Z)=0.$ Combined with $I(B;Z)=0$, we have
$I(B;Z,\Pi_{\mathrm{pub}})= 0,$ which proves \gls{sfb} with $\delta_B=0$.

Now we prove \gls{sfa}. Fix \(b\in\{0,1\}\) and write \(\bar b:=1-b\). To simplify the notation, we let
\begin{align}\label{EQ_Def_MKCE}
  M:=M_{\bar b},\qquad
  K:=K_{\bar b},\qquad
  C:=C_{\bar b},\qquad
  E:=\bigl(Y^n,\Pi_{\mathrm{sel}},S_b,C_b\bigr),  
\end{align}
so that Bob's completely observed information conditioned on \(B=b\) is \((E,C,S_{\bar b})=(Y^n,\Pi_{\mathrm{pub}})\).

To prove \gls{sfa}, it suffices to show
\begin{equation}\label{eq:SfA_cond_goal}
d_{\mathrm{var}}\!\big(P_{M,C,E\mid B=b},\ P_M\times P_{C,E\mid B=b}\big)\ \le\ 2\varepsilon_b .
\end{equation}
Note that since $S_{\bar b}$ is independent of $(M,E,B)$, it can be appended to $E$. However, $S_{\bar b}$ is not independent of $C_{\bar b}$. Let \(\widetilde C\sim\Unif(\{0,1\}^{\ell})\) be an auxiliary random variable, independent of \((M,E,B)\),
and define the product measure
\(
Q:=P_M\times P_{\widetilde C}\times P_{E\mid B=b}.
\)
Then, by the triangle inequality,
\begin{align}
d_{\mathrm{var}}\!\big(P_{M,C,E\mid B=b},\ P_M\times P_{C,E\mid B=b}\big)
\le
d_{\mathrm{var}}\!\big(P_{M,C,E\mid B=b},\ Q\big)
+
d_{\mathrm{var}}\!\big(Q,\ P_M\times P_{C,E\mid B=b}\big).
\label{eq:tri_split}
\end{align}

To derive the second term on the \gls{rhs} of \eqref{eq:tri_split}, we first derive $P_{C,E\mid M,B}(c,\mathcal{A}|m,b)$ given $c\in\{0,1\}^\ell$ and a measurable set $\mathcal{A}$, then marginalize it over $M$:
\begin{align}
P_{C,E\mid M,B}(c,\mathcal{A}|m,b)
&= \Pr\!\big[C=c,\ E\in\mathcal{A}\ \big|\ M=m,\ B=b\big] \\
&\overset{(b)}= \Pr\!\big[M\oplus K=c,\ E\in\mathcal{A}\ \big|\ M=m,\ B=b\big] \\
&= \Pr\!\big[m\oplus K=c,\ E\in\mathcal{A}\ \big|\ M=m,\ B=b\big] \\
&= \Pr\!\big[K=m\oplus c,\ E\in\mathcal{A}\ \big|\ M=m,\ B=b\big] \\
&= P_{K,E\mid M,B}(m\oplus c,\mathcal{A}|m,b) \\
&\overset{(c)}{=} P_{K,E\mid B}(m\oplus c,\mathcal{A}|b),\label{EQ_PCE_PKE}
\end{align}
where (b) is from \eqref{EQ_Def_MKCE}, $(c)$ uses the fact $(K,E)\pperp M \mid (B=b)$, which is derived as follows. 
Fix $b\in\{0,1\}$. 
Following the OT construction we know that:
\begin{equation}\label{eq:msg-indep-model}
(M_0,M_1)\ \pperp\ (B,\,U^n,\,S_0,\,S_1,\,N,\,R_{\mathrm B}),
\qquad\text{and}\qquad
M_0\ \pperp\ M_1,
\end{equation}
where $N$ denotes the channel noise and $R_{\mathrm B}$ denotes all of Bob's local randomness
that may affect $(\Pi_{\mathrm{sel}},Y^n)$.

Define $  G := (U^n,\Pi_{\mathrm{sel}},S_0,S_1,N,R_{\mathrm B}).$ By construction of the protocol given the fixed $b$, there exist deterministic maps $f_b$ and $g_b$, such that
\begin{equation}\label{eq:KE-as-functions}
  K_{\bar b} = f_b(G),
  \qquad
  E := g_b(G,M_b).
\end{equation}
In particular, $(K_{\bar b},E)$ is a deterministic function of $(G,M_b)$. It is clear that
$  M_{\bar b}\ \pperp\ (G,M_b,B).$ Then conditioning \(P_{M_{\bar b},G,M_b,B}=P_{M_{\bar b}}\,P_{G,M_b,B}\) on \(B=b\) gives
$P_{M_{\bar b},G,M_b\mid B=b}=P_{M_{\bar b}}\,P_{G,M_b\mid B=b}.$ Since \(M_{\bar b}\pperp B\), we also have \(P_{M_{\bar b}}=P_{M_{\bar b}\mid B=b}\), hence
\(P_{M_{\bar b},G,M_b\mid B=b}=P_{M_{\bar b}\mid B=b}\,P_{G,M_b\mid B=b}\). Together with \eqref{eq:KE-as-functions}, we have $ (K_{\bar b},E)\ \pperp\ M_{\bar b}$ given $ B=b.$

% --- Measure-theoretic form (to connect to the TV integral) ---
Let $E:=(Y^n,\Pi_{\mathrm{sel}},S_b,C_b)$ be a measurable mapping into a measurable space $(\mathcal{E},\mathscr{E})$. Because $Y^n\in\mathbb{R}^n$ and $(\Pi_{\mathrm{sel}},S_b,C_b)$ are discrete, we take the following product measure as a reference measure
\[
\mu := \lambda^n \otimes \#_{\Pi} \otimes \#_{\{0,1\}^{\ell}},
\]
where $\lambda^n$ is Lebesgue measure and $\#$ denotes counting measure.

Assume $P_{E\mid B=b}\ll\mu$ and $P_{K,E\mid B=b}(k,\cdot)\ll\mu$ for all $k\in\{0,1\}^{\ell}$, and define Radon--Nikodym derivatives 
\[
p_{E\mid B=b}(e):=\frac{\mathrm d P_{E\mid B=b}}{\mathrm d \mu}(e),
\qquad
p_{K,E\mid B=b}(k,e):=\frac{\mathrm d P_{K,E\mid B=b}(k,\cdot)}{\mathrm d \mu}(e).
\]
Similarly, for fixed $m\in\{0,1\}^{\ell}$ assume $P_{C,E\mid M=m,B=b}(c,\cdot)\ll\mu$ and set
\[
p_{C,E\mid M=m,B=b}(c,e):=\frac{\mathrm d P_{C,E\mid M=m,B=b}(c,\cdot)}{\mathrm d \mu}(e).
\]

By \eqref{EQ_PCE_PKE}, for every $m,c\in\{0,1\}^{\ell}$ and every $\mathcal{A}\in\mathscr{E}$,
\begin{equation}\label{eq:PCE-setwise}
P_{C,E\mid M,B}(c,\mathcal{A}|m,b)
=
P_{K,E\mid B}(m\oplus c,\mathcal{A}|b).
\end{equation}
Since both measures on the RHS/LHS are absolutely continuous w.r.t.\ $\mu$, the uniqueness of the
Radon--Nikodym derivative implies
\begin{equation}\label{eq:density-CE-from-KE}
p_{C,E\mid M=m,B=b}(c,e)=p_{K,E\mid B=b}(m\oplus c,e)
\qquad \text{for }\mu\text{-a.e.\ }e\in\mathcal{E}.
\end{equation}

% (optional one-line reminder right before the TV step; see note below)
Let $\nu:=\#_{\{0,1\}^{\ell}}\otimes \mu$ be the product reference measure on $\{0,1\}^{\ell}\times\mathcal{E}$.

Applying total variation via densities\footnote{Let $P,Q$ be probability measures on $(\mathcal{S},\mathscr{S})$ and let $\nu$ be $\sigma$-finite with
$P\ll\nu$ and $Q\ll\nu$. Writing $p:=\frac{\mathrm dP}{\mathrm d\nu}$ and $q:=\frac{\mathrm dQ}{\mathrm d\nu}$, we have
\[
d_{\mathrm{var}}(P,Q)=\tfrac12\int_{\mathcal{S}} |p(s)-q(s)|\,\nu(\mathrm d s).
\]} with $\nu$ yields
\begin{align}
d_{\mathrm{var}}\!\big(P_{C,E\mid M=m,B=b},\ \Unif\times P_{E\mid B=b}\big)
&=\tfrac12\int_{\{0,1\}^{\ell}\times\mathcal{E}}
\Big|
\frac{\mathrm d P_{C,E\mid M=m,B=b}}{\mathrm d\nu}(c,e)
-
\frac{\mathrm d(\Unif\times P_{E\mid B=b})}{\mathrm d\nu}(c,e)
\Big|\,\nu(\mathrm d(c,e))\notag\\
&=\tfrac12\sum_{c\in\{0,1\}^{\ell}}\int_{\mathcal{E}}
\Big|p_{C,E\mid M=m,B=b}(c,e)-2^{-\ell}p_{E\mid B=b}(e)\Big|\,\mu(\mathrm d e)\notag\\
&\overset{(b)}{=}\tfrac12\sum_{c\in\{0,1\}^{\ell}}\int_{\mathcal{E}}
\Big|p_{K,E\mid B=b}(m\oplus c,e)-2^{-\ell}p_{E\mid B=b}(e)\Big|\,\mu(\mathrm d e)\notag\\
&\overset{(c)}{=}\tfrac12\sum_{k\in\{0,1\}^{\ell}}\int_{\mathcal{E}}
\Big|p_{K,E\mid B=b}(k,e)-2^{-\ell}p_{E\mid B=b}(e)\Big|\,\mu(\mathrm d e)\notag\\
&=d_{\mathrm{var}}\!\big(P_{K,E\mid B=b},\ \Unif\times P_{E\mid B=b}\big),
\label{eq:tv_CE_eq_tv_KE_mu}
\end{align}
where (b) uses \eqref{eq:density-CE-from-KE}, and (c) is the bijective re-indexing $k:=m\oplus c$
on $\{0,1\}^{\ell}$.

For the first term on the \gls{rhs} in \eqref{eq:tri_split}, recall \(
Q:=P_M\times P_{\widetilde C}\times P_{E\mid B=b}.
\) and we can derive the following
\begin{align}
d_{\mathrm{var}}\!\big(P_{M,C,E\mid B=b},\ Q\big)
&=\sum_{m}P_{M\mid B=b}(m)\,
d_{\mathrm{var}}\!\big(P_{C,E\mid M=m,B=b},\ \Unif\times P_{E\mid B=b}\big)\notag\\
&\overset{(d)}{=}\sum_{m}P_{M}(m)\,
d_{\mathrm{var}}\!\big(P_{C,E\mid M=m,B=b},\ \Unif\times P_{E\mid B=b}\big),
\label{eq:first_term_as_avg_mu}
\end{align}
where (d) uses $M\pperp B$. For the second term on the \gls{rhs} in \eqref{eq:tri_split}, using
$P_{C,E\mid B=b}=\sum_m P_M(m)\,P_{C,E\mid M=m,B=b}$ and convexity of total variation in each argument, we can derive
\begin{align}
d_{\mathrm{var}}\!\big(Q,\ P_M\times P_{C,E\mid B=b}\big)
&=d_{\mathrm{var}}\!\big(\Unif\times P_{E\mid B=b},\ P_{C,E\mid B=b}\big)\notag\\
&\le \sum_m P_M(m)\,
d_{\mathrm{var}}\!\big(\Unif\times P_{E\mid B=b},\ P_{C,E\mid M=m,B=b}\big)\notag\\
&=\sum_m P_M(m)\,
d_{\mathrm{var}}\!\big(P_{C,E\mid M=m,B=b},\ \Unif\times P_{E\mid B=b}\big).
\label{eq:second_term_as_avg_mu}
\end{align}

Let \(V:=V_{\bar b}\) be the hash-input random variable for the unchosen key, and let
\(K:=h_{S_{\bar b}}(V)\), where \(h_{S_{\bar b}}\) is drawn uniformly at random from a \gls{uhf} family
with public seed \(S_{\bar b}\), generated independently of \((V,Y^n,\Pi_{\mathrm{sel}},S_B,N,R_{\mathrm B},M_0,M_1,B)\).
Apply Corollary~\ref{Corollary_LHL} with \(X=V\), \(Z=(E,B=b)\), and the random mapping
\(F:=h_{S_{\bar b}}\), where \(S_{\bar b}\sim\Unif(\mathcal{S})\) and \(\{h_s:s\in\mathcal{S}\}=\mathcal{F}\),
and set the extra leakage variable in Corollary~\ref{Corollary_LHL} as null.
Then for any \(\varepsilon_{\mathrm{sm}}\in(0,1)\),
\begin{equation}\label{eq:LHL_general_form_used}
d_{\mathrm{var}}\!\big(P_{K,E,F\mid B=b},\ \Unif(\{0,1\}^{\ell})\times P_{E\mid B=b}\times P_{F}\big)
\ \le\ 2\varepsilon_{\mathrm{sm}} \;+\; \tfrac12\sqrt{2^{\ell - H_{\min}^{\varepsilon_{\mathrm{sm}}}(V\mid E,B=b)}}.
\end{equation}
By marginalizing out \(F\), we get
\begin{equation}\label{eq:LHL_marginal_used_mu}
d_{\mathrm{var}}\!\big(P_{K,E\mid B=b},\ \Unif(\{0,1\}^{\ell})\times P_{E\mid B=b}\big)
\ \le\ 2\varepsilon_{\mathrm{sm}} \;+\; \tfrac12\sqrt{2^{\ell - H_{\min}^{\varepsilon_{\mathrm{sm}}}(V\mid E,B=b)}}.
\end{equation}
In particular, if \(\ell\) is chosen so that
\begin{equation}\label{eq:LHL_len_condition_general}
\ell\ \le\ H_{\min}^{\varepsilon_{\mathrm{sm}}}(V\mid E,B=b)
\ -\ 2\log\!\Big(\frac{1}{2(\varepsilon_b-2\varepsilon_{\mathrm{sm}})}\Big),
\end{equation}
for some target \(\varepsilon_b>2\varepsilon_{\mathrm{sm}}\), then \eqref{eq:LHL_marginal_used_mu} yields
\(d_{\mathrm{var}}(P_{K,E\mid B=b},\Unif\times P_{E\mid B=b})\le\varepsilon_b\).

Combining \eqref{eq:LHL_marginal_used_mu} with \eqref{eq:tv_CE_eq_tv_KE_mu} yields, for every $m$,
\begin{equation}\label{eq:tv_CE_bound_each_m_mu}
d_{\mathrm{var}}\!\big(P_{C,E\mid M=m,B=b},\ \Unif\times P_{E\mid B=b}\big)\le \varepsilon_b.
\end{equation}
Substituting \eqref{eq:tv_CE_bound_each_m_mu} into \eqref{eq:first_term_as_avg_mu} and \eqref{eq:second_term_as_avg_mu},
and then into \eqref{eq:tri_split}, we have the following upper bound:
\[
d_{\mathrm{var}}\!\big(P_{M,C,E\mid B=b},\ P_M\times P_{C,E\mid B=b}\big)\le \varepsilon_b+\varepsilon_b=2\varepsilon_b.
\]
Applying expectation over $B$, we complete the proof of \gls{sfa}.

\end{proof}

\section{Proof of Lemma~\ref{lem:polar-incidence}}
\label{APP_proof_lem:polar-incidence}

\begin{proof}
For each \(m\ge 1\), let
\(
  \mathbf{T}^{(m)} := \bT_0^{\otimes m}
\)
with rows and columns indexed by \(\mathcal{X}_m:=\{0,1\}^m\).
We prove by induction on \(m\) that
\begin{equation}\label{eq:T-encodes-order-app}
  \mathbf{T}^{(m)}_{x,y}
  \;=\;
  \mathds{1}\{\,y \le_b x\,\},
  \qquad x,y\in\mathcal{X}_m,
\end{equation}
where \(\le\) is the bit-wise order \(x\le_b y \mbox{ iff } x_i\le y_i\) for all \(i\).

For \(m=1\),
\[
  \mathbf{T}^{(1)} = \bT_0
  :=
  \begin{bmatrix}
    1 & 0\\
    1 & 1
  \end{bmatrix},
\]
and a direct check shows
\(
  \bT_{0,x,y}=\mathds{1}\{\,y\le_b x\,\}
\)
for \(x,y\in\{0,1\}\), so \eqref{eq:T-encodes-order-app} holds for \(m=1\).

Assume \eqref{eq:T-encodes-order-app} holds for some \(m\ge 1\), i.e.,
\[
  \mathbf{T}^{(m)}_{x',y'}
  = \mathds{1}\{\,y'\le_b x'\,\},
  \qquad x',y'\in\mathcal{X}_m.
\]
For \(m+1\), let any \(x,y\in\mathcal{X}_{m+1}\) as $  x=(x_1,x'),\quad y=(y_1,y'),$ with \(x_1,y_1\in\{0,1\}\) and \(x',y'\in\mathcal{X}_m\).
Using the Kronecker-product rule with \(\mathbf{T}^{(m+1)}=\bT_0\otimes \mathbf{T}^{(m)}\), we have
\begin{equation}\label{eq:Tm+1-entry}
  \mathbf{T}^{(m+1)}_{(x_1,x'),(y_1,y')}
  = \bT_{0,x_1,y_1}\,\mathbf{T}^{(m)}_{x',y'}.
\end{equation}
By the base case \(m=1\) and the induction hypothesis,
\[
  \bT_{0,{x_1,y_1}}=\mathds{1}\{\,y_1\le_b x_1\,\},
  \qquad
  \mathbf{T}^{(m)}_{x',y'}=\mathds{1}\{\,y'\le_b x'\,\},
\]
so \eqref{eq:Tm+1-entry} yields
\[
  \mathbf{T}^{(m+1)}_{(x_1,x'),(y_1,y')}
  = \mathds{1}\{\,y_1\le_b x_1\,\}\,\mathds{1}\{\,y'\le_b x'\,\}
  = \mathds{1}\bigl\{\,y_1\le_b x_1 \text{ and } y'\le_b x'\,\bigr\}.
\]
Since the bit-wise order on \(\mathcal{X}_{m+1}\) shows the equivalence between $
  y \le_b x$ and $   y_1\le_b x_1 \text{ and } y'\le_b x',$
\eqref{eq:T-encodes-order-app} holds for \(m+1\).
By induction, it holds for all \(m\ge 1\), which proves Lemma~\ref{lem:polar-incidence}.
\end{proof}
%%%%%%%%%%%%%%%%%%%%%%%%%%%%%%%%%%%%%%%%%%%%%%%%%%%%%%%%%%%%%%%%%%%%%%%%%%%%%%%%%%%%%%%%%%%%%%%%%%%%%%%%%%%%%%%%%%%%%%%%%%%%%

\section{Proof of Lemma~\ref{lem:T-poset-auto-merged}}
\label{APP_proof_lem:T-poset-auto-merged}

\begin{proof}
Recall that \(\mathcal{X}=\{0,1\}^m\) with bit-wise order
\(x\le_b y \mbox{ iff } x_i\le y_i\) for all \(i\in[m]\), and that by
Lemma~\ref{lem:polar-incidence},
\(
  \mathbf{T}_{x,y} = \mathds{1}\{\,y\le_b x\,\}
\)
for all \(x,y\in\mathcal{X}\). Let \(\pi:\mathcal{X}\to\mathcal{X}\) be a bijection, and let
\(\mathbf{P}_\pi\) be the corresponding permutation matrix, whose
\(x\)-th column is \(e_{\pi(x)}\), i.e., \eqref{EQ_P_pi-compact}. 
Then \((\mathbf{P}_\pi^{\top})_{x,u} = (\mathbf{P}_\pi)_{u,x}\).

For any \(x,y\in\mathcal{X}\), we can derive
\begin{align}
  \bigl(\mathbf{P}_\pi^{\top}\mathbf{T}\,\mathbf{P}_\pi\bigr)_{x,y}
  &= \sum_{u,v}
     (\mathbf{P}_\pi^{\top})_{x,u}\,\mathbf{T}_{u,v}\,(\mathbf{P}_\pi)_{v,y}\notag\\
  &= \sum_{u,v}
     (\mathbf{P}_\pi)_{u,x}\,\mathbf{T}_{u,v}\,(\mathbf{P}_\pi)_{v,y}\notag\\
  &= \sum_{u,v}
     \mathds{1}\{u=\pi(x)\}\,\mathbf{T}_{u,v}\,\mathds{1}\{v=\pi(y)\}\label{EQ_Aut_expansion}\\
  &= \mathbf{T}_{\pi(x),\pi(y)}\label{EQ_Aut_expansion2}\\
  &= \mathds{1}\{\,\pi(y)\le_b \pi(x)\,\}.  \label{eq:PpiTPpi-entry}
\end{align}

Assume \(\mathbf{P}_\pi^{\top}\mathbf{T}\,\mathbf{P}_\pi=\mathbf{T}\), then for all \(x,y\in\mathcal{X}\), we have
\[
  \mathds{1}\{\,\pi(y)\le_b \pi(x)\,\}
  = \bigl(\mathbf{P}_\pi^{\top}\mathbf{T}\,\mathbf{P}_\pi\bigr)_{x,y}
  = \mathbf{T}_{x,y}
  = \mathds{1}\{\,y\le_b x\,\},
\]
where the first equality is from \eqref{eq:PpiTPpi-entry}. Hence, we have $ y\le_b x  \,\mbox{ iff }\,  \pi(y)\le_b \pi(x)  \,\forall\,x,y\in\mathcal{X}.$ 

Conversely, assume  $  x\le_b y \mbox{ iff } \pi(x)\le_b \pi(y),  \,\forall\,x,y\in\mathcal{X}.$
Then for all \(x,y\), we have $  \mathds{1}\{\,\pi(y)\le_b \pi(x)\,\}  = \mathds{1}\{\,y\le_b x\,\}  = \mathbf{T}_{x,y}.$
Comparing with \eqref{eq:PpiTPpi-entry}, we obtain
\((\mathbf{P}_\pi^{\top}\mathbf{T}\,\mathbf{P}_\pi)_{x,y} = \mathbf{T}_{x,y}\)
for all \(x,y\), i.e.,
\(
  \mathbf{P}_\pi^{\top}\mathbf{T}\,\mathbf{P}_\pi = \mathbf{T}.
\)

Finally, the map \(\pi\mapsto \mathbf{P}_\pi\) is one-to-one and onto between the two sets:
each bijection \(\pi\) corresponds to exactly one permutation matrix \(\mathbf{P}_\pi\), and vice versa.
Therefore, the equivalence above gives a one-to-one correspondence between \(\Aut(\mathbf{T})\)
and \(\Aut(\mathcal{X},\le_b)\).

\end{proof}

\section{Proof of Theorem~\ref{thm:AutT}}
\label{APP_proof_thm:AutT}

\begin{proof}
Let \(\mathcal{X}=\{0,1\}^m\) with the bit-wise order, and index rows and
columns of \(\mathbf{T}=\mathbf{F}^{\otimes m}\) by \(\mathcal{X}\). For each 
index permutation \(\sigma\in\mathcal{S}_m\), define the induced
bit-permutation \(\pi_\sigma:\mathcal{X}\to\mathcal{X}\) by
\[
  (\pi_\sigma(x))_j := x_{\sigma^{-1}(j)},
  \qquad x\in\mathcal{X},\ j\in[m],
\]
and let \(\mathbf{P}_\sigma\) be the permutation matrix associated with
\(\pi_\sigma\), as in~\eqref{EQ_P_pi-compact}. Define $  \mathcal{P}_{perm}
  := \bigl\{\,\mathbf{P}_\sigma : \sigma\in\mathcal{S}_m\,\bigr\},$ as the set of permutation matrices induced
by bit-permutations of the binary index vectors. To show
\(\Aut(\mathbf{T})=\mathcal{P}_{perm}\), we first prove \(\Aut(\mathbf{T})\subseteq\mathcal{P}_{perm}\). Let \(\mathbf{P}_\pi\in\Aut(\mathbf{T})\), i.e.,
\(
  \mathbf{P}_\pi^{\top}\mathbf{T}\,\mathbf{P}_\pi = \mathbf{T}.
\)
By Lemma~\ref{lem:T-poset-auto-merged}, this is equivalent to \(\pi\) being
a poset automorphism of \((\mathcal{X},\le)\), i.e., $ x\le_b y \mbox{ iff } \pi(x)\le_b \pi(y)  \;\forall\,x,y\in\mathcal{X}.$
By Lemma~\ref{lem:bool-aut}, such \(\pi\) are exactly the bit-permutations,
i.e., there exists a unique \(\sigma\in\mathcal{S}_m\) such that
\[
  (\pi(x))_j = x_{\sigma^{-1}(j)},
  \qquad\forall\,x\in\mathcal{X},\,j\in[m].
\]
The permutation matrix associated with \(\pi\) is precisely
\(\mathbf{P}_\sigma\), so \(\mathbf{P}_\pi=\mathbf{P}_\sigma\in\mathcal{P}_{perm}\).
Hence
\(
  \Aut(\mathbf{T})\subseteq\mathcal{P}_{perm}.
\)

Now we prove \(\Aut(\mathbf{T})\supseteq\mathcal{P}_{perm}\). Let \(\sigma\in\mathcal{S}_m\) and consider the induced
bit-permutation \(\pi_\sigma\). Being a coordinate permutation,
\(\pi_\sigma\) clearly preserves the bit-wise order:
\[
  y\le_b x \mbox{ iff } \pi_\sigma(y)\le_b \pi_\sigma(x),
  \qquad\forall\,x,y\in\mathcal{X}.
\]
By Lemma~\ref{lem:T-poset-auto-merged}, this is equivalent to
\(
  \mathbf{P}_{\pi_\sigma}^{\top}\mathbf{T}\,\mathbf{P}_{\pi_\sigma}
  = \mathbf{T},
\)
i.e., \(\mathbf{P}_\sigma\in\Aut(\mathbf{T})\). Thus
\(
  \mathcal{P}_{perm}\subseteq\Aut(\mathbf{T}).
\)

\medskip
Combining both inclusions yields \(\Aut(\mathbf{T})=\mathcal{P}_{perm}\), which completes the proof. 
% In addition, the map
% \[
%   \varphi:\mathcal{S}_m\to\Aut(\mathbf{T}),\qquad
%   \sigma\mapsto\mathbf{P}_\sigma,
% \]
% is bijective and satisfies
% \(
%   \varphi(\sigma_1\sigma_2)
%   = \mathbf{P}_{\sigma_1\sigma_2}
%   = \mathbf{P}_{\sigma_1}\mathbf{P}_{\sigma_2}
% \),
% so \(\Aut(\mathbf{T})\cong \mathcal{S}_m\) and therefore
% \(|\Aut(\mathbf{T})| = m!\).
\end{proof}

%%%%%%%%%%%%%%%%%%%%%%%%%%%%%%%%%%%%%%%%%%%%%%%%%%%%%%%%%%%%%%%%%%%%%%%%%%%%%%%%%%%%%%%%%%%%%%%%%%%%%%%%%%%%%%%%%%%%%%%%%%%%%
\section{Proof of Corollary~\ref{cor:one-sided-iso-clean}}
\label{APP_proof_cor:one-sided-iso-clean}

\begin{proof}
Let $\mathbf P\in\Aut(\mathbf T)$ and define $\mathbf T_{\mathbf P}:=\mathbf P\mathbf T$.
We aim to prove $\Aut(\mathbf T_{\mathbf P})
=\bigl\{\,\mathbf Q\in\Aut(\mathbf T):\ \mathbf Q\mathbf P=\mathbf P\mathbf Q\,\bigr\}.$

We first prove
\(
\Aut(\mathbf{T}_{\mathbf{P}})
\supseteq
\bigl\{\,\mathbf{Q}\in\Aut(\mathbf{T}):\ \mathbf{Q}\mathbf{P}
  = \mathbf{P}\mathbf{Q}\,\bigr\}.
\)
Assume \(\mathbf{Q}\mathbf{P}=\mathbf{P}\mathbf{Q}\) and \(\mathbf{Q}\in\Aut(\mathbf{T})\), i.e.,
\(\mathbf{Q}^{\top}\mathbf{T}\mathbf{Q}=\mathbf{T}\).
Then \(\mathbf{T}\mathbf{Q}=\mathbf{Q}\mathbf{T}\), and
\(\mathbf{Q}^{\top}\mathbf{P}=\mathbf{P}\mathbf{Q}^{\top}\) since
\(\mathbf{Q}^{\top}=\mathbf{Q}^{-1}\) and \(\mathbf Q\mathbf P=\mathbf P\mathbf Q\).
Therefore,
\[
  \mathbf{Q}^{\top}\mathbf{T}_{\mathbf{P}}\mathbf{Q}
  = \mathbf{Q}^{\top}(\mathbf{P}\mathbf{T})\mathbf{Q}
  = (\mathbf{Q}^{\top}\mathbf{P})(\mathbf{T}\mathbf{Q})
  = (\mathbf{P}\mathbf{Q}^{\top})(\mathbf{Q}\mathbf{T})
  = \mathbf{P}(\mathbf{Q}^{\top}\mathbf{Q})\mathbf{T}
  = \mathbf{P}\mathbf{T}
  = \mathbf{T}_{\mathbf{P}}.
\]
Thus \(\mathbf{Q}\in\Aut(\mathbf{T}_{\mathbf{P}})\), and hence
$\bigl\{\,\mathbf{Q}\in\Aut(\mathbf{T}):\ \mathbf{Q}\mathbf{P}  = \mathbf{P}\mathbf{Q}\,\bigr\}
\subseteq\Aut(\mathbf{T}_{\mathbf{P}}).$

We now prove $\Aut(\mathbf T_{\mathbf P})\subseteq
\bigl\{\,\mathbf Q\in\Aut(\mathbf T):\ \mathbf Q\mathbf P=\mathbf P\mathbf Q\,\bigr\},$ i.e., starting from \(\mathbf Q^{\top}\mathbf T_{\mathbf P}\mathbf Q=\mathbf T_{\mathbf P}\), our goal is to show that
\(\mathbf Q\) must (1) already be an automorphism of the original \(\mathbf T\), and
(2) commute with the permutation \(\mathbf P\).
To obtain these two properties from \(\mathbf Q^{\top}\mathbf T_{\mathbf P}\mathbf Q=\mathbf T_{\mathbf P}\),
for the ease of derivation, we first rewrite the matrix identity entrywise via \(\mathbf T_{x,y}=\mathds{1}\{y\le_b x\}\), which yields the two-permutation order equivalence \eqref{eq:star-short}.
To prove (1), we use \eqref{eq:star-short} to show that the permutation \(\tau\) induced by \(\mathbf Q\)
preserves \(\le_b\), hence \(\mathbf Q\in\Aut(\mathbf T)\), due to Lemma \ref{lem:T-poset-auto-merged}. To prove (2), we use the same equivalence to compare, for each \(j\in\mathcal X\), the set of elements above \(j\)
with the sets of elements above \(\tau(j)\) and the conjugate \(\rho(j)\), where \(\rho:=\sigma^{-1}\tau\sigma\).
Since an element is uniquely determined by the collection of elements lying above it, we must have
\(\rho(j)=\tau(j)\) for all \(j\in\mathcal X\). Indeed, if \(a\neq b\), then either \(a\not\le_b b\) or \(b\not\le_b a\), which forces their upper-bound sets to differ.
Therefore \(\rho=\tau\), i.e., \(\tau\sigma=\sigma\tau\), equivalently \(\mathbf Q\mathbf P=\mathbf P\mathbf Q\).

Let \(\mathbf Q\in\Aut(\mathbf T_{\mathbf P})\), i.e.,
$\mathbf Q^{\top}\mathbf T_{\mathbf P}\mathbf Q=\mathbf T_{\mathbf P}$.
Let \(\sigma\) and \(\tau\) be the permutations of \(\mathcal X\) induced by \(\mathbf P\) and \(\mathbf Q\),
respectively, i.e., \(\mathbf P=\mathbf P_\sigma\) and \(\mathbf Q=\mathbf P_\tau\).
Then, by the definition of permutation matrices,
\begin{align}
  (\mathbf T_{\mathbf P})_{x,y}
  =(\mathbf P\mathbf T)_{x,y}
  =\mathbf T_{\sigma^{-1}(x),y}
  =\mathds{1}\{\,y\le_b \sigma^{-1}(x)\,\},
  \quad x,y\in\mathcal X,  \label{eq:TP-twisted}
\end{align}
where the second equality follows from \eqref{EQ_Aut_expansion} and the third equality follows from
Lemma~\ref{lem:polar-incidence}.
Plugging \eqref{eq:TP-twisted} into $\mathbf Q^{\top}\mathbf T_{\mathbf P}\mathbf Q=\mathbf T_{\mathbf P}$ and reading entrywise yields, for all \(x,y\in\mathcal X\),
\begin{equation}\label{eq:entrywise-twisted}
  \mathds{1}\{\,\tau(y)\le_b \sigma^{-1}(\tau(x))\,\}
  =
  \mathds{1}\{\,y\le_b \sigma^{-1}(x)\,\},
\end{equation}
where the left-hand side is from \eqref{EQ_Aut_expansion2} together with \eqref{eq:TP-twisted}, and the right-hand
side is from \eqref{eq:TP-twisted}.

After substituting \(x:=\sigma(i)\) and \(\rho:=\sigma^{-1}\tau\sigma\) into \eqref{eq:entrywise-twisted},
we have the following two-permutation order relation: for all \(i,y\in\mathcal X\),
\begin{equation}\label{eq:star-short}
  \tau(y)\le_b \rho(i)
  \mbox{ iff }
  y\le_b i.
\end{equation}

We first prove that \(\tau\) preserves \(\le_b\) from \eqref{eq:star-short}. Let \(y_1\le_b y_2\) and choose \(i^\star:=\rho^{-1}(\tau(y_2))\), which exists since \(\rho\) is a permutation.
Then \(\tau(y_2)\le_b \rho(i^\star)\) holds trivially, so \eqref{eq:star-short} gives \(y_2\le_b i^\star\),
hence \(y_1\le_b i^\star\). Applying \eqref{eq:star-short} again yields
\(\tau(y_1)\le_b \rho(i^\star)=\tau(y_2)\).
Therefore \(\tau\) is order-preserving. Since \(\tau\) is bijective, \(\tau\) is a poset automorphism, and
Lemma~\ref{lem:T-poset-auto-merged} implies
\begin{equation}\label{eq:QinAutT}
\mathbf Q=\mathbf P_\tau\in\Aut(\mathbf T).
\end{equation}

We now prove \(\mathbf Q\mathbf P=\mathbf P\mathbf Q\), i.e., \(\tau\sigma=\sigma\tau\).
Fix \(j\in\mathcal X\). From \eqref{eq:star-short}, for each \(i\in\mathcal X\), we have
$j\le_b i\mbox{ iff }\tau(j)\le_b \rho(i).$ Hence, with fixed \(j\in\mathcal X\), the sets of upper bounds satisfy
$\{i\in\mathcal X:\ j\le_b i\}=\{i\in\mathcal X:\ \tau(j)\le_b \rho(i)\}.$
Applying the map \(\rho\) to both sides and using that \(\rho\) is an order-preserving bijection, we obtain
\begin{align}\label{EQ_poset_transformation}
    \{\,\rho(i):\ j\le_b i\,\}=\{\,\rho(i):\ \tau(j)\le_b \rho(i)\,\}.    
\end{align}
Let \(u:=\rho(i)\), then the \gls{rhs} of \eqref{EQ_poset_transformation} becomes
\(\{\,u\in\mathcal X:\ \tau(j)\le_b u\,\}\).
On the \gls{lhs}, order preservation implies
\(j\le_b i \mbox{ iff } \rho(j)\le_b \rho(i)\), hence
\(\{\,\rho(i):\ j\le_b i\,\}=\{\,u\in\mathcal X:\ \rho(j)\le_b u\,\}\).
Therefore, \eqref{EQ_poset_transformation} becomes
\begin{equation}\label{eq:upper2}
\{\,u\in\mathcal X:\ \rho(j)\le_b u\,\}
=
\{\,u\in\mathcal X:\ \tau(j)\le_b u\,\}.
\end{equation}

In any poset, an element is uniquely determined by its set of upper bounds \cite[Ch.~1]{DaveyPriestley2002}.
Therefore, \(\rho(j)=\tau(j)\) for all \(j\in\mathcal X\), i.e., \(\rho=\tau\).
Recalling \(\rho=\sigma^{-1}\tau\sigma\), we obtain \(\sigma^{-1}\tau\sigma=\tau\), i.e.,
\(\tau\sigma=\sigma\tau\). In matrix form this is exactly $\mathbf Q\mathbf P=\mathbf P\mathbf Q.$
Combining the two parts, we conclude that \(\mathbf Q\in\Aut(\mathbf T)\) and
\(\mathbf Q\mathbf P=\mathbf P\mathbf Q\). This proves the reverse inclusion. Combining both directions completes the proof.
\end{proof}

%%%%%%%%%%%%%%%%%%%%%%%%%%%%%%%%%%%%%%%%%%%%%%%%%%%%%%%%%%%%%%%%%%%%%%%%%%%%%%%%%%%%%%%%%%%%%%%%%%%%%%%%%%%%%%%%%%%%%%%%%%%%%
\section{Proof of Lemma \ref{lem:relabel-invariance}}\label{APP_proof_lem:relabel-invariance}
\begin{proof}
Fix $\pi_{\rm rel}\in\mathcal{S}_n$ and let $\bP_{\rm rel}$ be its permutation matrix.
In this proof, we express a permutation by its permutation matrix, so that
products and powers are taken in the matrix form and correspond to composition.

Define the relabeling map as follows:
\begin{align}\label{EQ_Phi_R}
    \Phi:\mathcal{S}_n\to\mathcal{S}_n,\qquad \Phi(\bU):= \bP_{\rm rel}\,\bU\,\bP_{\rm rel}^{-1}.    
\end{align}
% This is useful because the assumptions in Lemma~\ref{lem:cyclic-hideB} and
% Corollary~\ref{cor:one-sided-iso-clean} are stated in terms of products and powers
% of permutations (e.g., $\bA^k$ and commutation relations), and we want these relations
% to be preserved after relabeling.
Recall (cf.\ Definition~\ref{Def_group_isomorphism}) that a map $\varphi:\mathcal{G}_1\to\mathcal{G}_2$ between groups $\mathcal{G}_1$ and $\mathcal{G}_2$ is called a
group homomorphism if $\varphi(g_1g_2)=\varphi(g_1)\varphi(g_2)$ for all
$g_1,g_2\in\mathcal{G}_1$. Then for all $\bU_1,\bU_2\in\mathcal{S}_n$, we can easily check that $\Phi$ is a group homomorphism by the following:
\begin{equation}\label{eq:Phi-hom}
\Phi(\bU_1\bU_2)
=\bP_{\rm rel}\bU_1\bU_2\bP_{\rm rel}^{-1}
=(\bP_{\rm rel}\bU_1\bP_{\rm rel}^{-1})(\bP_{\rm rel}\bU_2\bP_{\rm rel}^{-1})
=\Phi(\bU_1)\Phi(\bU_2).
\end{equation}
Moreover, $\Phi$ is bijective with inverse
$\Phi^{-1}(\widetilde{\bU})=\bP_{\rm rel}^{-1}\,\widetilde{\bU}\,\bP_{\rm rel}.$ Thus, by Definition~\ref{Def_group_isomorphism}, $\Phi$ is a group isomorphism
from $\mathcal{S}_n$ onto $\mathcal{S}_n$. After applying \eqref{eq:Phi-hom} repeatedly yields, for every
$k\in\mathbb{Z}_{\ge 0}$,
\begin{equation}\label{eq:Phi-powers}
\Phi(\bU^k)=\Phi(\bU)^k.
\end{equation}

Define the relabeled polarization matrix $\widetilde{\bT}:= \bP_{\rm rel}\,\bT\,\bP_{\rm rel}^{-1}.$
Recall the following equivalence
\begin{equation}\label{eq:Aut-commute-equiv}
\bU\in\Aut(\bT)
\ \mbox{ iff }\
\bU^{\top}\bT\,\bU=\bT
\ \mbox{ iff }\
\bU\bT=\bT\bU.
\end{equation}
Hence if $\bU\in\Aut(\bT)$, conjugating by $\bP_{\rm rel}$ gives
\begin{align}
     (\bP_{\rm rel}\bU\bP_{\rm rel}^{-1})(\bP_{\rm rel}\bT\bP_{\rm rel}^{-1})
     =(\bP_{\rm rel}\bT\bP_{\rm rel}^{-1})(\bP_{\rm rel}\bU\bP_{\rm rel}^{-1}),\label{EQ_Phi_R_Aut}
\end{align}
i.e., $\Phi(\bU)\,\widetilde{\bT}=\widetilde{\bT}\,\Phi(\bU)$. Using \eqref{eq:Aut-commute-equiv} again (with $\widetilde{\bT}$ in place of $\bT$),
this implies that $\Phi(\bU)\in\Aut(\widetilde{\bT}).$ We have already shown that $\bU\in\Aut(\bT)$ implies $\Phi(\bU)\in\Aut(\widetilde{\bT})$,
hence $\Phi(\Aut(\bT))\subseteq \Aut(\widetilde{\bT})$. For the converse inclusion,
take an arbitrary $\widetilde{\bU}\in\Aut(\widetilde{\bT})$. By
\eqref{eq:Aut-commute-equiv}, we have
$\widetilde{\bU}\,\widetilde{\bT}=\widetilde{\bT}\,\widetilde{\bU},$ and 
substitute $\widetilde{\bT}=\bP_{\rm rel}\bT\bP_{\rm rel}^{-1}$ into
it and conjugating by $\bP_{\rm rel}^{-1}$ yields
$\bP_{\rm rel}^{-1}\widetilde{\bU}\,(\bP_{\rm rel}\bT\bP_{\rm rel}^{-1})\,\bP_{\rm rel}
=\bP_{\rm rel}^{-1}(\bP_{\rm rel}\bT\bP_{\rm rel}^{-1})\,\widetilde{\bU}\,\bP_{\rm rel}$, which implies $
(\bP_{\rm rel}^{-1}\widetilde{\bU}\bP_{\rm rel})\,\bT
=\bT\,(\bP_{\rm rel}^{-1}\widetilde{\bU}\bP_{\rm rel}).$ Define $\bU:=\bP_{\rm rel}^{-1}\widetilde{\bU}\bP_{\rm rel}$, we then 
have $\bU\bT=\bT\bU$, i.e., $\bU\in\Aut(\bT)$ by \eqref{eq:Aut-commute-equiv}. Moreover,
$\widetilde{\bU}=\bP_{\rm rel}\bU\bP_{\rm rel}^{-1}=\Phi(\bU)$, hence
$\widetilde{\bU}\in \Phi(\Aut(\bT))$. Therefore,
$\Aut(\widetilde{\bT})\subseteq \Phi(\Aut(\bT))$, and combining both inclusions yields
\begin{equation}\label{eq:Aut-relabel}
\Aut(\widetilde{\bT})=\Phi(\Aut(\bT))=\bP_{\rm rel}\,\Aut(\bT)\,\bP_{\rm rel}^{-1}.
\end{equation}

Recall that Lemma~\ref{lem:cyclic-hideB} uses $\bP_1=\bA^K$, $\bA\in\Aut(\bT)$ with
$K\sim\mathrm{Unif}(\{0,\dots,\textsf{N}-1\})$ and
$\mathcal{P}=\{\bA^k:0\le k<\textsf{N}\}$.
Define $\widetilde{\bA}:=\Phi(\bA)$ and $\widetilde{\bP}_1:=\Phi(\bP_1)$.
Then by \eqref{eq:Phi-powers}, we can easily see
\[
\widetilde{\bP}_1=\Phi(\bA^K)=\Phi(\bA)^K=\widetilde{\bA}^K,
\qquad
\widetilde{\mathcal{P}}
:=\{\widetilde{\bA}^k:0\le k<\textsf{N}\}
=\{\Phi(\bA^k):0\le k<\textsf{N}\}.
\]
Since $\Phi$ is injective, the map $k\mapsto \widetilde{\bA}^k$ is injective on
$\{0,\dots,\textsf{N}-1\}$, so $\widetilde{\bP}_1=\widetilde{\bA}^K$ is uniform on
$\widetilde{\mathcal{P}}$. Thus the uniformity assumption in
Lemma~\ref{lem:cyclic-hideB} is preserved.

Now let $(a_1\,\dots\,a_k)$ be any cycle of $\bA$. Since
$\widetilde{\bA}=\Phi(\bA)=\bP_{\rm rel}\bA\bP_{\rm rel}^{-1}$ corresponds to the
conjugate permutation $\pi_{\rm rel}\bA\pi_{\rm rel}^{-1}$, the standard
conjugation rule for cycle decompositions \cite[Proposition~10, p.~125]{DummitFoote2004}
implies that this cycle becomes
$(\pi_{\rm rel}(a_1)\,\dots\,\pi_{\rm rel}(a_k))$ under relabeling. Equivalently,
letting $b_j:=\pi_{\rm rel}(a_j)$, we have
$\widetilde{\bA}(b_j)=b_{j+1}$ for $j=1,\dots,k-1$ and $\widetilde{\bA}(b_k)=b_1$.
Hence relabeling only renames the elements inside each cycle and does not change
cycle lengths.

We now verify that the conditions related to set operations used in Lemma~\ref{lem:cyclic-hideB}
are invariant under relabeling. Define the relabeled sets $\widetilde{\mathcal{I}_{\mathcal{G}}}:=\pi_{\rm rel}(\mathcal{I}_{\mathcal{G}}),\;\widetilde{\mathcal{I}_{\mathcal{B}}}:=\pi_{\rm rel}(\mathcal{I}_{\mathcal{B}}).$ Since $\pi_{\rm rel}:[n]\to[n]$ is a bijection, it preserves the standard set
operations: for all subsets $\mathcal{U},\mathcal{V}\subseteq[n]$,
\begin{align}\label{eq:image-setops}
\pi_{\rm rel}(\mathcal{U}\cap \mathcal{V})&=\pi_{\rm rel}(\mathcal{U})\cap \pi_{\rm rel}(\mathcal{V}),\\
\pi_{\rm rel}(\mathcal{U}\cup \mathcal{V})&=\pi_{\rm rel}(\mathcal{U})\cup \pi_{\rm rel}(\mathcal{V}),\\
\pi_{\rm rel}(\mathcal{U}\setminus \mathcal{V})&=\pi_{\rm rel}(\mathcal{U})\setminus \pi_{\rm rel}(\mathcal{V}),
\end{align}
and also $\bigl|\pi_{\rm rel}(\mathcal{U})\bigr|=|\mathcal{U}|$. Let $(i_1\,i_2\,\dots\,i_\ell)$ be one cycle in the disjoint cycle decomposition
of $\bA$, and define its associated cycle index set
$ \mathcal{C}\;:=\;\{i_1,i_2,\dots,i_\ell\}\subseteq[n].$
Under relabeling, this cycle becomes
$(\pi_{\rm rel}(i_1),\,\pi_{\rm rel}(i_2),\,\dots,\,\pi_{\rm rel}(i_\ell))$ and its
associated index set becomes $\widetilde{\mathcal{C}}:=\pi_{\rm rel}(\mathcal{C})$.
Applying \eqref{eq:image-setops} with the substitutions $(\mathcal{U},\mathcal{V})=(\mathcal{C},\mathcal{I}_{\mathcal{G}})$
and $(\mathcal{U},\mathcal{V})=(\mathcal{C},\mathcal{I}_{\mathcal{B}})$ yields respectively
$\pi_{\rm rel}(\mathcal{C}\cap\mathcal{I}_{\mathcal{G}})=\widetilde{\mathcal{C}}\cap\widetilde{\mathcal{I}_{\mathcal{G}}},
\; \pi_{\rm rel}(\mathcal{C}\cap\mathcal{I}_{\mathcal{B}})=\widetilde{\mathcal{C}}\cap\widetilde{\mathcal{I}_{\mathcal{B}}}.$
Consequently, any condition in Lemma~\ref{lem:cyclic-hideB} that is formulated
purely in terms of the cycle index sets $\mathcal{C}$ (of $\bA$) and the
partition $(\mathcal{I}_{\mathcal{G}},\mathcal{I}_{\mathcal{B}})$ via set operations is preserved
under relabeling.

\medskip
Finally, we aim to show the invariance of the commutation-based operation under relabeling in
Corollary~\ref{cor:one-sided-iso-clean}, i.e., 
\begin{equation}\label{eq:AutTP-relabel-goal}
\bP_{\rm rel}\,\Aut(\bT_{\bP})\,\bP_{\rm rel}^{-1}
=\Aut(\widetilde{\bT}_{\widetilde{\bP}}).
\end{equation}

Fix $\bP\in\Aut(\bT)$ and recall
$\bT_{\bP}:=\bP\,\bT$ and $\Aut(\bT_{\bP})=\{\bQ\in\Aut(\bT): \bQ\bP=\bP\bQ\}$.
Define $\widetilde{\bT}:=\bP_{\rm rel}\,\bT\,\bP_{\rm rel}^{-1},\;
\widetilde{\bP}:=\bP_{\rm rel}\,\bP\,\bP_{\rm rel}^{-1},\;
\widetilde{\bQ}:=\bP_{\rm rel}\,\bQ\,\bP_{\rm rel}^{-1}$. To prove \eqref{eq:AutTP-relabel-goal}, again we prove the two inclusions. First, we take an arbitrary $\bQ\in\Aut(\bT_{\bP})$. Similar to \eqref{eq:Aut-commute-equiv}, conjugating $\bQ\bP=\bP\bQ$ by $\bP_{\rm rel}$ gives $\widetilde{\bQ}\,\widetilde{\bP}=\widetilde{\bP}\,\widetilde{\bQ}.$
Moreover, from $\bQ\in\Aut(\bT)$ and \eqref{eq:Aut-commute-equiv} we have $\bQ\bT=\bT\bQ$,
and conjugating by $\bP_{\rm rel}$ yields $\widetilde{\bQ}\,\widetilde{\bT}=\widetilde{\bT}\,\widetilde{\bQ},$
i.e., $\widetilde{\bQ}\in\Aut(\widetilde{\bT})$ by \eqref{eq:Aut-commute-equiv} again. Hence
$\widetilde{\bQ}\in\Aut(\widetilde{\bT})$ and $\widetilde{\bQ}\widetilde{\bP}=\widetilde{\bP}\widetilde{\bQ}$, so by Corollary~\ref{cor:one-sided-iso-clean}, we have
\[
\widetilde{\bQ}\in\Aut(\widetilde{\bT}_{\widetilde{\bP}}),
\qquad\text{where}\qquad
\widetilde{\bT}_{\widetilde{\bP}}:=\widetilde{\bP}\,\widetilde{\bT}.
\]
Therefore, we have
\[
\bP_{\rm rel}\,\Aut(\bT_{\bP})\,\bP_{\rm rel}^{-1}
\subseteq
\Aut(\widetilde{\bT}_{\widetilde{\bP}}).
\]

Conversely, take an arbitrary $\widetilde{\bQ}\in\Aut(\widetilde{\bT}_{\widetilde{\bP}})$ and define
$\bQ := \bP_{\rm rel}^{-1}\,\widetilde{\bQ}\,\bP_{\rm rel}.$
Applying Corollary~\ref{cor:one-sided-iso-clean} to $(\widetilde{\bT},\widetilde{\bP})$ yields
$\widetilde{\bQ}\in\Aut(\widetilde{\bT})$ and $\widetilde{\bQ}\widetilde{\bP}=\widetilde{\bP}\widetilde{\bQ}$.
By \eqref{eq:Aut-relabel} we have $\Aut(\widetilde{\bT})=\bP_{\rm rel}\,\Aut(\bT)\,\bP_{\rm rel}^{-1}.$
Thus $\widetilde{\bQ}\in\Aut(\widetilde{\bT})$ implies that there exists some
$\bQ\in\Aut(\bT)$ such that $\widetilde{\bQ}=\bP_{\rm rel}\,\bQ\,\bP_{\rm rel}^{-1},
\;\text{equivalently}\; \bQ=\bP_{\rm rel}^{-1}\,\widetilde{\bQ}\,\bP_{\rm rel}.$
Moreover, conjugating the commutation relation
$\widetilde{\bQ}\widetilde{\bP}=\widetilde{\bP}\widetilde{\bQ}$ by $\bP_{\rm rel}^{-1}$ yields
\[
(\bP_{\rm rel}^{-1}\widetilde{\bQ}\bP_{\rm rel})(\bP_{\rm rel}^{-1}\widetilde{\bP}\bP_{\rm rel})
=
(\bP_{\rm rel}^{-1}\widetilde{\bP}\bP_{\rm rel})(\bP_{\rm rel}^{-1}\widetilde{\bQ}\bP_{\rm rel}),
\]
i.e., $\bQ\bP=\bP\bQ$. By definition of $\Aut(\bT_{\bP})$, from the above derivation, we conclude that $\bQ\in\Aut(\bT_{\bP})$. Therefore,
\[
\widetilde{\bQ}=\bP_{\rm rel}\,\bQ\,\bP_{\rm rel}^{-1}
\in
\bP_{\rm rel}\,\Aut(\bT_{\bP})\,\bP_{\rm rel}^{-1}.
\]
Thus, we have the other inclusion
\[
\Aut(\widetilde{\bT}_{\widetilde{\bP}})
\subseteq
\bP_{\rm rel}\,\Aut(\bT_{\bP})\,\bP_{\rm rel}^{-1}.
\]
Combining both inclusions yields
\[
\bP_{\rm rel}\,\Aut(\bT_{\bP})\,\bP_{\rm rel}^{-1}
=
\Aut(\widetilde{\bT}_{\widetilde{\bP}}),
\qquad\text{where}\quad
\widetilde{\bT}_{\widetilde{\bP}}:=\widetilde{\bP}\,\widetilde{\bT}.
\]
In particular, the commutation condition required in Corollary~\ref{cor:one-sided-iso-clean} is invariant under relabeling by $\bP_{\rm rel}$.

Combining the invariance of (i) the randomization \(K\sim\Unif(\{0,\dots,\textsf{N}-1\})\), \(\bP_1=\bA^{K}\) (hence \(\bP_1\) is uniform on \(\mathcal{P}\)) and the induced \(\bF=\bP_1^{\mathsf T}\bT_{B}\) in Lemma~\ref{lem:cyclic-hideB},
(ii) the cycle/partition conditions expressed via \eqref{eq:image-setops}, and
(iii) the commutation characterization in Corollary~\ref{cor:one-sided-iso-clean}, we
complete the proof of Lemma~\ref{lem:relabel-invariance}.
\end{proof}

\section{Proof of Lemma\ref{Lemma_Hmin_H_gap}}\label{APP_proof_LHL_cond_entropy}
\begin{proof}
Fix $r>1$ and define $
  \mathcal A_{r}:=
     \bigl\{\, {z}^{n}\in\mathbb R^{n}\;:\;
             | {z}_i|\le r \;\text{for all } i\bigr\}.$
For each $i$, since the received signal at Eve is \( {Z}_i = X_i+N_i,\,\mathds{E}[ {Z}_i]=1\), we have
\(
  \Pr(| {z}_i|>r) = Q\!\bigl((r-1)/\sigma\bigr).
\)
Define the $
  \varepsilon_{r}
  := \Pr\bigl[\mathcal A_{r}^{\mathrm c}\bigr]
  \;\le\;
  2n\,Q\!\bigl((r-1)/\sigma\bigr)$ as the tail probability and we select $r$ such that $\varepsilon_{r}=\varepsilon$. To simplify the notation, we let $\tilde{z}:= {z}^n$ and fix \( \tilde{z}\in\mathcal E\). Then inside $\mathcal A_{r}$, \(v_{ \tilde{z}}=|\supp X^{n}\big|_{ \tilde{z}}|\le 2^{n}\) and \(t_{ \tilde{z}}\le 2^{-n}(2\pi\sigma^{2})^{-n/2}\). Let \(m:=v_{ \tilde{z}}\), \(t:=t_{ \tilde{z}}\). Then conditions $v_{z}\!:=\!\bigl|\supp_x p_{X\mid Z=z}\bigr|\!<\!\infty,   \,   t_{z}:=\max_{x}p_{X\mid Z}(x\mid z)\!<\!\infty$ hold. Since Shannon's entropy is Schur-concave, and
$   q=(t,\tfrac{1-t}{m-1},\ldots,\tfrac{1-t}{m-1})$ majorizes all $p$, the posterior \gls{pmf}, then $q$ maximizes the entropy. Hence
\begin{align}\label{EQ_schur_convex}
    H(p)\le H(q)=H_{\mathrm b}(t)+(1-t)\log_2(m-1).    
\end{align}

With \(H_{\min}(p)=-\log_{2}t\) and recall $\psi_m(t) \;:=\; H_b(t) + (1-t)\log_2(m-1) + \log_2 t$, we can get the following:
\begin{align}
  -H_{\min}(p)&=\log_2\frac1t                                             \label{eq:1a}\\
   &=\psi_{m}(t)-H_{\mathrm b}(t)-(1-t)\log_{2}(m-1)                        \notag\\
  &\le -H(p)+\psi_{m}(t). \label{EQ_Hmin_UB_p}
\end{align}
  where  
\eqref{EQ_Hmin_UB_p} uses $H(p)\le H_b(t)+(1-t)\log_2(m-1)$ from \eqref{EQ_schur_convex}.  
From the definition
\begin{align}
  H(X\!\mid  \tilde{Z},\mathcal E)
            &:=\mathds E[H(p_{\tilde{z}})\mid  \tilde{Z}\in\mathcal E],          \\
  H_{\min}(X\!\mid  \tilde{Z},\mathcal E)
            &:=\mathds E[H_{\min}(p_{\tilde{Z}})\mid  \tilde{Z}\in\mathcal E],
\end{align}
after taking expectations of \eqref{EQ_Hmin_UB_p}, we have:
    \begin{align}
    -H_{\min}(X\mid  \tilde{Z},\mathcal E)&=        -\mathds E[H_{\min}(p_{\tilde{Z}})\mid\mathcal E]\\
         &\leq -\mathds E[H(p_{\tilde{Z}})\mid\mathcal E]
              +\mathbb{E}\bigl[\psi_{M}(T)\bigr]\label{eq:2a}\\
     &= -H(X\mid  \tilde{Z},\mathcal E)+\mathbb{E}\bigl[\psi_{M}(T)\bigr].          \label{EQ_Hmin_H}
   \end{align}

Now we expand $
P_{X \tilde{Z}}= (1-\varepsilon)P_{X \tilde{Z}}^{(\mathcal E)}
        \;+\;
        \varepsilon\,P_{X \tilde{Z}}^{(\mathcal E^{\mathrm c})},$
where 
\(
   P_{X \tilde{Z}}^{(\mathcal E)}(\,\cdot\,):=
      \dfrac{P_{X \tilde{Z}}(x, \tilde{z})\,\mathds{1}_{\{ \tilde{z}\in\mathcal E\}}}{1-\varepsilon},
\,
   P_{X \tilde{Z}}^{(\mathcal E^{\mathrm c})}(\,\cdot\,)
      :=\dfrac{P_{X \tilde{Z}}(x, \tilde{z})\,\mathds{1}_{\{ \tilde{z}\notin\mathcal E\}}}{\varepsilon}.
\)

For any measurable $(x, \tilde{z})\in\mathcal X\times\mathcal E$,
\begin{align}
P_{X\mid  \tilde{Z}}(x\mid  \tilde{z})
   &=\frac{P_{X \tilde{Z}}(x, \tilde{z})}{P_{\tilde{Z}}( \tilde{z})}
     \\
   &=\frac{(1-\varepsilon)\,P_{X \tilde{Z}}^{(\mathcal E)}(x, \tilde{z})
            +\varepsilon\,P_{X \tilde{Z}}^{(\mathcal E^{\mathrm c})}(x, \tilde{z})}
           {(1-\varepsilon)\,P_{ \tilde{Z}}^{(\mathcal E)}( \tilde{z})
            +\varepsilon\,P_{ \tilde{Z}}^{(\mathcal E^{\mathrm c})}( \tilde{z})}  \\
   &=\frac{(1-\varepsilon)\,P_{X \tilde{Z}}^{(\mathcal E)}(x, \tilde{z})}
           {(1-\varepsilon)\,P_{ \tilde{Z}}^{(\mathcal E)}( \tilde{z})}\\
   &=\frac{1}{1-\varepsilon}\,P_{X\mid  \tilde{Z},\mathcal E}(x\mid  \tilde{z}),\label{EQ_cond_P_smooth_cond_P}
\end{align}
where the third equality is because $P_{ \tilde{Z}}^{(\mathcal E^{\mathrm c})}( \tilde{z})=0\text{ for } \tilde{z}\in\mathcal E\ $. Therefore, after substituting \eqref{EQ_cond_P_smooth_cond_P} into min entropy by taking \(\max_x\) and \(-\log_2\), we have
\begin{align}
   H_{\min}(X\mid  \tilde{Z})
   \ge H_{\min}(X\mid  \tilde{Z},\mathcal E)-\log_2\!\frac1{1-\varepsilon}. \label{EQ_Hmin_Hmin}
\end{align}

Now we want to show that
\begin{align}
  H(X\mid  \tilde{Z},\mathcal E)
  \ge H(X\mid  \tilde{Z})-\frac{\varepsilon}{1-\varepsilon}\,H_{\max}(X).  
\end{align}
  
By convexity of entropy with the expansion \(P_{X \tilde{Z}}=(1-\varepsilon)P^{(\mathcal E)}_{X \tilde{Z}}
          +\varepsilon P^{(\mathcal E^{\mathrm c})}_{X \tilde{Z}}\), we have 
\begin{align}
H(X\mid  \tilde{Z})
&=\sum_{x, \tilde{z}}P_{X \tilde{Z}}(x, \tilde{z})
        \log_2\frac{1}{P_{X\mid  \tilde{Z}}(x\mid  \tilde{z})}                     \notag\\
   &=(1-\varepsilon)\!
        \sum_{x, \tilde{z}}P_{X \tilde{Z}}^{(\mathcal E)}(x, \tilde{z})
        \log_2\frac{1}{P_{X\mid  \tilde{Z},\mathcal E}(x\mid  \tilde{z})} +\varepsilon\!
        \sum_{x, \tilde{z}}P_{X \tilde{Z}}^{(\mathcal E^{\mathrm c})}(x, \tilde{z})
        \log_2\frac{1}{P_{X\mid  \tilde{Z},\mathcal E^{\mathrm c}}(x\mid  \tilde{z})}\notag\\
   &= (1-\varepsilon)H(X\mid  \tilde{Z},\mathcal E)
   +\varepsilon H(X\mid  \tilde{Z},\mathcal E^{\mathrm c})\\
&\le (1-\varepsilon)H(X\mid  \tilde{Z},\mathcal E)
    +\varepsilon H_{\max}(X).
\end{align}

After rearrangement, we have
\begin{align}
H(X\mid  \tilde{Z},\mathcal E)
  &\ge \frac{H(X\mid  \tilde{Z})}{1-\varepsilon}-\frac{\varepsilon}{1-\varepsilon}\,H_{\max}(X)\\
  &\ge H(X\mid  \tilde{Z})-\frac{\varepsilon}{1-\varepsilon}\,H_{\max}(X).\label{EQ_cond_H_cond_H}
\end{align}

Recall the definition of smooth min entropy in \eqref{Def_smooth_min_entropy2}. Then we can choose
\(  P_{X \tilde{Z}}:=P_{X \tilde{Z}}\bigl[\cdot\cap\{ \tilde{Z}\in\mathcal E\}\bigr]\)
in \eqref{Def_smooth_min_entropy2} results in 
\begin{align}
   H_{\min}^{\varepsilon}(X\mid  \tilde{Z})
   \ge H_{\min}(X\mid  \tilde{Z};  P_{X \tilde{Z}})
   = H_{\min}(X\mid  \tilde{Z},\mathcal E).                          \label{EQ_smoth_H_min_H}
\end{align}

Combining \eqref{EQ_smoth_H_min_H}, \eqref{EQ_Hmin_Hmin},  \eqref{EQ_Hmin_H},  and \eqref{EQ_cond_H_cond_H}, we complete the proof.
\end{proof}

\section{Proof of Lemma\ref{lem:hidden-set-ell-vs-leakage}}\label{APP_proof_hidden-set-ell-vs-leakage}
\begin{proof}
Recall $\tilde{\mathcal{J}}_{\bar{B}}=\mathcal J_1$. Let $\mathcal J_1=\{i_1<\cdots<i_k\}$ with $k:=|\mathcal J_1|$.
Recall $\Pi_{\mathrm{sel}}:=(\mathbf F,\mathcal J_0,\mathcal J_1)$ and $S_b$ is the public hash seed used to form $K_b$.
By the chain rule, we have
\begin{align}
H(U_{\mathcal J_1}\mid Y^n,\Pi_{\mathrm{sel}},S_B,B=b)
&=\sum_{t=1}^k
H\!\bigl(U_{i_t}\mid Y^n,\Pi_{\mathrm{sel}},S_B,B=b,U_{i_1},\dots,U_{i_{t-1}}\bigr)\notag\\
&\ge
\sum_{t=1}^k
H\!\bigl(U_{i_t}\mid Y^n,\Pi_{\mathrm{sel}},S_B,B=b,U^{i_t-1}\bigr),
\label{eq:cond-reduce-step-hidden}
\end{align}
where the inequality holds because $(U_{i_1},\dots,U_{i_{t-1}})\subseteq U^{i_t-1}$ and conditioning reduces entropy.

Fix $t\in\{1,\dots,k\}$ and use $i:=i_t$. Then
\begin{align}
H\!\bigl(U_i\mid Y^n,\Pi_{\mathrm{sel}},S_B,B=b,U^{i-1}\bigr)
&=
H\!\bigl(U_i\mid \Pi_{\mathrm{sel}},S_B,B=b,U^{i-1}\bigr)
-
I\!\bigl(U_i;Y^n\mid \Pi_{\mathrm{sel}},S_B,B=b,U^{i-1}\bigr).
\label{eq:split-step}
\end{align}

Consider the first term on the \gls{rhs} of \eqref{eq:split-step}. For any $i\in \mathcal J_0\cup \mathcal J_1$, conditioned on $\Pi_{\mathrm{sel}}$ (hence on the randomized index sets),
Alice generates $U_i\sim \mathrm{Bern}(\tfrac12)$ independently of $(B,S_B)$ and independently of $U^{i-1}$.
Therefore, for all $(\pi,b,s,u^{i-1})$,
\[
\Pr\!\Bigl(U_i=1 \,\Big|\, \Pi_{\mathrm{sel}}=\pi,\; B=b,\; S_B=s,\; U^{i-1}=u^{i-1}\Bigr)=\frac12,
\]
which implies
\begin{equation}\label{eq:first-term-1}
H\!\bigl(U_i \mid \Pi_{\mathrm{sel}},S_B,B=b,U^{i-1}\bigr)=1 .
\end{equation}

Now consider the second term on the \gls{rhs} of \eqref{eq:split-step}.
Since $U_i$ is conditionally independent of $U^{i-1}$ given $(\Pi_{\mathrm{sel}},S_B,B=b)$, we have
$I(U_i;U^{i-1}\mid \Pi_{\mathrm{sel}},S_B,B=b)=0$, and hence
\begin{equation}\label{eq:add-U-prev}
I\!\bigl(U_i;Y^n\mid \Pi_{\mathrm{sel}},S_B,B=b,U^{i-1}\bigr)
=
I\!\bigl(U_i;Y^n,U^{i-1}\mid \Pi_{\mathrm{sel}},S_B,B=b\bigr).
\end{equation}

In the proposed protocol, recall $(\mathcal J_0,\mathcal J_1)=(\mathcal I_B,\mathcal I_{1-B}),$
so conditioned on $(\mathbf F,B=b)$ the pair $(\mathcal J_0,\mathcal J_1)$ is deterministic.
Therefore, for all $i\in[n]$, we have
\begin{align}\label{eq:drop-J-in-MI}
I\!\bigl(U_i;Y^n,U^{i-1}\mid \Pi_{\mathrm{sel}},S_B,B=b\bigr)
=
I\!\bigl(U_i;Y^n,U^{i-1}\mid \mathbf F,S_B,B=b\bigr).
\end{align}

By Lemma~\ref{lem:markov-B-F-UY}, we have
$(U^n,Y^n)\pperp B\mid \mathbf F$, and since $S_B$ is public seed chosen independently of all other random variables,
we also have $S_B\pperp (U^n,Y^n,B,\mathbf F)$. Then, by Bayes rule, we can derive
\begin{align}\label{EQ_UY_perp_BS_give_F}
(U^n,Y^n)\ \pperp\ (B,S_B)\ \big|\ \mathbf F.
\end{align}
From \eqref{EQ_UY_perp_BS_give_F}, we have, for every \(f\) and every \((b,s)\) with positive probability,
\begin{equation}\label{eq:condlaw-1}
P_{U^n,Y^n\mid \mathbf F=f,\;B=b,\;S_B=s}
=
P_{U^n,Y^n\mid \mathbf F=f}.
\end{equation}
Taking marginals of \eqref{eq:condlaw-1} with respect to $U_{i+1}^n$ gives
\begin{equation}\label{eq:condlaw-2}
P_{U_i,\;Y^n,\;U^{i-1}\mid \mathbf F=f,\;B=b,\;S_B=s}
=
P_{U_i,\;Y^n,\;U^{i-1}\mid \mathbf F=f}.
\end{equation}

We now derive \eqref{eq:drop-J-in-MI}. Fix \(b\in\{0,1\}\) and define
\(g(f):=I\!\bigl(U_i;\,Y^n,U^{i-1}\mid \mathbf F=f\bigr)\). Then
\begin{align}
I\!\bigl(U_i;\,Y^n,U^{i-1}\mid \mathbf F,S_B,B=b\bigr)
&=
\mathbb{E}_{\mathbf F,S_B\,\mid\,B=b}\!\Big[
I\!\bigl(U_i;\,Y^n,U^{i-1}\mid \mathbf F=f,S_B=s,B=b\bigr)
\Big]\notag\\
&\overset{(a)}=
\mathbb{E}_{\mathbf F,S_B\,\mid\,B=b}\!\big[g(\mathbf F)\big]
\overset{(b)}=
\mathbb{E}_{\mathbf F\,\mid\,B=b}\!\big[g(\mathbf F)\big]\notag\\
&\overset{(c)}=
\mathbb{E}_{\mathbf F}\!\big[g(\mathbf F)\big]
=
I\!\bigl(U_i;\,Y^n,U^{i-1}\mid \mathbf F\bigr)
\;=:\;
I_i^{(n)}(\mathbf F),
\label{eq:mi-drop-SB}
\end{align}
where (a) follows from \eqref{eq:condlaw-2}, (b) is due the independence of $S_B$ on all other random variables, (c) uses \(\mathbf F\pperp B\) from Lemma \ref{lem:cyclic-hideB}.

Combining \eqref{eq:split-step}, \eqref{eq:first-term-1}, \eqref{eq:add-U-prev}, \eqref{eq:drop-J-in-MI}, and \eqref{eq:mi-drop-SB}, we obtain
\begin{align}\label{EQ_H_1_I}
H\!\bigl(U_i\mid Y^n,\Pi_{\mathrm{sel}},S_B,B=b,U^{i-1}\bigr)=1-I_i^{(n)}(\mathbf F).
\end{align}
Substituting \eqref{EQ_H_1_I} into \eqref{eq:cond-reduce-step-hidden} and summing over
$t=1,\dots,k$, we have
\begin{align}\label{EQ_H_U_given_Y_Pi_B}
H(U_{\mathcal J_1}\mid Y^n,\Pi_{\mathrm{sel}},S_B,B=b)\;\ge\;
\sum_{i\in\mathcal J_1}\Bigl(1-I_i^{(n)}(\mathbf F)\Bigr)
=
|\mathcal J_1|-\sum_{i\in\mathcal J_1} I_i^{(n)}(\mathbf F).
\end{align}

After averaging \eqref{EQ_H_U_given_Y_Pi_B} over $B$ and using $\Pr(B=0)=\Pr(B=1)=\tfrac12$, we obtain
\begin{align}
H(U_{\mathcal J_1}\mid Y^n,\Pi_{\mathrm{sel}},S_B,B)
&\ge |\mathcal J_1|-\sum_{i\in\mathcal J_1} I_i^{(n)}(\mathbf F).\label{eq:H-hidden-lb-step1}
\end{align}
Finally, combining \eqref{eq:H-hidden-lb-step1} with the design rule \eqref{EQ_ell}, we complete the proof.
\end{proof}

\section{Proof of Lemma\ref{lem:sw-hhat-discreteX-contY}}\label{APP_proof_sw-hhat-discreteX-contY}
\begin{proof}
To proceed, we first introduce the finite blocklength result of \gls{swc} where the side information at Bob is continuous, which is extended from the main result in \cite{Hayashi_SWC_Entropy20} (the side information at Bob is discrete).

\begin{lemma}\label{lem:sw_continuous_SI_second_order}
Let $X$ take values in a finite set $\mathcal{X}$.  Assume that $(X_i,Y_i)_{i=1}^n$ are i.i.d.\ generated from $P_{XY}$ and that
$P_{X|Y}(x|y)$ exists for $\mu$-a.e.\ $y$, so that the extension in \cite[Remark~1]{Hayashi_SWC_Entropy20} to continuous $\mathcal{Y}$ applies. Define
\[
  \imath_{X|Y}(X;Y) := -\log_2 P_{X|Y}(X|Y),\qquad
  H(X|Y) := \mathbb{E}[\imath_{X|Y}(X;Y)],\qquad
  \textsf{V}(X|Y) := \mathrm{Var}\!\bigl(\imath_{X|Y}(X;Y)\bigr),
\]
and assume $\mathbb{E}[\imath_{X|Y}(X;Y)^2]<\infty$ and $\textsf{V}(X|Y)>0$.
Let $\textsf{M}(n,\varepsilon)$ and $\overline{\textsf{M}}(n,\varepsilon)$ be defined as in \cite[Sec.~3.1]{Hayashi_SWC_Entropy20}.
Then, for every $0<\varepsilon<1$,
\begin{align}
  \log_2 \textsf{M}(n,\varepsilon)
  \;=\;
  \log_2 \overline{\textsf{M}}(n,\varepsilon) + o(\sqrt{n})
  \;=\;
  n H(X|Y) + \sqrt{n\,\textsf{V}(X|Y)}\,\Phi^{-1}(1-\varepsilon) + o(\sqrt{n}),
  \label{eq:sw_second_order_continuousY}
\end{align}
where $\Phi$ is the standard normal CDF.
\end{lemma}

The proof is relegated to Section \ref{APP_proof_sw_continuous_SI_second_order}. Recall that $\textsf{M}(n,\varepsilon)$ is the optimal (smallest) message alphabet size $\textsf{M}_n$ such that there exists a length-$n$ encoder/decoder pair with error probability at most $\varepsilon$. 
In contrast, $\overline{\textsf{M}}(n,\varepsilon)$ is the optimal message size when we restrict to a specific achievability scheme, namely via a two-universal hash as defined in \cite[Sec.~3.1]{Hayashi_SWC_Entropy20}. 
Hence typically $\overline{\textsf{M}}(n,\varepsilon)\ge \textsf{M}(n,\varepsilon)$, and \eqref{eq:sw_second_order_continuousY} states that their logarithms coincide up to an $o(\sqrt{n})$ term. Now fix $\mathcal{A}_{\rm SI}\subseteq\mathcal J_0$ and define
\[
\mathcal{A}_{\rm SI}^c \;:=\; \mathcal J_0\setminus \mathcal{A}_{\rm SI}.
\]
Let
$  Z := (Y^n,U_{\mathcal{A}_{\rm SI}}).$ Set the source to be recovered as
$  \tilde{X} := U_{\mathcal{A}_{\rm SI}^c}\in\{0,1\}^{|\mathcal{A}_{\rm SI}^c|},$ so $\tilde{X}$ takes values in a finite alphabet. Note that $Y^n$ may be continuous while $\tilde{X}$ is discrete, $P_{\tilde{X}|Z}(x|z)$ is well-defined and the Slepian--Wolf bounds used in \cite[Sec.~3.1--Sec.~3.2]{Hayashi_SWC_Entropy20}
remain valid for the pair $(\tilde{X},Z)$, due to \cite[Remark~1]{Hayashi_SWC_Entropy20}.
Applying Lemma~\ref{lem:sw_continuous_SI_second_order} with side information $Z$ and target error
$\varepsilon_{\rm sw}$, there exists a Slepian--Wolf encoder that outputs a public message
$M_p\in\{1,\dots,\textsf{M}\}$ such that Bob can reconstruct $\tilde{X}=U_{\mathcal{A}_{\rm SI}^c}$ from $(Z,M_p)$ with
error probability at most $\varepsilon_{\rm sw}$, provided that
\begin{equation}\label{eq:ell_ge_Hplusbeta}
  \ell_{\rm SWC} \;\ge\; \log_2 \textsf{M}(n,\varepsilon_{\rm sw})
  \;=\; H(\tilde{X}|Z) \;+\; \beta_n(\varepsilon_{\rm sw}),
\end{equation}
where $  \beta_n(\varepsilon_{\rm sw})  :=  \sqrt{n\,\textsf{V}(\tilde{X}|Z)}\,\Phi^{-1}(1-\varepsilon_{\rm sw}) + o(\sqrt{n}),\,\textsf{V}(\tilde{X}|Z)
  :=
  \Var\!\left[-\log_2 P_{\tilde{X}|Z}(\tilde{X}|Z)\right].$

It remains to upper bound the first-order term $H(\tilde{X}|Z)=H(U_{\mathcal{A}_{\rm SI}^c}\,|\,Y^n,U_{\mathcal{A}_{\rm SI}})$
by the sum on the RHS of \eqref{eq:hhat-lemma-b}.

Let $\mathcal{A}_{\rm SI}^c=\{i_1<i_2<\cdots<i_m\}$.
By the chain rule, we have
\begin{align}
  H(U_{\mathcal{A}_{\rm SI}^c}\,|\,Y^n,U_{\mathcal{A}_{\rm SI}})
  &=
  \sum_{t=1}^m
  H\!\left(
    U_{i_t}\,\middle|\,
    Y^n, U_{\mathcal{A}_{\rm SI}}, U_{i_1},\ldots,U_{i_{t-1}}
  \right).
  \label{eq:chain_rule_subset}
\end{align}

In the following, we split all past bits $U^{i_t-1}$
and future side-information bits with indices $\ge i_t$. By such a way, we can drop the additional future bits and upper bound
the term by $H(U_{i_t}\,|\,Y^n,U^{i_t-1})$.

By definition of $\mathcal{A}_{\rm SI}^c$, it is clear that
\begin{align}\label{EQ_i_set_A_and_i-1_set}
    \{i_1,\ldots,i_{t-1}\}=\mathcal{A}_{\rm SI}^c\cap[i_t-1].    
\end{align}

Then we can derive the following
\begin{align}
\mathcal{A}_{\rm SI}\cup\{i_1,\ldots,i_{t-1}\}
&\overset{(a)}=\mathcal{A}_{\rm SI}\cup\bigl(\mathcal{A}_{\rm SI}^c\cap[i_t-1]\bigr)\\
&\overset{(b)}=\bigl(\mathcal{A}_{\rm SI}\cup\mathcal{A}_{\rm SI}^c\bigr)\cap\bigl(\mathcal{A}_{\rm SI}\cup[i_t-1]\bigr)\\
&\overset{(c)}=\mathcal J_0\cap\bigl(\mathcal{A}_{\rm SI}\cup[i_t-1]\bigr)\\
&\overset{(d)}=\bigl(\mathcal J_0\cap[i_t-1]\bigr)\ \cup\ \mathcal{A}_{\rm SI}\\
&\overset{(e)}=\bigl(\mathcal J_0\cap[i_t-1]\bigr)\ \cup\ \bigl(\mathcal{A}_{\rm SI}\cap\{i_t,\ldots,n\}\bigr)\ \cup\ \bigl(\mathcal{A}_{\rm SI}\cap[i_t-1]\bigr)\\
&\overset{(f)}=\bigl(\mathcal J_0\cap[i_t-1]\bigr)\ \cup\ \bigl(\mathcal{A}_{\rm SI}\cap\{i_t,\ldots,n\}\bigr),\label{EQ_AUI_IUA}
\end{align}
where (a) is due to \eqref{EQ_i_set_A_and_i-1_set}, (b) is due to $\mathcal A\cup(\mathcal B\cap \mathcal C)=(\mathcal A\cup \mathcal B)\cap(\mathcal A\cup \mathcal C)$,
(c) uses $\mathcal{A}_{\rm SI}\cup\mathcal{A}_{\rm SI}^c=\mathcal J_0$,
(d) uses $\mathcal{A}_{\rm SI}\subseteq \mathcal J_0$,
(e) is by splitting $\mathcal{A}_{\rm SI}$ into two subsets,
(f) uses $\mathcal{A}_{\rm SI}\cap[i_t-1]\subseteq \mathcal J_0\cap[i_t-1]$.

From \eqref{EQ_AUI_IUA}, we have
\begin{align}
  \{ U_{\mathcal{A}_{\rm SI}},U_{i_1},\ldots,U_{i_{t-1}}\}
  = \{U_{\mathcal J_0\cap[i_t-1]},\,U_{\mathcal{A}_{\rm SI}\cap\{i_t,\ldots,n\}}\}.  
\end{align}
Since the indices in $[i_t-1]\setminus \mathcal J_0$ are frozen in the protocol, conditioning on $U_{\mathcal J_0\cap[i_t-1]}$
is equivalent to conditioning on $U^{i_t-1}$. Hence
\begin{align}
  \{ U_{\mathcal{A}_{\rm SI}},U_{i_1},\ldots,U_{i_{t-1}}\}
  = \{U^{i_t-1},\,U_{\mathcal{A}_{\rm SI}\cap\{i_t,\ldots,n\}}\}.  
\end{align}
  
Consequently, we can derive 
\begin{align}
  H\!\left(
    U_{i_t}\,\middle|\,
    Y^n, U_{\mathcal{A}_{\rm SI}}, U_{i_1},\ldots,U_{i_{t-1}}
  \right)
  &=
  H\!\left(
    U_{i_t}\,\middle|\,
    Y^n, U^{i_t-1}, U_{\mathcal{A}_{\rm SI}\cap\{i_t,\ldots,n\}}
  \right) \notag\\
  &\le
  H\!\left(U_{i_t}\,\middle|\,Y^n,U^{i_t-1}\right).
  \label{eq:cond_reduce_for_upper_bound}
\end{align}

Combining \eqref{eq:chain_rule_subset} and \eqref{eq:cond_reduce_for_upper_bound}, we have
\[
  H(U_{\mathcal{A}_{\rm SI}^c}\,|\,Y^n,U_{\mathcal{A}_{\rm SI}})
  \leq \sum_{i\in\mathcal{A}_{\rm SI}^c} H(U_i\,|\,Y^n,U^{i-1}).
\]
By definition of the polarized bit-channel mutual information, $I_i^{(n)}(\bF):=I(U_i;\,Y^n,U^{i-1})$,
and since $U_i$ is uniform on $\{0,1\}$, we have
\[
  H(U_i\,|\,Y^n,U^{i-1})
  =
  H(U_i) - I(U_i;\,Y^n,U^{i-1})
  =
  1 - I_i^{(n)}(\bF).
\]
Therefore,
\[
   H(U_{\mathcal{A}_{\rm SI}^c}\,|\,Y^n,U_{\mathcal{A}_{\rm SI}})
  \leq \sum_{i\in\mathcal{A}_{\rm SI}^c}\bigl(1-I_i^{(n)}(\bF)\bigr).
\]
Thus, if $\ell_{\rm SWC}$ satisfies
\[
  \ell_{\rm SWC}
  \;\ge\;
  \sum_{i\in\mathcal{A}_{\rm SI}^c}\bigl(1-I_i^{(n)}(\bF)\bigr)
  \;+\;
  \beta_n(\varepsilon_{\rm sw}),
\]
then it also satisfies $\ell_{\rm SWC}\ge H(\tilde{X}|Z)+\beta_n(\varepsilon_{\rm sw})$ by \eqref{eq:ell_ge_Hplusbeta}.
Hence the Slepian--Wolf encoder guaranteed by Lemma~\ref{lem:sw_continuous_SI_second_order} exists, and we complete the proof.
\end{proof}

\section{Proof of Lemma\ref{lem:sw_continuous_SI_second_order}}\label{APP_proof_sw_continuous_SI_second_order}
\begin{proof}
Although the main result in \cite{Hayashi_SWC_Entropy20} is written for discrete alphabets, \cite[Remark~1]{Hayashi_SWC_Entropy20} explains how to extend the entropy-based quantities to the case where $X$ is discrete and $Y$ may be continuous. In the following, we show the derivation for this extension. Let $X$ take values in a finite alphabet $\mathcal{X}$, and let $Y$ take values in a continuous alphabet $\mathcal{Y}$ equipped with a reference measure $\mu$. Assume $P_Y\ll \mu$ and define the Radon--Nikodym derivative $p_Y:=\frac{\mathrm{d}P_Y}{\mathrm{d}\mu}$, i.e., $P_Y(\mathrm{d}y)=p_Y(y)\,\mu(\mathrm{d}y)$. Similarly, assume $Q_Y\ll \mu$ and define $q_Y:=\frac{\mathrm{d}Q_Y}{\mathrm{d}\mu}$, i.e., $Q_Y(\mathrm{d}y)=q_Y(y)\,\mu(\mathrm{d}y)$, so that
$P_{XY}(x,\mathrm{d}y)=P_{X|Y}(x|y)\,p_Y(y)\,\mu(\mathrm{d}y)$.
Consider a Slepian--Wolf source code $\Psi=(e,d)$ with encoder
$e:\mathcal{X}\to\{1,\dots,\textsf{M}\}$ and decoder
$d:\{1,\dots,\textsf{M}\}\times\mathcal{Y}\to\mathcal{X}$.
Define the decoding error probability
$P_s[\Psi] := \Pr( X \neq d(e(X),Y))$ and the optimal error at message size $\textsf{M}$ as
$P_s(\textsf{M}) := \inf_{\Psi} P_s[\Psi]$ as in \cite[(109)--(112)]{Hayashi_SWC_Entropy20}.
Moreover, let $\overline{P}_s(\textsf{M})$ denote the corresponding error criterion for the hash-based construction in \cite[(113)--(114)]{Hayashi_SWC_Entropy20}, and let $\textsf{M}(n,\varepsilon)$ and
$\overline{\textsf{M}}(n,\varepsilon)$ be the encoder output size under blocklength-$n$ defined in
\cite[Eqs.~(109)--(116)]{Hayashi_SWC_Entropy20}. With this convention, any expression in \cite[Sec.~3.2]{Hayashi_SWC_Entropy20} written as an expectation or
probability under $P_{XY}$ carries over with the same algebra. The only change is that sums over $y$ are
replaced by integrals, i.e., $\sum_{y}$ becomes $\int_{\mathcal{Y}} \mu(\mathrm{d}y)$.
Consequently, the corresponding bounds in \cite[Sec.~3.2]{Hayashi_SWC_Entropy20} remain valid for our setting
(discrete $\mathcal{X}$ and continuous $\mathcal{Y}$).

Recall the following achievability bound from \cite[Lemma~13]{Hayashi_SWC_Entropy20}:
for any message size $\textsf{M}$,
\begin{align}
  \overline{P}_s(\textsf{M})
  \;\le\;
  \inf_{\gamma\ge 0}\Bigl[
    P_{XY}\!\Bigl(\log_2\frac{1}{P_{X|Y}(X|Y)} > \gamma\Bigr)
    + \frac{e^\gamma}{\textsf{M}}
  \Bigr].
  \label{eq:HW_L13}
\end{align}
This statement depends only on the random variable $\log_2 \frac{1}{P_{X|Y}(X|Y)}$ under $P_{XY}$ and
therefore remains valid when $Y$ is continuous by \cite[Remark~1]{Hayashi_SWC_Entropy20}.

Define the conditional information density $  \imath_{X|Y}(x;y) \;:=\; \log_2\frac{1}{P_{X|Y}(x|y)}.$
After applying \eqref{eq:HW_L13} to the $n$-fold i.i.d.\ model, by memorylessness we have:
\[
  -\log_2 P_{X^n|Y^n}(X^n|Y^n)
  =
  \sum_{i=1}^n \imath_{X|Y}(X_i;Y_i).
\]
Let $S_n := \sum_{i=1}^n \imath_{X|Y}(X_i;Y_i)$.  By the \gls{clt} under finite second moment assumption, we have the convergence in distribution
\[
  \frac{S_n - nH(X|Y)}{\sqrt{n\textsf{V}(X|Y)}} \xrightarrow{d} \mathcal{N}(0,1),
\]
where $\textsf{V}(X|Y):=\Var\!\bigl[\log_2\frac{1}{P_{X|Y}(X|Y)}\bigr]$ is the variance of information density
defined in \cite[(9)]{Hayashi_SWC_Entropy20}.

Fix $0<\varepsilon<1$ and set $  \textsf{R} := \sqrt{\textsf{V}(X|Y)}\,\Phi^{-1}(1-\varepsilon).$
Choose
\[
  \textsf{M} := \exp\!\bigl(nH(X|Y)+\sqrt{n}\,\textsf{R}\bigr),
  \mbox{ and }
  \gamma := nH(X|Y)+\sqrt{n}\,\textsf{R} - n^{1/4}
\]
as in \cite[Proof of Theorem~11]{Hayashi_SWC_Entropy20}. Then we have:
\begin{align}
  \Pr(S_n>\gamma)
  &=
  \Pr\!\left(
      \frac{S_n-nH(X|Y)}{\sqrt{n\,\textsf{V}(X|Y)}}
      >
      \frac{\sqrt{n}\,\textsf{R}-n^{1/4}}{\sqrt{n\,\textsf{V}(X|Y)}}
    \right)
  \to \varepsilon
  \label{EQ_Pr_S_ge_gamma_approach_epsilon}
\end{align}
by the \gls{clt} and the choice of $\textsf{R}$.
Moreover, the penalty term in \eqref{eq:HW_L13} satisfies
\begin{align}
  \frac{e^\gamma}{\textsf{M}}
  \;=\;
  \frac{\exp\!\bigl(nH(X|Y)+\sqrt{n}\,\textsf{R}-n^{1/4}\bigr)}
       {\exp\!\bigl(nH(X|Y)+\sqrt{n}\,\textsf{R}\bigr)}
  \;=\;
  \exp(-n^{1/4})
  \;\to\;
  0.
  \label{EQ_penalty_to_zero}
\end{align}

Substitute the above choice of $\gamma$ into \eqref{eq:HW_L13}, we have:
\begin{align}
  \overline{P}_s^{(n)}(\textsf{M})
  &\le
  \Pr\!\Bigl(\log_2\frac{1}{P_{X^n|Y^n}(X^n|Y^n)}>\gamma\Bigr)
  +\frac{e^\gamma}{\textsf{M}}=
  \Pr(S_n>\gamma)
  +\frac{e^\gamma}{\textsf{M}}.
  \label{EQ_substitute_gamma_into_HW13}
\end{align}
% Taking $\limsup_{n\to\infty}$ on both sides and using
% \eqref{EQ_Pr_S_ge_gamma_approach_epsilon} and \eqref{EQ_penalty_to_zero} yields
% \begin{align}
%   \limsup_{n\to\infty}\overline{P}_s^{(n)}(\textsf{M})
%   \;\le\;
%   \varepsilon.
%   \label{EQ_limsup_Pbar_le_eps}
% \end{align}

By definition of $\overline{\textsf{M}}(n,\varepsilon)$ (cf.\ \cite[Eqs.~(109)--(116)]{Hayashi_SWC_Entropy20}),
\eqref{EQ_substitute_gamma_into_HW13} implies that for any fixed $\delta>0$, there exists $n_0(\delta)$ such that
for all $n\ge n_0(\delta)$,
\[
  \overline{P}_s^{(n)}(\textsf{M}) \le \varepsilon+\delta,
  \qquad
  \text{with }\ \log \textsf{M}=nH(X|Y)+\sqrt{n}\,\textsf{R}.
\]
Equivalently, for all sufficiently large $n$, a message size $\textsf{M}=\exp(nH(X|Y)+\sqrt{n}\,\textsf{R})$ is achievable up to an $o(\sqrt{n})$ gap in the exponent. Hence, we have
\begin{align}\label{EQ_bar_M_UB}
  \log_2 \overline{\textsf{M}}(n,\varepsilon)  \le   nH(X|Y)+\sqrt{n}\textsf{R} + o(\sqrt{n}).
\end{align}

A converse is given in \cite[Lemma~18]{Hayashi_SWC_Entropy20}: for any $Q_Y\in\mathcal{P}(\mathcal{Y})$,
\begin{align}
  P_s(\textsf{M})
  \;\ge\;
  \sup_{\gamma\ge 0}\Bigl[
    P_{XY}\!\Bigl(\log_2\frac{Q_Y(Y)}{P_{XY}(X,Y)} > \gamma\Bigr)
    - \frac{\textsf{M}}{e^\gamma}
  \Bigr].
  \label{eq:HW_L18}
\end{align}
Choosing $Q_Y=P_Y$ and applying \eqref{eq:HW_L18} to the $n$-fold i.i.d.\ model, the random term inside the probability becomes
$  \log_2\frac{Q_{Y^n}(Y^n)}{P_{X^nY^n}(X^n,Y^n)}  = S_n$. In other words, the converse bound
\eqref{eq:HW_L18} depends on the same normalized sum $S_n$ that appeared in the achievability part.
Again, validity for continuous $Y$ is ensured by \cite[Remark~1]{Hayashi_SWC_Entropy20}.

Now take the same $\textsf{M}=\exp(nH(X|Y)+\sqrt{n}\textsf{R})$ as the achievability part, but choose
$\gamma := nH(X|Y)+\sqrt{n}\textsf{R} + n^{1/4}$ as in \cite[Proof of Theorem~11]{Hayashi_SWC_Entropy20}.
Then the same CLT argument used to justify \eqref{EQ_Pr_S_ge_gamma_approach_epsilon} yields $\Pr(S_n>\gamma)\to\varepsilon$, while the penalty term satisfies $ \frac{\textsf{M}}{e^\gamma}=\exp(-n^{1/4})\to 0.$
Substituting these into \eqref{eq:HW_L18} gives $\liminf_{n\to\infty} P_s^{(n)}(\textsf{M})\ge \varepsilon$.
As in \cite[Proof of Theorem~11]{Hayashi_SWC_Entropy20}, by taking $\textsf{R}$ arbitrarily close to $\sqrt{\textsf{V}(X|Y)}\,\Phi^{-1}(1-\varepsilon)$ from below, we obtain that
for all sufficiently large $n$, $P_s^{(n)}(\textsf{M})>\varepsilon$. By the definition of $\textsf{M}(n,\varepsilon)$ in
\cite[Eqs.~(115)--(116)]{Hayashi_SWC_Entropy20}, for all sufficiently large $n$, we have
\begin{align}\label{EQ_SWC_FBL_converse}
  \log_2 \textsf{M}(n,\varepsilon) \ge nH(X|Y)+\sqrt{n}\textsf{R} + o(\sqrt{n}).
\end{align}
  After matching \eqref{EQ_bar_M_UB} and \eqref{EQ_SWC_FBL_converse}, we complete the proof.
\end{proof}

%\input{chapters/Appendix2}

% \input{chapters/99_template}

% If you want to *print* the list of acronyms at the end:
% \printglossary[type=\acronymtype]

\bibliographystyle{IEEEtran}
\bibliography{ref2}

% \appendix
% \input{chapters/App1}

\end{document}